\documentclass[opre, nonblindrev]{informs4} % current default for manuscript submission

\OneAndAHalfSpacedXI % current default line spacing

\usepackage{endnotes}
\usepackage{diagbox}
\usepackage{tikz}
\usetikzlibrary{shapes.geometric} % For geometric shapes like stars
\usetikzlibrary{calc}

\usepackage{amsmath}
\usetikzlibrary{decorations.pathreplacing}
\usetikzlibrary{arrows.meta}
\allowdisplaybreaks
\usepackage{dsfont, pifont, mathrsfs}
\usepackage{multirow}
\usepackage{setspace}
\usepackage[hidelinks, colorlinks=true, citecolor=blue]{hyperref}
\let\footnote=\endnote

\usepackage{natbib}
 \bibpunct[, ]{(}{)}{,}{a}{}{,}%
 \def\bibfont{\small}%
\TheoremsNumberedThrough     % Preferred (Theorem 1, Lemma 1, Theorem 2)
\ECRepeatTheorems

\EquationsNumberedThrough    % Default: (1), (2), ...
\newcommand{\mbP}{\mathbb P}
\newcommand{\mbE}{\mathbb E}

\newcommand{\prn}[1]{\left({#1}\right)} % parentheses
\newcommand{\brk}[1]{\left[{#1}\right]} % bracket
\newcommand{\ex}[2]{\mathbb{E}_{#1}\left[#2\right]}
\newcommand{\WT}{\W^{(T)}}
\newcommand{\quadd}{\quad\ }

\newcommand{\cD}{\mathcal{D}}

\newcommand{\E}{\mathbb{E}}

\renewcommand{\P}{\mathbb{P}}

\newcommand{\real}{\mathbb{R}}
\newcommand{\W}{\mathcal W}

\newcommand{\Buffer}{\text{\rm Buffer}}
\newcommand{\rmax}{r_{\text{\rm max}}}

\newcommand{\st}{\text{s.t.}}

\newcommand{\pwa}{\textit{PWA}}
\newcommand{\pwr}{\textit{PWR}}

\newcommand{\low}{\textit{low}}
\newcommand{\midd}{\textit{mid}}
\newcommand{\high}{\textit{high}}

\newcommand{\DP}{\textbf{DP}}

\newcommand{\MLB}{\textbf{MLB}}
\newcommand{\SG}{\textbf{SG}}
\newcommand{\FR}{\textbf{FR}}
\newcommand{\IRT}{\textbf{IRT}}
\newcommand{\FRT}{\textbf{FRT}}
\newcommand{\Bayes}{\textbf{Bayes}}

\newcommand{\HOany}{\textbf{HO\textsubscript{any}}}
\newcommand{\HOanyL}{\textbf{HO\textsubscript{any,L}}}
\newcommand{\HOfix}{\textbf{HO\textsubscript{fix}}}

\newcommand{\DLP}{\text{\textbf{DLP}}}

\newcommand{\mix}{\text{\textbf{MIX}}}

\newcommand{\pow}{f}

\definecolor{arc}{RGB}{128,0,128}

\newcommand*{\circled}[1]{\lower.7ex\hbox{\tikz\draw (0pt, 0pt)%
    circle (.5em) node {\makebox[1em][c]{\small #1}};}}
\usepackage{algorithm}
\usepackage{algorithmicx}
\usepackage{algpseudocode}
\algnewcommand{\algorithmicand}{\textbf{and }}
\algnewcommand{\algorithmicor}{\textbf{or }}
\algnewcommand{\OR}{\algorithmicor}
\algnewcommand{\AND}{\algorithmicand}
\makeatletter
\newenvironment{breakablealgorithm}
  {% \begin{breakablealgorithm}
  \begin{center}
     \refstepcounter{algorithm}% New algorithm
     \hrule height.8pt depth0pt \kern2pt% \@fs@pre for \@fs@ruled
     \renewcommand{\caption}[2][\relax]{% Make a new \caption
      {\raggedright\textbf{\ALG@name~\thealgorithm} ##2\par}%
      \ifx\relax##1\relax % #1 is \relax
         \addcontentsline{loa}{algorithm}{\protect\numberline{\thealgorithm}##2}%
      \else % #1 is not \relax
         \addcontentsline{loa}{algorithm}{\protect\numberline{\thealgorithm}##1}%
      \fi
      \kern2pt\hrule\kern2pt
     }
  }{% \end{breakablealgorithm}
     \kern2pt\hrule\relax% \@fs@post for \@fs@ruled
  \end{center}
  }

\begin{document}
%%%%%%%%%%%%%%%%

%\setcounter{page}{1}
%\DOI{10.1287/xxxx.0000.0000}%

\RUNAUTHOR{Ao, Chen, Simchi-Levi and Zhu}

\RUNTITLE{Online Resource Allocation with Average Budget Constraints}

% \TITLE{Bundling Information Goods of Decreasing Value}

% \TITLE{Bayesian Online Multiple Testing: \texorpdfstring{\\}{} A Resource Allocation Approach}

\TITLE{Online Resource Allocation with Average Budget Constraints}

\ARTICLEAUTHORS{%
\AUTHOR{Ruicheng Ao\textsuperscript{1} \quad\quad Hongyu Chen\textsuperscript{1} \quad\quad David Simchi-Levi\textsuperscript{1,2,3} \quad\quad Feng Zhu\textsuperscript{1}}
\AFF{\textsuperscript{1}Institute for Data, Systems, and Society, Massachusetts Institute of Technology, Cambridge, MA 02139\texorpdfstring{\\}{}\textsuperscript{2}Department of Civil and Environmental Engineering, Massachusetts Institute of Technology, Cambridge, MA 02139\texorpdfstring{\\}{}\textsuperscript{3}Operations Research Center, Massachusetts Institute of Technology, Cambridge, MA 02139\texorpdfstring{\\}{}\EMAIL{\texttt{\{aorc, chenhy, dslevi, fengzhu\}@mit.edu}}} %, \URL{}}
} % end of the block

\ABSTRACT{%
% Enter your abstract
We consider the problem of online resource allocation with average budget constraints. At each time point the decision maker makes an irrevocable decision of whether to accept or reject a request before the next request arrives with the goal to maximize the cumulative rewards. In contrast to existing literature requiring the total resource consumption is below a certain level, we require the average resource consumption per accepted request does not exceed a given threshold. This problem can be casted as an online knapsack problem with exogenous random budget replenishment, and can find applications in various fields such as online anomaly detection, sequential advertising, and per-capita public service providers. We start with general arrival distributions and show that a simple policy achieves a $O(\sqrt{T})$ regret. We complement the result by showing that such a regret growing rate is in general not improvable. We then shift our focus to discrete arrival distributions. We find that many existing re-solving heuristics in the online resource allocation literature, albeit achieve bounded loss in canonical settings, may incur a $\Omega(\sqrt{T})$ or even a $\Omega(T)$ regret. With the observation that canonical policies tend to be too optimistic and over accept arrivals, we propose a novel policy that incorporates budget safety buffers. It turns out that a little more safety can greatly enhance efficiency --- small additional logarithmic buffers suffice to reduce the regret from $\Omega(\sqrt{T})$ or even $\Omega(T)$ to $O(\ln^2 T)$. From a practical perspective, we extend the policy to the scenario with continuous arrival distributions, time-dependent information structures, as well as unknown $T$. We conduct both synthetic experiments and empirical applications on a time series data of New York City taxi passengers to validate the performance of our proposed policies. Our results emphasize how effective policies should be designed to reach a balance between circumventing wrong accept and reducing wrong reject in online resource allocation problems with average budget constraints.
}%

%\KEYWORDS{deterministic inventory theory; infinite linear programming duality;
%  existence of optimal policies; semi-Markov decision process; cyclic schedule}

% \HISTORY{\today.}

\maketitle
%%%%%%%%%%%%%%%%%%%%%%%%%%%%%%%%%%%%%%%%%%%%%%%%%%%%%%%%%%%%%%%%%%%%%%

% Samples of sectioning (and labeling) in OPRE
% NOTE: (1) \section and \subsection do NOT end with a period
%       (2) \subsubsection and lower need end punctuation
%       (3) capitalization is as shown (title style).
%
%\section{Introduction.}\label{intro} %%1.
%\subsection{Duality and the Classical EOQ Problem.}\label{class-EOQ} %% 1.1.
%\subsection{Outline.}\label{outline1} %% 1.2.
%\subsubsection{Cyclic Schedules for the General Deterministic SMDP.}
%  \label{cyclic-schedules} %% 1.2.1
%\section{Problem Description.}\label{problemdescription} %% 2.

% Text of your paper here
\section{Introduction}

Resource allocation problems are fundamental in numerous fields, including operations management, computer science, and economics. In an online environment where each task arrives sequentially, the decision-maker has to make an irrevocable decision of whether to accept or reject the task every time a task arrives. In such a setting, feasibility is commonly enforced through a \emph{total budget} constraint that caps aggregate spending by the end of a planning horizon. This formulation has proved effective in many short-run applications where funds are fixed ex ante and the horizon is well defined. For example, a digital advertiser allocates a preset monthly spend across impressions that arrive stochastically; a retailer procures inventory against a quarterly cash cap; and a cloud team schedules jobs under a fixed pool of credits or reserved instances. In each case, the total-budget view provides a transparent way to trade off near-term rewards against the need to remain within an aggregate resource envelope.

At the same time, there are settings in which it is natural to frame feasibility through a \emph{running average} (possibly weighted) cost rather than a terminal total. In such settings, the decision-maker has to make sure that the average cost for the accepted cases are below a preset threshold at every period. A canonical example is the per-capita public service provision: refugee resettlement agencies operating under the U.S.\ Reception \& Placement program receive a fixed per-person allotment to cover initial services offered to the refugees including employment assistance, family reunion, etc., capitated healthcare arrangements prepay plans a per-member-per-month rate (with prospective risk adjustment) for covered medical services. In such systems, the operational requirement for financial viability would be that the running average cost across accepted cases not exceed the per-capita payment.  Moreover, the total budget would not even be available ex ante since the organizations get paid for every case they accepted.

In other settings where the total budget is large or adjustable, imposing an average budget constraint provides a straightforward control for the overall quality of the accepted task. For example, an online advertiser with ample media funds need not buy every campaign—some will be systematically inefficient. A disciplined rule is to admit campaigns only while the \emph{running average} cost-per-acquisition (CPA) stays below a target (e.g., a fraction of expected customer lifetime value). This maintains portfolio quality in real time without requiring a hard end-of-period spending cap.

Generally speaking, the average budget constraint offers another unique perspective on efficient management of resources and can be seen as a complementary way to the traditional total budget constraint. It has several advantages for resource management. First, average budget thresholds are often easier to calibrate from historical unit costs or policy targets than forecasting a full-horizon aggregate budget.  Second, average constraints are independent of the planning horizon and thus suitable for longer horizon resource management.  Third, by enforcing feasibility at every time point, they can encourage smoother allocations over time. We summarize the comparisons to total budget constraints in Table~\ref{tab:comparison}.

\begin{table}[ht]
    \centering
    \renewcommand{\arraystretch}{1.2}
    \begin{tabular}{p{0.45\linewidth}|p{0.45\linewidth}}
        \multicolumn{1}{c|}{\textbf{Total Budget Constraint}} & \multicolumn{1}{c}{\textbf{Average Budget Constraint}} \\ \hline
        $\bullet$ Input: aggregate budget fixed ex ante. & $\bullet$ Input: average (per-unit) budget threshold. \\
        $\bullet$ Feasibility checked at the \textit{final} period. & $\bullet$ Feasibility maintained at \textit{every} time. \\
        $\bullet$ Depends on horizon length. & $\bullet$ Horizon-independent; can operate indefinitely. \\
        \hline
        $\bullet$ Often used in short-run, fixed-fund settings. & $\bullet$ Natural for flow-based, per-capita settings. \\
        $\bullet$ May concentrate adjustments near horizon boundaries. & $\bullet$ Tends to support smoother allocations over time.
    \end{tabular}
    \caption{Comparison of total vs.\ average budget constraints.}
    \label{tab:comparison}
\end{table}

In this paper, we consider the simplest setting in online resource allocation: the online knapsack problem. In this formulation, there is only a single resource for which each task will consume. At each time step, a task arrives characterized by three parameters: its expected reward, expected cost, and an associated weight. The decision-maker must decide whether to accept or reject the task. Accepting the task incurs the cost and yields the reward; rejecting it results in no cost or reward. The decision-maker is given a threshold for average cost in advance and the objective is to maximize the cumulative reward while ensuring that the (possibly weighted) average cost of accepted tasks remains below this threshold. 

The performance measure we choose is the regret of an algorithm, which measures the cumulative difference in the objective value between a particular algorithm and the optimal online algorithm. Interestingly, under this new average budget constraint framework, standard online resource allocation algorithms typically fail to achieve constant or logarithm regret in time even in the single-resource knapsack setting. To address this issue, we propose a simple yet powerful algorithm based on the idea of maintaining a modest budget buffer, enabling our approach to achieve provable logarithmic regret. This approach could potentially provide more insight on how to balance safety and efficiency in online decision-making.

In the rest of this section, we first illustrate two representative application scenarios of the average budget constraint framework. We then discuss our primary theoretical and methodological contributions, laying the groundwork for more efficient and economically intuitive resource allocation solutions. Finally, we discuss related literature and works.

\subsection{Application of the Average Budget Constraint}
Below, we discuss two main applications of the average budget constraint.

\textbf{Per-capita Public Service Organizations.} These are the organizations that receive a fix payment for every person they serve. For example, \textit{Refugee resettlement} offers a concrete illustration: under the U.S. Reception \& Placement (R\&P) program,\footnote{\url{https://2017-2021.state.gov/refugee-admissions/reception-and-placement/?safe=1}} local agencies receive a one-time per-refugee allocation to cover initial services, despite wide variation in case needs. \textit{Capitated healthcare payment} provides another example. Capitation is a model in which the payer prepays a fixed per-member-per-month amount to a plan or provider for a defined set of services, regardless of the actual utilization of any individual enrollee; rates are set prospectively and often incorporate risk adjustment. Capitation is foundational in U.S. Medicaid managed care (where states pay Managed Care Organizations a capitated amount for covered benefits)\footnote{\url{https://www.macpac.gov/subtopic/medicaid-managed-care-payment}} and in Medicare Advantage (where the Centers for Medicare \& Medicaid Services pays Medicare Advantage plans on a risk-adjusted capitated basis).\footnote{\url{https://www.cms.gov/newsroom/fact-sheets/2025-medicare-advantage-and-part-d-rate-announcement}}  

In both cases, the organizational objective is not profit maximization but to maximize the number of individuals served(refugees admitted, patients covered) while ensuring financial viability. Two distinctive features of this setting make it different from standard budget-constrained problems:
\begin{enumerate}
    \item Agencies receive a fixed per-person payment conditional on providing the service.
    \item Their objective is to serve as many people as possible, not to maximize monetary profit.
\end{enumerate}

Because total funding is endogenous to the number of individuals served, a total budget constraint is not meaningful ex ante. Instead, the relevant constraint is that the running average cost across all accepted cases remains below the per-capita payment. The second point, which is the distinction in the objective, also implies that it can be rational to admit individuals whose costs exceed the average payment, so long as they are balanced by lower-cost cases; purely comparing each case’s cost to the per-capita payment would therefore yield a sub-optimal service policy.

\textbf{Online False Discovery Rate (FDR) Control.} Another motivation for us to consider a knapsack problem with an average budget constraints is {online false discovery rate (FDR) control} in statistics (\cite{benjamini1995controlling}, \cite{efron2001empirical}).  Suppose we are doing an initial screening of a large number of potential algorithms that can improve revenue sequentially. At each time point, we obtain a posterior probability on how likely the algorithm is beneficial for the revenue. Then we must make a decision of whether to take the algorithm into secondary evaluation (which requires more resources) or not. 

In this setup, the goal is not to miss any working algorithms. That is, to take as many algorithms into secondary screening as possible. However, this would take up too many resources and a natural idea is to control the error rate of the selected subgroup. The appropriate error rate is defined as the average posterior probability of each algorithm being inefficient, which is the Bayesian posterior of the FDR \citep{whittemore2007bayesian}.

Thus, this problem is naturally a knapsack problem with average budget constraints. Specifically, at each time period, we obtain the posterior probability of being inefficient for the corresponding algorithm. Our action is whether to take this algorithm into secondary evaluation (equivalent to accept the task) or not. The goal is to maximizing the number of secondary evaluations while keeping the average posterior probability of inefficiency below a designated threshold.

\subsection{Main Contributions}
We discuss our primary theoretical and methodological contributions in the following three points.

1. For general arrival distributions, we propose a simple policy called the \textbf{Static Greedy (SG)} policy. This policy makes use of the solution from a deterministic optimization problem where the random cost is replaced by its expectations. For this policy, we prove that it achieves a $O(\sqrt{T})$ regret relative to the offline counterpart uniformly across a set of distributions. We accompany the result by showing that in general, any online policy will suffer a $\Omega(\sqrt T)$ loss.

\begin{figure}[ht]
\centering
\begin{tikzpicture}[scale=0.6, transform shape]
    % Draw axes
    \draw[thin,->] (4,-10) -- (4,2.7) node[anchor=south] {\large Expected rewards};
    
    \filldraw [black] (4,2) circle (2pt) node[anchor=east,xshift=-6pt] (DLP) {\DLP};

    \filldraw [black] (4,0) circle (2pt) node[anchor=east,xshift=-5pt] (HOfix) {\HOfix};

    \filldraw [black] (4,-2) circle (2pt) node[anchor=east,xshift=-5pt] (HOany) {\HOany};

    \filldraw [black] (4,-3) circle (2pt) node[anchor=east,xshift=-5pt] (DP) {\textbf{DP}};
    
    \filldraw [red] (3.9,-4) rectangle ++(0.2,0.2) node[anchor=west] {};
    \node at (4,-4)[font=\normalsize,anchor=west,xshift=5pt] (MLB) {\MLB};

    \filldraw [black] (4,-3) circle (2pt) node[anchor=east,xshift=-5pt] (DP) {\textbf{DP}};

    \filldraw [black] (4,-5.5) circle (2pt) node[anchor=east,xshift=-5pt] (IRT) {\textbf{IRT}};

    \filldraw [black] (4,-9) circle (2pt) node[anchor=east,xshift=-5pt] (bayes) {\textbf{Bayes}};

    \draw [decorate, decoration={brace, amplitude=10pt,mirror, raise=4pt}]
    ($(DLP.east)+(-1.2cm,0)$) -- ($(HOfix.east)+(-1.2cm,0)$) node[anchor=east,midway, xshift=-30pt] {$\Theta(\sqrt{T})$};
    \node at (0.85,0.45)[font=\normalsize] (A) {[BW20]};

    \draw [decorate, decoration={brace, amplitude=10pt,mirror, raise=4pt}]
    ($(HOfix.east)+(-1.2cm,0)$) -- ($(HOany.east)+(-1.2cm,0)$) node[anchor=east,midway, xshift=-30pt] {0};

    \draw [decorate, decoration={brace, amplitude=10pt,mirror, raise=4pt}]
    ($(HOany.east)+(-1.2cm,0)$) -- ($(IRT.east)+(-1.2cm,0)$) node[anchor=east,midway, xshift=-30pt] {$O(1)$};
    \node at (0.95,-4.25)[font=\normalsize] (A) {[BW20]};

    \draw [decorate, decoration={brace, amplitude=10pt,mirror, raise=4pt}]
    ($(HOany.east)+(-3.5cm,0)$) -- ($(bayes.east)+(-3.5cm,0)$) node[anchor=east,midway, xshift=-30pt] {$O(1)$};
    \node at (-1.25,-6)[font=\normalsize] (A) {[VB21]};
    
    \draw [decorate, decoration={brace, amplitude=10pt, raise=4pt}]
    ($(DLP.east)+(3.5cm,0)$) -- ($(HOfix.east)+(3.5cm,0)$) node[anchor=west,midway, xshift=30pt] {$\Omega(\sqrt{T})$};
    \node at (9.1,0.5)[font=\normalsize] (A) {[Prop \ref{prop:sqrt}]};

    \draw [decorate, decoration={brace, amplitude=10pt, raise=4pt}]
    ($(HOfix.east)+(3.5cm,0)$) -- ($(HOany.east)+(3.5cm,0)$) node[anchor=west,midway, xshift=30pt] {$\Omega(\sqrt{T})$};
    \node at (9.1,-1.5)[font=\normalsize] (A) {[Prop \ref{prop:sqrt}]};
    
    \draw [decorate, decoration={brace, amplitude=10pt, raise=4pt}]
    ($(HOany.east)+(3.5cm,0)$) -- ($(IRT.east)+(3.5cm,0)$) node[anchor=west,midway, xshift=30pt] {$\Omega(\sqrt{T})$};
    \node at (9.1,-4.25)[font=\normalsize] (A) {[Prop \ref{prop:canonical-bad}]};
    
    \draw [decorate, decoration={brace, amplitude=10pt, raise=4pt}]
    ($(HOany.east)+(6.5cm,0)$) -- ($(bayes.east)+(6.5cm,0)$) node[anchor=west,midway, xshift=30pt] {$\Omega(T)$};
    \node at (12.1,-5.95)[font=\normalsize] (A) {[Prop \ref{prop:canonical-bad}]};
    
    \draw [decorate, decoration={brace, amplitude=10pt, raise=4pt}]
    ($(HOany.east)+(1.55cm,0)$) -- ($(MLB.west)+(1.2cm,0)$) node[anchor=west,midway,xshift=30pt] {$\widetilde O(1)$};
    \node at (6.95,-3.5)[font=\normalsize] (A) {[Thm \ref{thm:regret_discrete_general}]};

    \node at (0.5,4)[font=\large] (previous) {\textbf{Total Budget Constraint}};
    \node at (8,4)[font=\large] (we) {\textbf{Average Budget Constraint}};
\end{tikzpicture}
\caption{A summary of results in the discrete case of the initial budget problem (previous work) and the exogenous replenishment problem (this paper). Here, {\DLP} refers to the deterministic linear programming upper bound, {\HOfix} and {\HOany} are two ways of calculating the hindsight optimal upper bound, {\DP} is the dynamic programming policy which is the optimal online policy. The three policies are {\MLB} (proposed in this paper), {\IRT} (proposed in [BW20]), and {\Bayes} (proposed in [VB21]). Here, [BW20] refers to \cite{bumpensanti2020re}, [VB21] refers to \cite{vera2021bayesian}, and $T$ is the time horizon.}
\label{fig:summary}
\end{figure}
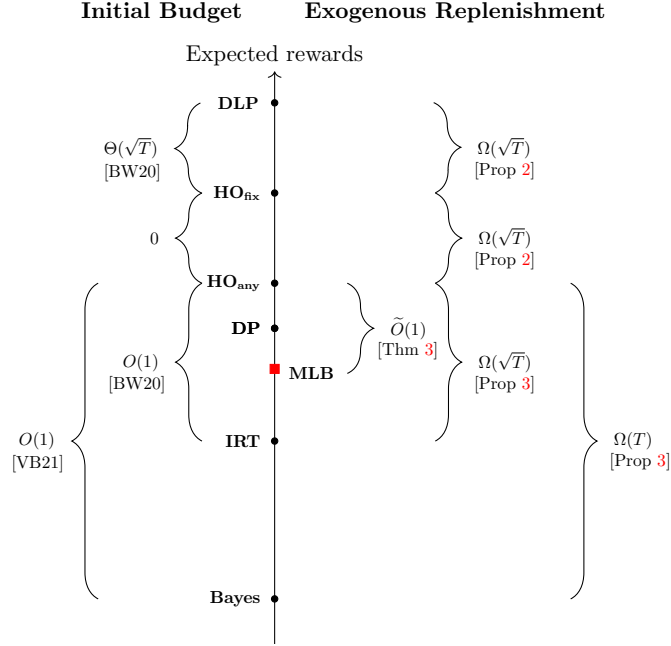

2. For discrete arrival distributions, we find out that many renowned policies in the online resource allocation literature will incur a $\Omega(\sqrt{T})$ (or even $\Omega(T)$) regret in our problem. Our analysis pinpoints the phenomenon that these policies tend to be over optimistic about future replenishment and may over claim discoveries even when the current budget level is low. To mitigate such an issue, we devise a new policy --- the \textbf{Multilevel Logarithmic Buffer ($\MLB$)} policy --- where a discovery is claimed only if the current budget is above a certain safety buffer. The threshold scales logarithmically with the total time horizon and is different for different types of arrivals. We show that a little more safety suffices to greatly enhance efficiency: $\MLB$ achieves a near-optimal performance with the regret growing at a rate of $O(\ln^2T)$. A comparison of our results versus those in the standard initial budget setting is summarized in Figure \ref{fig:summary}.

3. Empirically, we validate the performance of the $\MLB$ policy through synthetic numerical experiments, which show significant improvement over existing policies. Furthermore, we extract the insights from $\MLB$ and construct a new heuristic --- Adaptive Multilevel Logarithmic Buffer for Continuous Distributions ($\MLB$-AC) --- to handle the case for unstable arrivals and continuous reward distributions. We also propose $\MLB$-AC-A, a modified version of $\MLB$-AC where we further relax the requirement of knowing the time horizon in advance. By using data of New York City taxi passengers, we demonstrate the superior performance of both policies compared to other policies in false discovery control.

\subsection{Related Work}
Our work is related to several areas of research in online resource allocation.

\textbf{Online resource allocation with continuous distributions.}
While formulating the problem as an online resource allocation problem, our local FDR control framework resembles the stochastic online knapsack problems with continuous weights \citep{lueker1998average, arlotto2020logarithmic,jiang2020online,jiang2022degeneracy}. Despite near-optimal regret achieved by policies in the classic settings, the theoretical guarantees in these literature may no longer be valid due to the existence of exogenous replenishment. Moreover, these works typically make strong smoothness assumptions on the cost distribution, while our analysis for the general case does not rely on specific properties of the distribution.

\textbf{Online resource allocation with discrete distributions.} When the cost distribution is discrete, a line of research closely related to ours investigates the impact of re-solving (see, e.g., \citealt{reiman2008asymptotically},
\citealt{jasin2012re}, \citealt{jasin2013analysis}, \citealt{ferreira2018online}, \citealt{bumpensanti2020re}, \citealt{banerjee2020uniform}, \citealt{vera2021bayesian}, \citealt{zhu2023assign}) in an asymptotic sense. We refer readers to \cite{balseiro2024survey} for a survey paper on dynamic resource allocation problems. One related area is the two-sided stochastic online matching problems with abandonment and arrivals on both sides \citep{kendall1951some,castro2020matching}. Many of them relied on techniques concerning fluid and diffusion approximation and asymptotic analysis \citep{liu2015diffusion,buke2017fluid}. Another related class of problems is the online reusable resource allocation problem, where the resources used will be available again after certain time periods \citep{levi2010provably,chen2017revenue,owen2018price,rusmevichientong2020dynamic,gong2022online,zhang2022online,simchi2025greedy}. There are two main differences between our problem and the online reusable resource allocation problems. First, the replenishment in our case is exogenous, independent of the decision-making and in line with settings of the sequential testing in practice. Second, existing policies for stochastic online reusable resource allocation problems mostly used competitive ratio as measure despite the assumption of i.i.d. arrivals. The gap between online policies and offline benchmarks typically grows at the $O(\sqrt{T})$ rate as the initial budget scales up to infinity, while in our problem we show that the gap can be up to some log factors under the discrete distribution case.

\textbf{Online knapsack with replenishment.} There has been some but not much work exploring the online knapsack problem with replenishment. Two recent works related to ours are \cite{kumar2022non} and \cite{bernasconi2023bandits}. \cite{kumar2022non} consider an online bandit problem with knapsack, assuming the existence of a null arm that allows the decision-maker to \textit{actively} increase the budget to make sure it never drops to negative. This means the replenishment is controllable, which is different from our setting. They achieve bounded regret compared to an LP relaxation upper bound, under the further assumption that the LP solution is non-degenerate. In the discrete case of our problem, we show that an $\tilde O(1)$ regret is in general not achievable except if we select a tighter benchmark, otherwise the regret has to be $\Omega(\sqrt{T})$ when the LP solution is degenerate. \cite{bernasconi2023bandits} extend \cite{kumar2022non} and obtain $\tilde O(\sqrt T)$ regret under more general settings. Again, the benchmark is a static fluid approximation, a natural generalization of the LP relaxation in \cite{kumar2022non}.

\subsection{Notations}
For integer $n\ge 1,$ we denote $[n]=\{1,2,\dots,n\}$ as the set of integers from $1$ to $n$. For $x\in\real$, denote $\lceil x\rceil$ as the smallest integer not smaller than $x$ and $\lfloor x\rfloor$ as the largest integer not greater than $x$. Denote $x_+=\max\{x, 0\}$. For set $S$, denote $|S|$ as its cardinality. For two functions $f(T)$ and $g(T)$, we use $f(T)=O(g(T))$ if there exists constant $c_1>0$ such that $f(T)\le c_1g(T)$ as $T\to+\infty$ and $f(T)=\Omega(g(T))$ if there exists constant $c_2>0$ such that $f(T)\ge c_2g(T)$ as $T\to +\infty$. We will point out explicitly if the constants above are absolute.

\section{Problem Setup}\label{sec:model}
Suppose we have a decision maker facing a sequence of $T$ tasks and has to decide whether to complete each task in an online fashion. To be more specific, at each time point $t\in[T]$, a task come with three parameters $(r^{(t)}, c^{(t)}, w^{(t)})$ where $r^{(t)}$ is the expected reward upon completion of the task, $c^{(t)}$ is the associated cost for completing the task, and $w^{(t)}$ is a weight factor. The decision maker can make the decision of whether to accept the task, which means paying $c^{(t)}$ cost and get $r^{(t)}$ in return, or to reject the task and wait until the next arrival. Without loss of generality, we assume that $r^{(t)}\ge 0$, meaning that the expected reward is always non-negative.

At the start of the planning horizon, the decision maker is also given a threshold $c_0$, for which he will try to control the average cost against. The goal for the decision maker is to maximize the cumulative reward while also ensuring the weighted average cost for each completed task is less than the threshold $c_0$ at every time point, formally,
\begin{equation}\label{eq:opt_average_cost}
    \begin{aligned}
        \max_{\pi\in\Pi} \quad & \E\left[\sum_{t=1}^Tr^{(t)}X^{(t)}\right]\\
        \text{s.t.} \quad & \frac{\sum_{i=1}^t c^{(i)}X^{(i)}}{\sum_{i=1}^tw^{(i)}X^{(i)}}\leq c_0, \quad \forall t\in[T] \quad (a.s.).
    \end{aligned}
\end{equation}
Here $X^{(t)}\in\{0, 1\}$ is the decision at time $t$ indicating whether a task is accepted ($X^{(t)}=1$) or rejected ($X^{(t)}=0$) and the policy class $\Pi$ denote all non-preempting policies. In the advertisement example, $r^{(t)}$ and $w^{(t)}$ could be the expected number of conversion for the $t$-th ad, $c^{(t)}$ could be the associated cost. As the result, the optimization problem \eqref{eq:opt_average_cost} represents the marketer trying to maximizing the expected number of conversions while controlling for the average cost-per-acquisition rate below $c_0$. In the investment risk control example, $r^{(t)}$ can be the expected net profit for the $t$-th project, $c^{(t)}$ could be the standard error for this profit, and $w^{(t)}$ is the investment cost. Thus, optimization problem \eqref{eq:opt_average_cost} indicated that the investor is trying to maximizing the expected net profit, while controlling for the standard error per invested dollars below a certain threshold.

With a simple transformation, we can re-write the optimization problem as
\begin{equation}\label{eq:knapsack}
    \begin{aligned}
        \max_{\pi\in\Pi} \quad & \E\left[\sum_{t=1}^Tr^{(t)}X^{(t)}\right]\\
        \text{s.t.} \quad & \sum_{i=1}^ta^{(i)}X^{(i)}\leq 0, \quad \forall t\in[T] \quad (a.s.).
    \end{aligned}
\end{equation}
where $a^{(i)}=c^{(i)}-c_0w^{(i)}$ can be seen as the \textit{adjusted cost} for each task. This is the actual cost $c^{(i)}$ minus the expected allocated cost for this task, which is the target average cost $c_0$ times the weight $w^{(i)}$. Thus, if $a^{(i)}<0$, it means the completing this task required less resources than the target cost level, which would bring the decision maker some extra resource that can be used in future tasks. 

Thus, the problem \eqref{eq:knapsack} can be viewed as a knapsack problem as follows. At time zero, we start with zero initial budget. Then at each time period $t\in[T]$, we observe an arrival with cost $a^{(t)}$ (which can be positive or negative) and reward $r^{(t)}$ and decide whether to accept the arrival. If an arrival is accepted, we will pay a cost of $a^{(t)}$ (in case when $a^{(t)}<0$, we gain more budget) and gain a reward of $r^{(t)}$. The goal is to maximize the cumulative reward while having the budget at each time period no less than zero. The only difference here with classical knapsack problem is that the cost $a^{(t)}$ can be either positive or negative. In case of a negative cost, it simply means a replenishment of budget. This kind of exogenous replenishment property brings two unique features to our problem.
\begin{itemize}
    \item The initial budget does not have to be at the same scale as the time horizon $T$ as in the classic knapsack problem. Instead, it can be very small or even zero as in our case because the budget will be replenished along the time horizon.
    \item The budget constraint satisfied at the final period does not imply it is satisfied at an earlier period. When the costs are all positive, if the total cost at the final period does not exceed the initial budget, then automatically the total cost at any time period is within the budget. However, when the weights can be negative, this property no longer holds and a total of $T$ constraints have to be imposed explicitly in the formulation: the cumulative cost does not exceed the budget up to \textit{any} time period $t$.
\end{itemize}

These additional structures will lead to a set of unique theoretical properties that are different from the classical online knapsack problem, as we will demonstrate in the following sections. Before that, we make the following assumption about the costs.
\begin{assumption}\label{asmp:common_dist}
    We assume the adjusted pairs $(a^{(1)},r^{(1)}), \cdots, (a^{(T)},r^{(T)})$ are independent and identically distributed (i.i.d.) random variables sampled from a common distribution $\cD$. Without loss of generality, we also assume $-1\leq a^{(t)}\leq 1$ and $0\leq r^{(t)}\leq 1$.
\end{assumption}
Assumption \ref{asmp:common_dist} further assumes that the costs are i.i.d. distributed, which is true if the tasks are similar in nature. While our policies are primarily designed for the i.i.d. case, we will relax this assumption in Section \ref{sec:discrete} by proposing an extension to the non-stationary case. The policy will also be tested using non-stationary real-world data in Section \ref{sec:experiment}. Assumption \ref{asmp:common_dist} also imposes constraint on the boundedness of the cost and reward, which is standard in online resource allocation (\cite{bumpensanti2020re}, \cite{vera2021bayesian}). We would like to point out the actual value of the boundary does not matter here as one can always scale the cost and reward to make it within $[-1,1]\times[0,1]$.

\subsection{Analysis Framework} \label{sec:continuous}
We first point out that the problem (\ref{eq:knapsack}) can be solved by dynamic programming (\DP). Let $h(t,B)$ denote the expected reward if the decision maker starts at time $t$ and have $B$ as the budget in hand. Then the Bellman equation can be written as
\begin{equation}\label{eq:DP}
    h(t,B) = \E\left[\max\{ h(t+1, B), r^{(t)} + h(t+1, B-a^{(t)}) \}\right]
    \tag{$\DP$}
\end{equation}
with boundary condition
\begin{equation*}
    h(T,B) = 0, \ \forall B\geq 0 ;\qquad h(t,B)=-\infty, \ \forall B < 0.
\end{equation*}
The solution to \eqref{eq:DP} is denoted as $f_\cD^T(\DP):=h(1, 0)$. While the $\eqref{eq:DP}$ problem is easy to solve in small-scale, solving (\ref{eq:DP}) requires discretization or enumeration on the state space of the budget $B$, which can be computationally hard when the time horizon and budget is large. Also, it requires knowing the time horizon $T$ in advance, which can be inaccessible in some applications given the long-term nature for average budget management. Moreover, the $\DP$ method does not offer enough intuition or insight on what properties or structures an effective policy should enjoy. The remainder of the paper is dedicated to proposing intuitive policies that achieve near optimal performance. Moreover, we will try to characterize and understand how the online sequential nature of our problem deviates from its offline counterpart where we have full knowledge to all arrivals in advance.

% In this section, we consider the general adjusted cost distributions and propose a policy that achieves a $O(\sqrt{T})$ performance gap to the best possible policy (the $\DP$ policy), accompanied with a lower bound result that the $O(\sqrt{T})$ regret is in general the best we can hope. In the next section, we will demonstrate that better performance can be achieved under discrete cost distributions. 

To evaluate the efficacy of our proposed model, we consider the following three offline upper bound approximations: the Deterministic Linear Program ({{\DLP}}) approximation, the Hindsight Optimal with fixed-time constraint at the final time period (\textbf{{\HOfix}}), and the Hindsight Optimal with any-time constraints (\textbf{{\HOany}}) at all time periods. 

The \textbf{{\DLP}} approach is a standard benchmark that replaces the sample-path constraint in \eqref{eq:knapsack} with its expectation, providing an upper bound for the objective function. The optimization problem for \textbf{\DLP} is as follows:
\begin{equation}   \label{eq:DLP}
    \begin{aligned}
        \pow_\cD^T(\DLP):= \max_{x: [-1,1]\times[0,1]\to [0,1]} \quad  & T\cdot\E_{(a,r)\sim\cD}[r\cdot x(a,r)]      \\ 
        \text{s.t.} \quad & \E_{(a,r)\sim\cD}[a\cdot x(a,r)] \le 0.
    \end{aligned}
    \tag{$\DLP$}
\end{equation}
Here, $x(a,r)$ can be interpreted as the expected frequency of accepting a task with adjusted-cost $a$ and reward $r$. Hence, the objective function here is the expected reward and the constraint means that the expected adjusted-cost is negative. 

The \textbf{{\HOfix}} approximation is designed as an oracle solution that optimally allocates resources with full knowledge of future outcomes subject to a \textit{fixed} constraint in the final period. On a sample path $\W^T=\{r^{(1)}, a^{(1)}, r^{(2)}, a^{(2)}, \dots, r^{(T)}, a^{(T)}\}$, the optimization problem for $\HOfix$ is defined as:
\begin{equation}
    \label{eq:HOfix_discrete}
    \begin{aligned}
        \pow(\HOfix|\W^{T}):=\max_{(X^{(1)},\dots,X^{(t)})\in\{0,1\}^T} & \quad \sum_{t=1}^{T} r^{(t)}X^{(t)} \\
        \text{s.t.} & \quad  \sum_{t=1}^{T}a^{(t)}X^{(t)} \le 0.
    \end{aligned}
  \tag{$\HOfix$}
\end{equation}

A more stringent upper bound compared to $\HOfix$ is to add $T-1$ additional constraints that account for budget control at \textit{any} time period within the decision-making horizon. We call this upper bound \textbf{{\HOany}}. 
For $\HOany$ on sample path $\W^T$, the optimization problem is expressed as:
\begin{equation}
    \label{eq:HOany_discrete}
    \begin{aligned}
        \pow(\HOany|\W^T):=\max_{(X^{(1)},\dots,X^{(t)})\in\{0,1\}^T} & \quad \sum_{t=1}^{T} r^{(t)}X^{(t)} \\
        \text{s.t.} &  \quad \sum_{i=1}^{t}a^{(i)}X^{(i)} \le 0, \quad \forall t \in [T].
    \end{aligned}\tag{\HOany}
\end{equation}

Furthermore, we denote 
\begin{equation}
    \begin{aligned}
        \pow_\cD^T(\HOfix) & = \ex{}{\pow(\HOfix|\W^T)}\\
        \pow_\cD^T(\HOany) & = \ex{}{\pow(\HOany|\W^T)}
    \end{aligned}
\end{equation}
as the total expected number of accepted arrivals given by $\HOfix$ and $\HOany$ over all possible sample paths $\W^T$ where each $a^{(t)}$ is sampled independently from a distribution $\cD$.
The following proposition shows the interrelationships among different offline upper bounds: \(\DLP\), \(\HOfix\), \(\HOany\), \(\DP\) become progressively tighter in a sequential order. The proof is standard and thus omitted.
\begin{proposition} \label{prop:simple}
For any distribution $\cD$ and any $T$, we have
\begin{equation*}
\begin{aligned}
    \pow_\cD^T(\DLP)\ge \pow_\cD^T(\HOfix)\ge \pow_\cD^T(\HOany)\ge\pow_\cD^T(\DP).
\end{aligned}
\end{equation*}
\end{proposition}

\subsection{Static Greedy Policy}
In this section, we propose an intuitive policy called the Static Greedy ($\SG$) policy designed for general cost and reward distributions and also does not necessarily require the prior knowledge of the time horizon $T$. Without loss of generality, we assume that $\P(a^{(t)}=0) = 0$. If in reality, the decision-maker sees certain arrivals with $a^{(t)}=0$, the decision-maker can always accept them with no impact on the decision-making process. 

To start with, consider solving the optimization problem \eqref{eq:DLP}. The optimal solution $x^*$ always enjoys special structures. Concretely, there exists a threshold $\overline{\rho} \geq 0$ such that $x^*(a,r)=1$ if $a/r < \overline{\rho}$ and $x^*(a,r) = 0$ if $a/r > \overline{\rho}$. That is, we first calculate the average cost per reward by $a/r$. Then we accept those task whose average cost is strictly below the threshold $\overline{\rho}$ and reject those that is strictly above the threshold. Meanwhile, as for $x^*(a, r)$ where $a/r=\overline{\rho}$, it is the maximum number $x^*(\overline{\rho})$ in $[0, 1]$ such that
\begin{align*}
    \ex{}{a^{(t)}\cdot\mathds 1\{a^{(t)}/r^{(t)} < \overline{\rho}\}} + a\cdot x^*(\overline{\rho})\cdot\P(a^{(t)}/r^{(t)}=\overline{\rho}) \leq 0.
\end{align*}
Upon obtaining the solution $x^*(a,r)$, we can interpret $x^*(a,r)$ as the probability to accept an arrival with cost-reward pair $(a, r)$. In particular, we should always accept the arrival when its unit cost $a/r$ is strictly lower than $\overline{\rho}$ and reject the arrival when its unit cost is strictly higher than $\overline{\rho}$. As for those arrivals with unit cost exactly equal to $\overline{\rho}$, we accept the arrival with probability $x^*(\overline{\rho})$. This is intuitive since it is always beneficial to accept arrivals with lower unit cost. If the distribution of the cost-reward pair is continuous, then the optimal solution $x^*(a,r)$ is either $0$ or $1$. If the distribution has discrete parts, it can be the case that a fraction of tasks with unit cost $\overline{\rho}$ should be accepted.

The above discussion motivates the Static Greedy ($\SG$) policy described in Algorithm \ref{alg:static}. Note that to ensure the budget constraint, we can only accept the $t$-the arrival when its $\alpha$-cost does not exceed the budget.

\bigskip
\begin{breakablealgorithm}
\caption{Static Greedy ($\SG$)}
\label{alg:static}
\begin{algorithmic}[1]
    \Require The distribution of cost-reward pair $\mathcal{D}$, time horizon $T$.
    \State Obtain the optimal solution to \eqref{eq:DLP} as $\{x^*(a,r)\}$. Let the boundary cost be $\overline{\rho}$. Let the initial budget $B^{(1)}=0$. Denote $x^*:=x^*(\overline{\rho})$
    \For{$t=1, 2, \cdots$}
        \If{$a^{(t)}/r^{(t)}=\overline{\rho}$}
            \State Sample an independent uniform random variable $u^{(t)}$.
        \EndIf
        \State Accept the arrival if and only if one of the followings are satisfied:
        \begin{equation}
            \left\{
            \begin{matrix}
                & a^{(t)} \leq 0. & & \\
                & 0 < a^{(t)}/r^{(t)} < \overline{\rho}, & & B^{(t)} \geq a^{(t)}. \\
                & a^{(t)} /r^{(t)}= \overline{\rho}, & u^{(t)} \leq x^*, & B^{(t)} \geq a^{(t)}.
            \end{matrix}
            \right.
        \end{equation}
        \State $B^{(t+1)} = B^{(t)}-a^{(t)}\mathds 1\{\text{the $t$-th task is accepted}\}$.
    \EndFor
\end{algorithmic}
\end{breakablealgorithm}
\bigskip

Before we demonstrate the theoretical performance of $\SG$, we introduce some additional notations. Let $A^{(t)} = a^{(t)}\left(\mathds 1\{a^{(t)}/r^{(t)} < \overline{\rho}\} + \mathds 1\{a^{(t)}/r^{(t)} = \overline{\rho}, u^{(t)}\leq x^*\}\right)$ be the cost depleted at time period $t$ if we have enough budget. Then we know that $A^{(1)}, \cdots, A^{(T)}$ are i.i.d. random variables with distributions fully determined by $\cD$. $A^{(1)} - \E[A^{(1)}], \cdots, A^{(T)} - \E[A^{(T)}]$ are i.i.d. random variables bounded within $[-1, 1]$. Denote 
\begin{equation*}
\begin{aligned}
    \sigma_\cD:=\ex{}{|A^{(t)}-\E[A^{(t)}]|^2}^{1/2}, \quad\rho_\cD:=\ex{}{|A^{(t)}-\E[A^{(t)}]|^3}^{1/3}, \quad\kappa_\cD:=\rho^6/\sigma^6.
\end{aligned}
\end{equation*}

Theorem \ref{thm:continuous-upper} shows that the gap between $\SG$ and the most relaxed upper bound $\DLP$ grows at the rate of at most $\sqrt{T}$. 

\begin{theorem} \label{thm:continuous-upper}
For any $T$ and any distribution $\cD$ for cost-reward pair, denote $f_{\cD}^T(\DLP)$ and $f_{\cD}^T(\SG)$ as the expected cumulative reward under the $\DLP$ and $\SG$ policy. Then we have
\begin{equation} \label{eq:continuous-upper}
    f_{\cD}^T(\DLP) - f_{\cD}^T(\SG) = O\left(\kappa_\cD\sqrt{T}\right),
\end{equation}
where $O(\cdot)$ is hiding an absolute constant. 
\end{theorem}

Combining Theorem \ref{thm:continuous-upper} and Proposition \ref{prop:simple}, we immediately know that the gap between the $\SG$ policy and the offline any-time benchmark $\HOany$ is at most $O(\sqrt{T})$ uniformly for \textit{any} distribution $\cD$ with bounded $\kappa_\cD$. The next theorem shows that such rate cannot be improvable, even if we choose $\HOany$, the tightest offline upper bound in this paper, as the benchmark.

\begin{theorem} \label{thm:continuous-lower}
There exists a sequence of distributions $\{\cD_T\}_{T\geq 1}$ such that for any $T$:
\begin{equation} \label{eq:continuous-lower}
\begin{aligned}
    f_{\cD_T}^T(\HOany) - f_{\cD_T}^T(\DP) = \Omega\left(\sqrt{T}\right),
\end{aligned}
\end{equation}
where in $\Omega(\cdot)$ we are hiding absolute constants.
\end{theorem}

To prove Theorem \ref{thm:continuous-lower}, we essentially construct a series of \textit{discrete} cost distributions $\{\cD_T\}_{T\geq 1}$ such that for any $T$ we have $\kappa_{\cD_T} = \Omega(1)$. For each task, the reward is one, and the cost consists of only three different values: a negative cost of $-1/5$ with probability $1/2+1\sqrt{T}$; a low cost of $1/5$ with probability $1/2-2\sqrt{T}$; and a high cost of $3/5$ with probability $1/\sqrt{T}$. We first show that it suffices to only consider policies that always accept arrivals with cost $-1/5$ or $1/5$. We then show that the total expected loss incurred by making wrong accept/reject decisions on the arrivals of $3/5$ scales at a rate of $\Omega(\sqrt{T})$. The probability scaling rate of $1/\sqrt{T}$ is critical --- roughly speaking, it makes the budget $B^{(t)}$ lie within the scale of $\Theta(\sqrt{t})$ with high probability for suitably chosen $t$. More proof details of Theorem \ref{thm:continuous-lower} will become clear in the next section after we build more comprehensive tools on analyzing the gap between any online policy and $\HOany$.

\section{Logarithmic Regret within Discrete Cost Distributions} \label{sec:discrete}

In the previous section, we show that the simple $\SG$ policy suffices to achieve $O(\sqrt{T})$ regret loss compared to $\HOany$, and such rate is not improvable in general for any online policy. This motivates us to consider special cases where the regret loss can be further reduced. In this section, we study the scenario where the distribution $\cD$ of the cost-reward pair is discrete and finite. Such case can occur when the task can be classified into different categories. For example, in service industry—such as in hospitality and airlines—where companies categorize customers into distinct clusters. By dividing customers into a limited number of groups, businesses can tailor their policies more effectively, enhancing both service quality and overall revenue. Limiting the number of groups also simplifies the decision-making process, reducing operational complexity.

% In this section, we consider the scenario where the distribution of $\alpha$-cost is discrete and finite. The main motivation behind this setting is that, apart from the posterior being a discrete distribution, the experimenter may actively discretize the posterior null probability to grids like $\{0\%, 5\%, 10\%, \dots, 100\%\}$ and transforms a general distribution into a discrete one. This discretization aims to reduce estimation errors --- since the estimation of posterior probabilities may not be entirely accurate due to empirical prior, data, computation error, etc, rounding the posteriors to discrete scores will potentially reduce the bias generated by the aforementioned error sources. We note that this is analogous to the approach in service industry such as hospitality and airlines where companies prefer applying clustering to different types of customers and divide them into several groups in order to apply different policies to improve the service quality as well as the total revenue --- the number of groups is typically limited in order to reduce the complexity of decision-making.

% This scenario can be practical when (i) the experimenter discretizes the posterior null probability in to grids like $\{0\%, 5\%, 10\%, \dots, 100\%\}$ and transforms the general distribution into a discrete one, or (ii) the summary statistics is essentially discrete. In applications such as inventory control and task scheduling, a discrete distribution corresponds to the situation when the number of types of demand or consumption is finite.

To start with, assume the state space for the cost-reward pairs is
\begin{align*}
    \{(a_{-m},r_{-m}), \ldots, (a_{-1},r_{-1}), (a_0,r_0), (a_1,r_1), \ldots, (a_{n},r_n)\}, 
\end{align*}
where $m \geq 0, n > 0$ and the sequence $\{(a_i,r_i)\}_{i=-m}^n$ satisfies $a_{-m}/r_{-m} \le \cdots \le a_{0}/r_0 \leq 0 \leq a_1/r_1 \leq \cdots < a_{n}/r_n$. For each index $i$, ranging from $-m$ to $n$, the probability that $(a^{(t)},r^{(t)})$ takes the value $(a_i,r_i)$ is denoted by $p_i = \P((a^{(t)},r^{(t)}) = (a_i,r_i))$. Importantly, we assume, without loss of generality, that there is a nonzero probability for $a^{(t)}$ to be greater than $0$. If no $a^{(t)}$ is above $0$, this becomes a trivial setting where the experimenter can accept all the arrivals and the budget constraint is always satisfied. 

In Section \ref{sec:model}, we have stated three offline benchmarks, $\DLP$, $\HOfix$ and $\HOany$, and showcase their relationship in Proposition \ref{prop:simple}. On the same route, we now demonstrate that the gap between any pair of the approximations can be as bad as $\Omega(\sqrt{T})$, even if under a simple discrete distribution.

\begin{proposition}\label{prop:sqrt}
There exists an instance $\cD$ such that
    \begin{equation}
        \left\{
        \begin{aligned}
            &\pow_\cD^T(\DLP)-\pow_\cD^T(\HOfix) = \Omega\prn{\sqrt{T}},\\
            &\pow_\cD^T(\HOfix) -\pow_\cD^T(\HOany) = \Omega\prn{\sqrt{T}}.
        \end{aligned}\right.
    \end{equation}
\end{proposition}

The proof of Proposition \ref{prop:sqrt} is completed using a simple random walk on the real line. Details are left to Appendix \ref{appendix:proof_general}. Proposition \ref{prop:sqrt} elucidates that when employing \( \DLP \) or \( \HOfix \) as benchmarks, one should not anticipate an online policy to attain a regret bound finer than \( O(\sqrt{T}) \). Given this understanding, the subsequent sections of this article will focus on comparing the performance of our policies with \( \HOany \) as the primary reference point.

\subsection{Canonical Re-solving Heuristics May Over Accept} \label{sec:discrete:resolve}

Before introducing our new policy, it is instructive to examine the state-of-the-art benchmark policies in the literature that employ the ``re-solving" technique. This examination is crucial for contextualizing our proposed methodology within the existing body of knowledge.

The ``re-solving'' technique stands as a significant strategy in the field of revenue management. It re-solves the $\DLP$ approximation at a set of specific time points utilizing the information in the past and provides an updated control policy for the future. A general framework for this class of policy adapted to our problem is shown in Algorithm \ref{alg:resolve_framework}. In Algorithm \ref{alg:resolve_framework}, we first specify the re-solving time step $\tau_{(0)}<\tau_{(1)}<\dots<\tau_{(K)}$ and then solve the \eqref{eq:DLP} problem at $\tau_{(u)}$ as 
\[\mathbf{x}^u=\argmax_{(x_{-m},\dots,x_n)}\left\{\sum_{i=-m}^{n}p_ir_ix_i\;\bigg|\;\sum_{i=-m}^{n}p_ia_ix_i\le B^{(\tau_{(u)})}/(T-\tau_{(u)}+1),\st \;0\le x_i\le 1,\forall i\right\}\] 
where $B^{(\tau_{(u)})}$ is the budget at time $\tau_{(u)}$. Here, $x^u_i$ can be interpreted as the accept probability for the type-$i$ arrival. Thus, the optimization problem shown above is to maximize the expected reward while controlling for the expected cost at every time step to be lower than the average budget across the remaining horizon.  In some cases, the actual acceptance probability for type-$i$ arrival in the algorithm is tweaked via a function $g(\cdot)$, i.e., $g(x_i^u)$. Below are $4$ canonical policies from the literature.

\begin{itemize}
    \item {\bf Frequent Re-solving ($\FR$).} \cite{jasin2012re} propose FR which re-solves the DLP problem at every time step and directly uses the solution $x_i$ as the rejection probability for the next hypothesis. That is, $g(x_i) = x_i$.
    \item {\bf Infrequent Re-solving with Thresholding (IRT).} \cite{bumpensanti2020re} propose IRT where the DLP problem is re-solved at a less frequent level. They use $x_i$ as the rejection probability, except that the $x_i$ is truncated to $0$ or $1$ when it's close to $0$ or $1$. They also propose a variant of $\IRT$, called \textbf{Frequent Re-solving with Thresholding} ($\FRT$), where the only difference is that the DLP problem is re-solved at every time step.
    \item {\bf Bayes Selector ($\Bayes$).} \cite{vera2021bayesian} propose a policy called ``Bayes Selector'' which re-solves the DLP problem at every time step  and sets the accept probability equals $g(x_i) = \mathds 1\{x_i\geq 1/2\}$.
\end{itemize}

\bigskip
\begin{breakablealgorithm}
    \caption{The Re-solving Framework}
    \label{alg:resolve_framework}
    \begin{algorithmic}[1] % The number tells where the line numbering should start
        \Require{ Cost-reward pairs $\{(a_i, r_i)\}_{i=-m}^n$, arrival probability $\{p_i\}_{i=-m}^n$, time horizon $T$, decision function $g:[0, 1]\to[0, 1]$.}
        \State{Initialize re-solving time $\tau_{(0)}<\tau_{(1)}<\dots<\tau_{(K)}$.}
        \For{$u=0,1,\dots,K$}
            \State{Re-solve the problem using current accumulated cost,  i.e., set
            \begin{equation*}
            \begin{aligned}
                \mathbf{x}^u=\argmax\left\{\left.\sum_{i=-m}^{n}p_ir_ix_i\right|\sum_{i=-m}^{n}a_ip_ix_i\le B^{(\tau_{(u)})}/(T-\tau_{(u)}+1),\st \;0\le x_i\le 1,\forall i\right\}.
            \end{aligned}
            \end{equation*}
            }
            \For{$t\in[\tau_{(u)},\tau_{(u+1)})$}
            \State Observe the arrival type $i$.
            \If{$B^{(t)}>a_i$}
                \State Accept $a^{(t)}$ with probability $g(x_i^u)$.
            \EndIf
            \State $B^{(t+1)} = B^{(t)} - a^{(t)}\mathds 1\{a^{(t)}\text{ is accepted}\}$.
            \EndFor
        \EndFor
    \end{algorithmic}
\end{breakablealgorithm}
\bigskip

Ideally, by using the most recent information, these policies would enjoy an improvement compared to policies like Algorithm \ref{alg:static} where only information at the start of the process is used. In fact, policies incorporating this technique have been shown to achieve an $O(1)$ regret in network revenue management problems \citep{jasin2012re,bumpensanti2020re,vera2021bayesian}. However, despite their superior performance in network revenue management problems, these policies may encounter limitations in the problem considered here. The following proposition shows that under some instances, the regret of these policies compared to $\HOany$ could be $\Omega(\sqrt{T})$, or even $\Omega(T)$. This is in contrast with \cite{jasin2012re, bumpensanti2020re,vera2021bayesian} in the network revenue management setting where an $O(1)$ regret is provable, suggesting that the any-time constraint for any $t\in[T]$ sets our problem significantly apart from the classical ones.

\begin{proposition}
\label{prop:canonical-bad}
There exists some $\cD$ such that 
\begin{equation} \label{eq:canonical-sqrt}
\begin{aligned}
    f_\cD^T(\HOany)-\max\left\{f_\cD^T(\FR), f_\cD^T(\IRT), f_\cD^T(\FRT), f_\cD^T(\Bayes)\right\}=\Omega(\sqrt{T}).
\end{aligned}       
\end{equation}
Moreover, there exists some $\cD$ such that
\begin{equation} \label{eq:canonical-linear}
\begin{aligned}
    f_\cD^T(\HOany)-f_\cD^T(\Bayes)=\Omega(T).
\end{aligned}
\end{equation}
\end{proposition}

The proof of Proposition \ref{prop:canonical-bad} is reserved for Appendix \ref{appendix:proof_general}. Proposition \ref{prop:canonical-bad} highlights the limitations of policies that directly utilize the solution of $\DLP$ in scenarios involving exogenous replenishment, even when the re-solving technique is employed. The core issue stems from the \textit{optimistic} consideration of future replenishment, leading the policies to \textit{over accept} arrivals even when the current budget level is relatively low. This observation underscores the necessity of implementing a \textit{safety buffer} mechanism to preemptively mitigate such occurrences, ensuring that the allocation strategy does not radically accept too many arrivals in anticipation of potential replenishment.

% We construct two instances separately for \eqref{eq:canonical-sqrt} and \eqref{eq:canonical-linear}. In the first instance, the $\DLP$ solution is \textit{degenerate}, while in the second instance the $\DLP$ solution is \textit{non-degenerate}. In both cases, any of the heuristics is reduced to the greedy policy that always accepts the arrival whenever the budget is able to cover the cost. The core proof idea is that the greedy policy incurs a non-negligible probability of dropping below a constant level at any $t\in[T/4, T/2)$. In the first instance, the probability is $\Omega(1/\sqrt{t})$; in the second instance, the probability is $\Omega(1)$. However, whenever the budget drops below the constant level, we show that with probability of $\Omega(1)$ the decision to accept the arrival at time $t$ is ``wrong'' (the precise meaning will be clear in Section \ref{ssec:analysis}). Summing over $t$ leads to the results.

\subsection{A Little More Safety Greatly Enhances Efficiency} \label{sec:discrete:MLB}

In this section, inspired by our previous finding that canonical re-solving heuristics may over accept arrivals, we present our novel Multilevel Logarithmic Buffer ($\MLB$) policy, detailed in Algorithm \ref{alg:MLB}.

\bigskip
\begin{breakablealgorithm}
\caption{Multilevel Logarithmic Buffer ({\MLB})}
\label{alg:MLB}
\begin{algorithmic}[1] % The number tells where the line numbering should start
\Require{ Cost-reward pairs $(a_{-m}, r_{-m}),\dots,(a_{n}, r_{n})$ arrival probability $p_{-m},\dots,p_{n}$, time horizon $T$.}
\State Initialize threshold coefficients $\{C_i\}_{i>1}$.
\State Compute $i_0 = \max_{i\leq n}\{\sum_{i=-m}^{i_0}p_ia_i< 0\}$.
\For{$t=1,\dots,T$}
    \State Observe the $t$-th arrival type $i$.
    \If{$B^{(t)} < a_i$}
        \State Reject the $t$-th arrival.
    \Else
    \State Accept the $t$-th arrival if and only if one of the followings is satisfied:
    \begin{equation} \label{eq:buffer}
        \left\{
        \begin{matrix}
            & i \leq 1. & B^{(t)} \geq 0. \\
            & 2 \leq i \leq i_0 + 1, & B^{(t)} \geq C_i\ln (T-t+1). \\
            & i \geq i_0 + 2, & B^{(t)} \geq K_i(T-t+1) +C_i\ln(T-t+1).
        \end{matrix}
        \right.
    \end{equation}
    \EndIf
    \State $B^{(t+1)} = B^{(t)}-a^{(t)}\mathds 1\{\text{the $t$-th arrival is accepted}\}$.
\EndFor 
\end{algorithmic}
\end{breakablealgorithm}
\bigskip

$\MLB$ stratifies incoming arrivals into several distinct categories based on their impact on the resource budget. Recalling that \( a_{-m}/r_{-m} \le\dots \le a_0 /r_0< 0 < a_1/r_1\dots \le a_{n}/r_n \). Define
\begin{equation*}
\begin{aligned}
    \Delta_i := \sum_{j=-m}^ip_ja_j.
\end{aligned}
\end{equation*}

The first category consists of arrivals with $a_i\leq 0$, which effectively serve as a ``replenishment" for the current budget. These arrivals are crucial in maintaining the balance of available resources and should always be accepted.

The second category includes ``low-cost" arrivals, characterized by their incremental expectation being lower than the expected replenishment (i.e. $\Delta_i<0$). For each $i$, we employ a carefully chosen value of $C_i\ln(T-t+1)$ as a budget safety buffer to prevent premature resource depletion. The additional $\ln(T-t+1)$ term seeks to balance between circumventing over accept and preventing over reject. The constant $C_i$ is dependent on $\cD$, and should not be too small such that the probability of over accept can be controlled. We would like to note that for the lowest cost type of arrival $a_1$, a buffer is not necessary because always accepting it does no harm.

The third category, referred to as the ``boundary'' arrivals, is identified by the index \( i = i_0+1 \) such that $\Delta_{i_0}<0\leq\Delta_{i_0+1}$, as the incremental expectation of cost up to $i_0$ surpasses the replenishment. The decision rule is also decided by the budget safety buffer. Albeit sharing the same decision structure as that of ``low cost'' arrivals, the analysis becomes more complicated, particularly when $\Delta_{i_0+1}=0$, i.e., the optimal solution to $\DLP$ is degenerate.

The final category includes arrivals deemed as ``high-cost". Acceptance of these arrivals is contingent upon the remaining resources being greater than a buffer that decays linearly with the remaining time length, plus an additional logarithmic buffer. It is noteworthy that although the linear buffer bears some similarity with the principle of the $\Bayes$ policy applied in the online knapsack problem without replenishment in \cite{vera2021bayesian}, as we will show in our choice of $K_i$, our buffers are smaller than those in the $\Bayes$ policy, suggesting that compared to standard heuristics, while our policy behaves more conservative when dealing with low-cost arrivals, it appears less conservative when facing with high-cost arrivals for large $i$. Moreover, the linear buffer for each $i$ is tweaked by a logarithmic buffer specifically applied to manage high-cost arrivals.

We now characterize the performance of our $\MLB$ policy in Theorem \ref{thm:regret_discrete_general}.
\begin{theorem}
\label{thm:regret_discrete_general}
Take
\begin{equation} \label{eq:discrete-C}
    K_i = \frac{\Delta_{i_0+1}+\Delta_i}{2}\ (\forall i > i_0+1), \quad
    C_i = 
    \left\{
    \begin{matrix}
        |\Delta_{i_0-1}|^{-1}\triangleq C_{\low}, & \text{if }1<i \leq i_0. \\
        |\Delta_{i_0-1}|^{-1}+|\Delta_{i_0}|^{-1}\triangleq C_{\midd}, & \text{if }i > i_0.
    \end{matrix}
    \right.
\end{equation}
We have:

\noindent 1. (Non-degenerate) If $\Delta_{i_0+1} > 0$, then
\begin{equation*}
\begin{aligned}
    \indent f_\cD^T(\HOany) - f_\cD^T(\MLB) & = O\prn{\frac{\ln T}{\Delta_{i_0}^2} + \frac{\ln T}{|\Delta_{i_0}\Delta_{i_0+1}|} + \frac{1}{\Delta_{i_0+1}^2} + \frac{\ln T}{|\Delta_{i_0}|p_{i_0+2}a_{i_0+2}} + \frac{\ln T}{(p_{i_0+2}a_{i_0+2})^2}}.
\end{aligned}
\end{equation*}

\noindent 2. (Degenerate) If $\Delta_{i_0+1} = 0$, then
\begin{equation*}
\begin{aligned}
    f_\cD^T(\HOany) - f_\cD^T(\MLB) & = O\prn{\frac{\ln^2 T}{\sigma_\cD^2\Delta_{i_0}^2} + \frac{\ln T}{|\Delta_{i_0}|p_{i_0+2}a_{i_0+2}} + \frac{\ln T}{(p_{i_0+2}a_{i_0+2})^2}}.
\end{aligned}
\end{equation*}
In $O(\cdot)$ we are hiding absolute constants.
\end{theorem}

In Theorem \ref{thm:regret_discrete_general} there are two parts in each of the upper bounds.
\begin{itemize}
    \item The last term can be interpreted as the regret incurred by wrongly accepting or rejecting high-cost arrivals. This part has some analogy to regret upper bounds of canonical re-solving heuristics in standard online knapsack problems (see, e.g., \citealt{bumpensanti2020re}, \citealt{vera2021bayesian}). However, the main difference lies in the fact that our choice of $K_i$ leads to bounds only concerned with the ``boundary'' high-cost arrival of type $i_0+2$, while standard re-solving heuristics such as $\Bayes$, if translated in our setting, yield a choice of $K_i'=\Delta_i-p_ia_i/2\geq K_i$ and leads to a bound of $O(\sum_{i>i_0+1}1/p_i)$ which is related to \textit{all} high-cost arrivals.

    \item The remaining part is the regret incurred by wrongly accepting or rejecting low-cost and ``boundary'' arrivals. To be more precise, under some carefully chosen safety buffer parameters, the $\MLB$ policy achieves an $\tilde O(1)$ regret. In fact, if the $\DLP$ solution is non-degenerate, then the $\MLB$ policy achieves a $O(\ln T)$ regret. Otherwise, $\MLB$ incurs a $O(\ln^2 T)$ regret. In fact, our proof further suggests that beyond the specifically chosen parameters in \eqref{eq:discrete-C}, as long as the buffer parameters are not too small:
    \begin{equation}
    \begin{aligned}
        C_{\low} \geq |\Delta_{i_0-1}|^{-1}, \quad C_{\midd} - C_{\low} \geq |\Delta_{i_0}|^{-1}, 
    \end{aligned}
    \end{equation}
    then Theorem \ref{thm:regret_discrete_general} still holds, though the absolute constant term can be varying according to our choice of buffer values.
\end{itemize}

\subsection{Main Idea of Proof} \label{ssec:analysis}

This section is dedicated to providing the main proof idea behind our main results on both regret lower bounds (Theorem \ref{thm:continuous-lower} and Proposition \ref{prop:canonical-bad}) and upper bounds (Theorem \ref{thm:regret_discrete_general}). 

To handle the heterogeneous cost-reward pairs with discrete distributions, we introduce a new offline benchmark --- {\HOanyL}. Recall that the benchmark {\HOany} in the general-reward setting can be formulated as:
\begin{equation}
    \label{eq:HOany_discrete_general}
    \begin{aligned}
        \pow(\HOany|\W^T):=\max_{(X^{(1)},\dots,X^{(t)})\in\{0,1\}^T} & \quad \sum_{t=1}^{T} r^{(t)}X^{(t)} \\
        \text{s.t.} &  \quad \sum_{i=1}^{t}a^{(i)}X^{(i)} \le 0, \quad \forall t \in [T].
    \end{aligned}\tag{\HOany}
\end{equation}
The benchmark {\HOany} solves an offline \textit{integer} programming, assuming the full knowledge of the sequence of arrivals. {\HOanyL} is the relaxation of $\HOany$, given by relaxing the constraints  $(X^{(1)},\dots,X^{(t)})\in\{0,1\}^T$ to $(X^{(1)},\dots,X^{(t)})\in[0,1]^T$, i.e.
\begin{equation}
    \label{eq:HOanyL_discrete_general}
    \begin{aligned}
        \pow(\HOanyL|\W^T):=\max_{(X^{(1)},\dots,X^{(t)})\in[0,1]^T} & \quad \sum_{t=1}^{T} r^{(t)}X^{(t)} \\
        \text{s.t.} &  \quad \sum_{i=1}^{t}a^{(i)}X^{(i)} \le 0, \quad \forall t \in [T].
    \end{aligned}\tag{\HOany}
\end{equation}
Denote
\begin{align*}
    f_\cD^T(\HOanyL) & =\ex{}{\pow(\HOanyL|\W^T)}
\end{align*}
as the total expected rewards given by $\HOany$ and $\HOanyL$ over all possible sample paths $\W^T$. Note that it holds that $f_\cD^T(\HOanyL)\ge f_\cD^T(\HOany)$. Moreover, the following lemma reveals that, the two benchmarks are equivalent up to a constant order.
\begin{lemma}
    \label{lem:HOanyL}
    For any distribution $\cD$ and any $T$, we have 
    \begin{equation*}
        f_\cD^T(\HOanyL) - f_\cD^T(\HOany) \le \sum_{i=1}^nr_i.
    \end{equation*}
\end{lemma}
By Lemma \ref{lem:HOanyL}, measuring the gap from online policies to benchmark $\HOanyL$ or $\HOany$ will lead to the same order of regret. However, bounding the regret between $\HOanyL$ and $\MLB$ will be easier, since $\HOanyL$ allows ``partial acceptance'', i.e., it allows $0<X^{(t)}<1$. Details are left in Appendix.

We then introduce the definition of a mixed coupling of any online policy $\pi$ and the hindsight optimal policy $\HOanyL$, which shares similar spirits with those appeared in the revenue management literature \citep{jasin2012re,bumpensanti2020re,vera2021bayesian}.
\begin{definition}
    \label{def:mixL}
    For $1\le t\le T-1$, we define $\mix^{(t)}$ as the policy that applying an online policy $\pi$ in time $[1,t]$ and applying the hindsight optimal policy $\HOanyL^{[t+1,T]}$ to the remaining time periods. As an example,  $\mix^{(0)}=\HOanyL^{[1,T]}$ is the policy that applying hindsight optimal throughout the process and $\mix^{(T)}=\pi^{[1,T]}$ is the policy that applying $\pi$ throughout the process.
\end{definition}

Note that here $\mix^{(t)}$ is dependent on $\pi$. We do not explicitly write such dependence for sake of notational simplicity. By definition, it holds that $\pow(\mix^{(t-1)}\mid \W^T)\ge \pow(\mix^{(t)}\mid \W^T),1\le t\le T$. We call \textit{$\pi$ makes a wrong decision at time $t$} if $\pow(\mix^{(t-1)}|\W^T) > \pow(\mix^{(t)}|\W^T)$. That is, a wrong decision happens if following $\pi$ until time $t$ can be inferior to following $\pi$ until $t-1$. We can thus decompose the regret (given any sample path $\W^T$) as follows:

\begin{equation} \label{eq:regret_decomp}
\begin{aligned}
    \pow(\HOanyL|\W^T) - \pow(\pi|\W^T) = \sum_{t=1}^{T}\brk{\pow(\mix^{(t-1)}|\W^T)-\pow(\mix^{(t)}|\W^T)}.
\end{aligned}
\end{equation}
Analyzing the regret is equivalent to bounding each term $\pow(\mix^{(t-1)}|\W^T)-\pow(\mix^{(t)}|\W^T)$ --- the incremental loss caused by making a wrong decision at time $t$ --- and add them up altogether. 

% We now categorize a wrong decision into two types: \textit{wrong accept} and \textit{wrong reject}. We call that \textit{$\pi$ wrongly accepts $a^{(t)}$} if at time $t$ $\pi$ accepts $a^{(t)}$ and $\pow(\mix^{(t-1)}|\W^T) > \pow(\mix^{(t)}|\W^T)$ (or in other words, $\mix^{(t-1)}$ rejects $a^{(t)}$). We also call that \textit{$\pi$ wrongly rejects $a^{(t)}$} if at time $t$ $\pi$ rejects $a^{(t)}$ and $\pow(\mix^{(t-1)}|\W^T) > \pow(\mix^{(t)}|\W^T)$ (or in other words, $\mix^{(t-1)}$ accepts $a^{(t)}$). The following lemma relates the incremental loss with the event of wrong accept/reject.
% \begin{lemma}
% \label{lem:there_is_gap1_general}
% For $1\le t\le T$, it holds that
% \begin{align*}
%     & \quadd \pow(\mix^{(t-1)}|\W^T)-\pow(\mix^{(t)}|\W^T) \\
%     & \le \left(L\wedge(T-t)\rmax\right)\mathds 1\{\pi\text{ wrongly accepts }a^{(t)}\}+\mathds 1\{\pi\text{ wrongly rejects }a^{(t)}\}, 
% \end{align*}
% where $L:= \max_{i\in[n]}(r_n/a_n-r_i/a_i)a_i$.
% \end{lemma}
% Analyzing the regret is equivalent to bounding each term $\pow(\mix^{(t-1)}|\W^T)-\pow(\mix^{(t)}|\W^T)$ --- the incremental loss caused by making a wrong decision at time $t$ --- and add them up altogether. 

We now categorize a wrong decision into two types: \textit{wrong accept} and \textit{wrong reject}. We call that \textit{$\pi$ wrongly accepts $a^{(t)}$} if at time $t$, $\pi$ accepts $a^{(t)}$ and $\pow(\mix^{(t-1)}|\W^T) > \pow(\mix^{(t)}|\W^T)$ (or in other words, $\mix^{(t-1)}$ rejects $a^{(t)}$). We also call that \textit{$\pi$ wrongly rejects $a^{(t)}$} if at time $t$, $\pi$ rejects $a^{(t)}$ and $\pow(\mix^{(t-1)}|\W^T) > \pow(\mix^{(t)}|\W^T)$ (or in other words, $\mix^{(t-1)}$ accepts $a^{(t)}$). The following lemma relates the incremental loss with the event of wrong accept/reject.

\begin{lemma}
\label{lem:there_is_gap1_general}
For $1\le t\le T$, it holds that
\begin{align*}
    & \quadd \pow(\mix^{(t-1)}|\W^T)-\pow(\mix^{(t)}|\W^T) \\
    & \ge \mathds 1\{\pi\text{ wrongly accepts }a^{(t)}\}+\mathds 1\{\pi\text{ wrongly rejects }a^{(t)}\}, \\\\
    & \quadd \pow(\mix^{(t-1)}|\W^T)-\pow(\mix^{(t)}|\W^T) \\
    & \le \left(L\wedge r_{\max}(T-t)\right)\mathds 1\{\pi\text{ wrongly accepts }a^{(t)}\}+\mathds 1\{\pi\text{ wrongly rejects }a^{(t)}\}, 
\end{align*}
where $L:= \max_{i\in[n]}(r_n/a_n-r_i/a_i)a_i$.
\end{lemma}

From Lemma \ref{lem:there_is_gap1_general}, by taking expectation over $\W^T$ we can obtain that
\begin{equation} \label{eq:bound_difference_general}
\begin{aligned}
    & \quadd \P\prn{\pi\text{ wrongly accepts }a^{(t)}} + \P\prn{\pi\text{ wrongly rejects }a^{(t)}} \\
    & \leq \pow_\cD^T(\mix^{(t-1)})-\pow_\cD^T(\mix^{(t)}) \\
    & \leq (L\wedge r_{\max}(T-t))\cdot\P\prn{\pi\text{ wrongly accepts }a^{(t)}} + \P\prn{\pi\text{ wrongly rejects }a^{(t)}}.
\end{aligned}
\end{equation}

\eqref{eq:bound_difference_general} serves as simple but quite powerful inequalities to obtain both regret lower bounds and upper bounds. The first inequality gives guidance to regret lower bounds. It suggests that the performance of a policy $\pi$ is intrinsically imposed by its ability to circumvent the probability of \textit{either wrong accept or wrong reject}.
\begin{itemize}
    \item Theorem \ref{thm:continuous-lower} is proved by showing that there is an intrinsic trade-off between wrong accept and wrong reject --- if we reduce the probability of wrong accept (reject), then the probability of wrong reject (accept) inevitably increases. We show that given $a^{(t)}$ realized as a high cost arrival, the sum of the two probabilities is essentially $\Omega(1)$. Since the probability of a high cost arrival is $1/\sqrt{T}$, the expected regret is at least $\Omega(\sqrt{T})$.

    \item Proposition \ref{prop:canonical-bad} is proved by showing that any one of the listed canonical re-solving heuristics, once reduced to a greedy policy, incurs a nontrivial probability of wrong accept. This again highlights the observation that canonical re-solving heuristics tend to over accept arrivals. Specifically, \eqref{eq:canonical-sqrt} is proved by constructing an instance with the $\DLP$ optimal solution being degenerate and showing that the probability of wrong accept decays slowly at a $\Omega(1/\sqrt{t})$ rate. \eqref{eq:canonical-linear} is proved by constructing an instance with the $\DLP$ optimal solution being non-degenerate and showing that such probability is $\Omega(1)$, thus causing a linear regret. 
\end{itemize}

The second inequality paves the way for regret upper bounds. It suggests that we can obtain a small regret if we manage to control \textit{both wrong accept and wrong reject}. However, controlling the two types of wrong decisions can be quite different: a wrong reject only causes a loss of at most $1$, but a wrong accept can cause much more --- wrongly accepting a high-cost arrival at time $t$ may cause a loss of budget to accept multiple (may be as bad as $T-t$) small-cost arrivals in the future. Let $\pi=\MLB$. The question is: how can we relate the probabilities of making a wrong decision at time $t$ to the budget $B^{(t)}$ and future arrivals after time $t$? It turns out that such probabilities are closely linked to the \textit{maximum} of potentially drifted random walks, as is documented in Lemma \ref{lem:there_is_gap2_general}.

\begin{lemma} \label{lem:there_is_gap2_general}
    For $1\leq t\leq T,$ we have 
    \begin{equation*}
        \begin{aligned}
            & \quadd \P(\MLB\text{ wrongly accepts }a^{(t)}) \\
            & = \sum_{i=1}^n p_i\cdot{{\P\prn{\MLB\text{ wrongly accepts }a^{(t)}|a^{(t)}=a_i}}} \\
            & \leq \sum_{i=1}^{n}p_i\cdot\P\prn{B^{(t)} - a_i < \max_{s\in[t+1,T]}\sum_{l=t+1}^sa^{(l)}\mathds 1\{a^{(r)}/r^{(l)}<a_i/r_i\}, B^{(t)} \geq \Buffer_i^{(t)}}
        \end{aligned}
    \end{equation*}
    and
    \begin{equation*}
        \begin{aligned}
            & \quadd \P(\MLB\text{ wrongly rejects }a^{(t)}) \\
            & = \sum_{i=1}^n p_i\cdot{{\P\prn{\MLB\text{ wrongly rejects }a^{(t)}|a^{(t)}=a_i}}} \\
            & \leq \sum_{i=1}^np_i\cdot\P\prn{B^{(t)}-a_i\geq\max_{s\in[t+1,T]}\sum_{l=t+1}^sa^{(l)}\mathds 1\{a^{(r)}/r^{(l)}\leq a_i/r_i\}, B^{(t)} < \Buffer_i^{(t)}},
        \end{aligned}
    \end{equation*}
    where $\Buffer_i^{(t)}$ is the safety buffer set as in \eqref{eq:buffer}.
\end{lemma}

Lemma \ref{lem:there_is_gap2_general} essentially bridges the two types of wrong online decisions with the sequential structure of the offline sample path $\W^T$. The proof builds on a simple yet powerful observation of the structure of the optimal offline decision (induced by $\HOany$) for any $\W^T$: when we fully know the sample path, it is always feasible and no worse to delay an accept to the future if possible. Briefly speaking, Lemma \ref{lem:there_is_gap2_general} tells us that a wrong accept happens only if there \textit{exists} some $s>t$ such that accepting all the arrivals from $t$ to $s$ with cost \textit{less than} $a^{(t)}$ will \textit{completely deplete} the current budget $B^{(t)}$. Meanwhile, a wrong reject happens only if for \textit{any} $s>t$ accepting all the arrivals from $t$ to $s$ with cost \textit{no greater than} $a^{(t)}$ will \textit{never exceed} the current budget $B^{(t)}$.

With \eqref{eq:regret_decomp}, Lemma \ref{lem:there_is_gap1_general}, \eqref{eq:bound_difference_general}, Lemma \ref{lem:there_is_gap2_general} at hand, the remaining technical steps are completed via probability bounds and stochastic properties of drifted random walks and Lindley processes. Details are left to Appendix \ref{appendix:proof_general}. We would like to note that the motivation behind setting a gap between $C_{\low}$ and $C_{\midd}$ in Theorem \ref{thm:regret_discrete_general} is to control the wrong reject of ``low-cost'' arrivals --- through the gap, the budget can stay around or above $C_{\midd}\ln(T-t+1)$ with high probability, making the probability of the budget staying below $C_{\low}\ln(T-t+1)$ decays exponentially with $t$.

\subsection{Two Practical Heuristics}

Acknowledging the complexities presented by the cost distribution of real-world data, which often (i) is characterized by a continuous distribution, (ii) shows time-dependent structures among different observations, and (iii) the total time horizon $T$ is not known a priori, we introduce two amended versions of the $\MLB$ policy: one is suitable for a continuous distribution and adapts to local information structures, and the other is built on the first one while further relaxing the requirement of knowing $T$ in advance. Note that although in Theorem \ref{thm:continuous-lower} we have shown that in the worst-case it is not possible to improve over the $\Omega(\sqrt{T})$ rate even if we take $\HOany$ as the benchmark, a new heuristic may have the potential to empirically improve over the simple policy $\SG$ by borrowing insights from the design and analysis of $\MLB$:

\begin{itemize}
    \item For very low-cost type arrivals, they can be regarded as $i=1$ in the discrete case. Thus, a greedy policy without safety buffers may work well. 
    \item For other low-cost type arrivals, it almost does no harm to add a logarithmic safety buffer. In particular, our analysis in the discrete case shows that for a low-cost arrival that is distant from the ``boundary'' cost $\overline{\rho}$, a logarithmic buffer can well control the probability of wrong accept. 
    \item For high-cost type arrivals, with the precise knowledge of $T$, our design shows that it is necessary to be less conservative and not wise to always reject them. A safety buffer linearly decreasing with $t$ may empirically help maximize the total number of acceptance, while also control for the average cost.
\end{itemize}

The discussion above leads to the following heuristic $\MLB$-AC.
\bigskip
\begin{breakablealgorithm}
\caption{Adaptive Multilevel Logarithmic Buffer for Continuous Distribution ({\MLB-AC})}
\label{alg:MLB-AC}
\begin{algorithmic}[1] % The number tells where the line numbering should start
\Require{Time horizon $T$. Initial budget $B^{(0)}=0$. Hyper-parameters $d, \underline{\rho}, C_1, C_2$.}
\For{$t=1,2,\dots,d$}
    \State Accept arrival $a^{(t)}$ only if $a^{(t)}\leq 0$.
    \State $B^{(t+1)} = B^{(t)}-a^{(t)}\mathds 1\{a^{(t)}\text{ is accepted}\}$.
\EndFor
\For{$t=d+1,\dots,T$}
    \State /* \texttt{Estimate the threshold $\overline{\rho}^{(t)}$ and gap $\Delta(a^{(t)})$} */
    \State Sort the observations from previous $d$ periods as $a_{(1)}^{(t)}/r_{(1)}^{(t)}\leq a_{(2)}^{(t)}/r_{(2)}^{(t)}\leq\dots\leq a_{(d)}^{(t)}/r_{(d)}^{(t)}$.
    \If{$a_{(1)}^{(t)}/r_{(1)}\geq 0$}
        \State Set $\overline{\rho}^{(t)}=\overline{\rho}^{(t-1)}$.
    \Else
        \State Set $\overline{\rho}^{(t)}=a_{(j^*)}^{(t)}/r_{(j^*)}^{(t)}$ where $j^*=\max\{j:\sum_{i=1}^ja_{(i)}^{(t)}\leq0\}$.
    \EndIf
    \State Calculate $\Delta^{(t)}=\sum_{i=t-d}^{t-1}a^{(i)}\mathds{1}\{a^{(i)}/r^{(i)}<a^{(t)}/r^{(t)}\}/\sum_{i=t-d}^{t-1}\mathds{1}\{a^{(i)}/r^{(i)}<a^{(t)}/r^{(t)}\}$.
    \State /* \texttt{Make accept/reject decision.} */
    \If{$B^{(t)} < a^{(t)}$}
        \State Reject $a^{(t)}$.
    \Else
    \State Accept the arrival if and only if one of the followings is satisfied:
    \begin{equation} \label{eq:buffer-C}
        \left\{
        \begin{matrix}
            & a^{(t)}/r^{(t)}\leq \underline{\rho}, & B^{(t)} \geq 0. \\
            & \underline{\rho}<a^{(t)}/r^{(t)}\leq \overline{\rho}^{(t)}, & B^{(t)} \geq C_1\ln (T-t+1). \\
            & a^{(t)}/r^{(t)}>\overline{\rho}^{(t)}, & B^{(t)} \geq \Delta^{(t)}/2\cdot(T-t+1) + C_2\ln(T-t+1).
        \end{matrix}
        \right.
    \end{equation}
    \EndIf
    \State $B^{(t+1)} = B^{(t)}-a^{(t)}\mathds 1\{a^{(t)}\text{ is accepted}\}$.
\EndFor 
\end{algorithmic}
\end{breakablealgorithm}
\bigskip

In $\MLB$-AC, for a given control level $\alpha$ at any time $t>d$, we divide the real axis into three parts: $(-\infty, \underline{\rho})$, $[\underline{\rho}, \overline{\rho}^{(t)})$ and $[\overline{\rho}^{(t)},+\infty)$, where $\underline{\rho}$ is a hyper-parameter and $\overline{\rho}^{(t)}$ satisfies $\E_{a\sim\cD^{(t)}}[a\cdot\mathds 1\{a<\overline{\rho}^{(t)}\}]=0$. Here, $\cD^{(t)}$ is an empirical cost distribution estimated from the observations in the previous $d$ time periods. We then decide $\overline{\rho}^{(t)}$ in the following way: we first sort the observations from time $[t-d, t-1]$ as $a_{(1)}^t/r_{(1)}^t\leq a_{(2)}^t/r_{(2)}^t\leq\dots\leq a_{(d)}^t/r_{(d)}^t$. Then we take the threshold $\overline{\rho}^{(t)}$ at time $t$ as $a_{(j^*)}^t/r_{(j^*)}^t$ where $j^*=\max\{j:
\sum_{i=1}^ja_{(i)}^t\leq 0\}$. This step is motivated by the potential time-dependent structure in the real-world data setting. For example, in many time series data, each single observation may follow the same known distribution, but there might be time dependence among adjacent observations. Therefore, the threshold is adapted to the local distribution structure by using the most recent observations. Our rule of deciding $\overline{\rho}^{(t)}$ also coincides with that in \cite{gang2023structure}.

We impose different safety buffers for an observation coming from different segments, echoing with that in Algorithm \ref{alg:MLB}. In particular, for $a^{(t)}<\underline{\rho}$, we do not impose any buffer and accept the arrival (or equivalently claim the discovery) whenever the $\alpha$-budget is enough. For $a^{(t)}\in[\underline{\rho},\overline{\rho}^{(t)})$, we add a buffer of $C_1\ln(T-t+1)$, and for $a^{(t)}>\overline{\rho}^{(t)}$, the buffer is $\Delta^{(t)}/2\cdot(T-t+1)+C_2\ln(T-t+1)$, where $\Delta^{(t)}=\E_{a\sim \cD^{(t)}}[a\cdot \mathds 1\{a<a^{(t)}\}]=\sum_{i=t-d}^{t-1}a^{(i)}\mathbf{1}\{a^{(i)}<a^{(t)}\}/\sum_{i=t-d}^{t-1}\mathbf{1}\{a^{(i)}<a^{(t)}\}$.

In the case when $T$ is uncertain, we propose the following modified version of \eqref{eq:buffer-C}:
\begin{equation} \label{eq:buffer-C-}
    \left\{
    \begin{matrix}
        & a^{(t)}\leq \underline{\rho}, & B^{(t)} \geq 0. \\
        & \underline{\rho}<a^{(t)}\leq \overline{\rho}^{(t)}, & B^{(t)} \geq C_1\ln t.
    \end{matrix}
    \right.
\end{equation}
That is, we (i) do not accept any arrival if $a^{(t)}>\overline{\rho}^{(t)}$, which is equivalent to taking $T=+\infty$ in \eqref{eq:buffer-C}, and (ii) accept low-cost (but not too low) arrivals if the budget is beyond some buffer $C_1\ln t$ which can be regarded as an analogy to $C_1\ln(T-t+1)$ in \eqref{eq:buffer-C}. We call the modified policy $\MLB$-AC-A (the letter ``A'' stands for ``any-time'' without knowing $T$).

We will show in Section \ref{sec:experiment} from a real-world data setting that our heuristics $\MLB$-AC and $\MLB$-AC-A can have superior performance compared to many other existing policies.

\section{Experiments} \label{sec:experiment}
\subsection{Synthetic Experiments in the Discrete Case}
In this section, we conduct numerical experiments to demonstrate the performance of the proposed $\MLB$ policy (Algorithm \ref{alg:MLB}) when the incoming cost-reward pair follows a discrete distribution. In particular, we compare it with five policies, namely Frequent Re-solving ($\FR$), Infrequent Re-solving ($\IRT$), Frequent Re-solving with Threshold ($\FRT$), Bayesian Selector ($\Bayes$), and Static Greedy ($\SG$). The $\FR$, $\IRT$, $\FRT$, and $\Bayes$ policies are four existing re-solving heuristics introduced in Section \ref{sec:discrete:resolve}, which have been proven to achieve a constant regret in canonical online resource allocation problems. $\SG$ is Algorithm \ref{alg:static} applied to the discrete case. 

\begin{figure}[ht]
    \centering
    \includegraphics[width=0.48\linewidth]{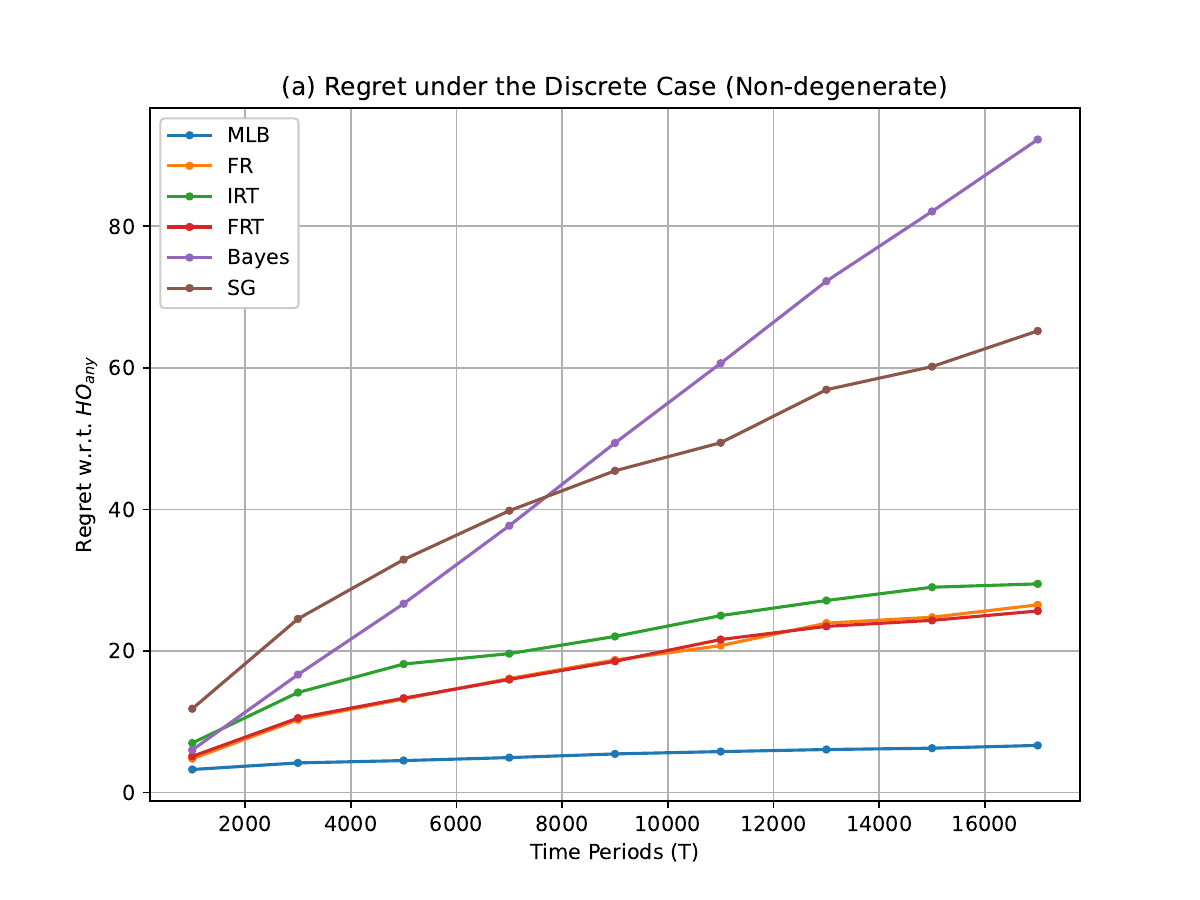}
    \includegraphics[width=0.48\linewidth]{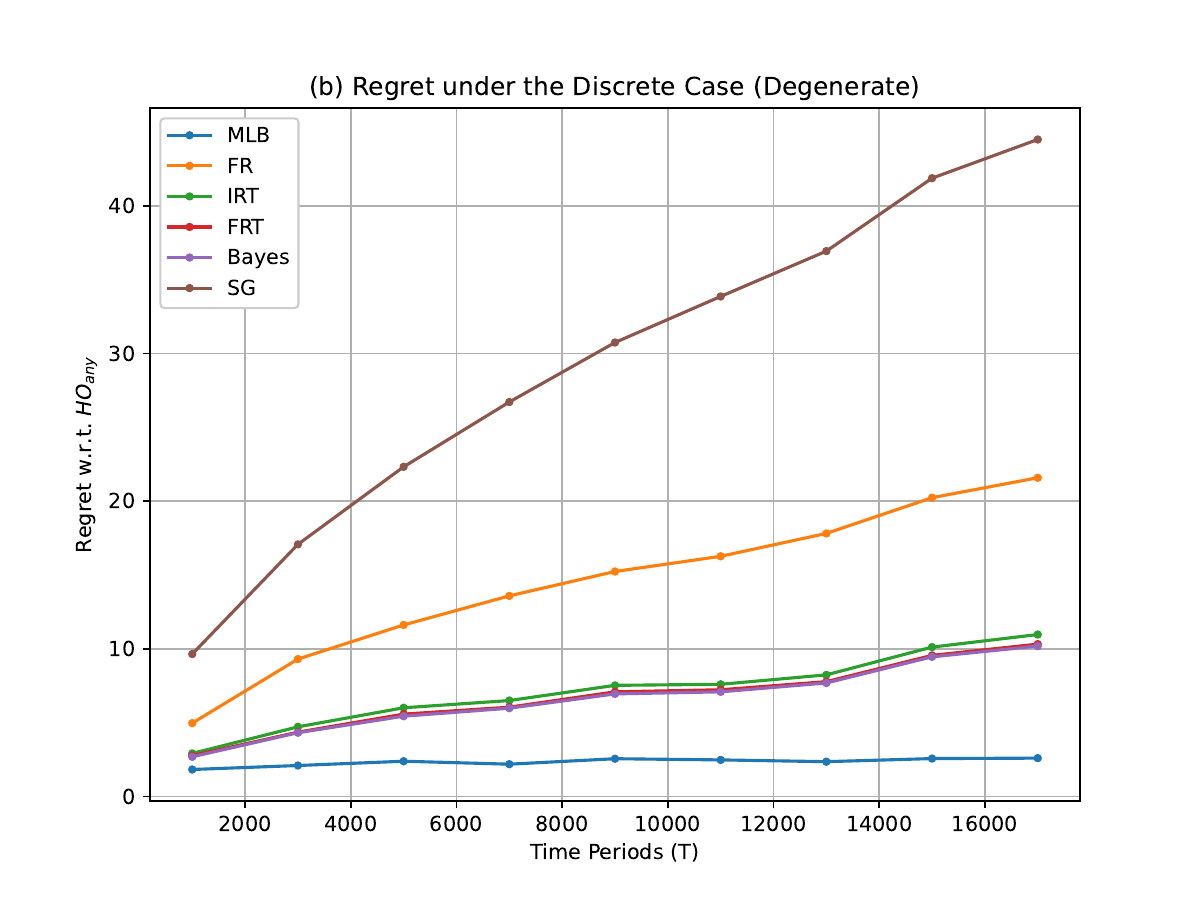}
    \caption{Regret with respect to the \eqref{eq:HOany_discrete} for different policies. For Panel (a), it is the non-degenerate case where the cost takes values in $\{-2, 3, 4\}$ with probability $\{0.6, 0.3, 0.1\}$. Panel (b) is the degenerate case where the the cost takes values in $\{-2, 1, 3, 6, 8\}$ with probability $\{0.5, 0.1, 0.1, 0.1, 0.2\}$. The reward for each arrival is set to be one.}
    \label{fig:discrete_upper}
\end{figure}

We conduct the experiment in two settings. In the first setting, we set the reward to be one and let the cost taking values in $\{-2, 3, 4\}$ with probability $\{0.6, 0.3, 0.1\}$. One can easily scale the cost to let it take values in $[-1,1]$ and we omit that for demonstration purpose. Here, the accumulated cost equals to $\Delta_1=-1.2, \Delta_2=-0.3, \Delta_3=0.1$, which are all non-zero, for which we call the example the \textit{non-degenerate} one. We report the regret of each of the six policy with respect to $\HOany$ averaged across 100 sample paths in Panel (a) of Figure \ref{fig:discrete_upper}. 

In the second experiment, we also let the reward to be one. However, in this case, we let the cost $a^{(t)}$ take values in $\{-2, 1, 3, 6, 8\}$ with probability $\{0.5, 0.1, 0.1, 0.1, 0.2\}$ respectively. Now the cumulative $\alpha$-cost becomes $\Delta_1=-1, \Delta_2=-0.9, \Delta_3=-0.6, \Delta_4=0, \Delta_5=1.6$. Note that now because $\Delta_4=0$, we call the case \textit{degenerate}. The problem in this case is inherently harder than the previous one as it is easier for the policy to over-accept the low-cost arrivals, leaving little budgets for future high-cost arrivals. We demonstrate the regret of each of the five policy with respect to $\HOany$ averaged across 100 sample paths in Panel (b) of Figure \ref{fig:discrete_upper}. Here, the regret of $\Bayes$ and $\FRT$ coincides as they both tend to reject high-cost arrivals more often.

From Figure \ref{fig:discrete_upper}, we make the following observations. (1) The regret of the $\MLB$ policy remains the lowest and does not grow much with the time horizon. This validates Theorem \ref{thm:regret_discrete_general} where we prove the regret grows at a rate of $O(\ln^2T)$. (2) The regret of all other policies grow at a faster rate, mostly at a $\Omega(\sqrt{T})$ rate, but some even grows linearly with respect to the time horizon ($\Bayes$ in the first setting). (3) The performance of some policies can be fragile to the cost distribution. For example, while $\Bayes$ performs well in the second setting, it has linear regret in the first one. Also, even though $\FR$ performs better in the non-degenerate case, it can have poor performance in the degenerate setting.

\subsection{Application to time series anomaly detection}
One practical application of the proposed average budget control is online False Discovery Rate (FDR) Control (\cite{benjamini1995controlling}, \cite{ramdas2017online}, \cite{gang2023structure}). In this section, we illustrate this through an example of real-time anomaly detection. At each time point, the decision-maker observes a new data point in a time series with its probability of being normal being $c^{(t)}$. Then the decision-maker has to make a decision of whether to label it as an abnormal data point ($X_t=1$) or not ($X_t=0$). Here, the control variable is the FDR, which is defined as  average error rate on the data that is deemed as abnormal, i.e., 
\[FDR^{(t)}=\frac{\sum_{i=1}^tX_tc^{(t)}}{\sum_{i=1}^tX_t}.\]
The goal for the decision maker is to identify as many abnormal points as possible while keeping the FDR rate below a preset threshold $\alpha$ at every time period, i.e.
\[ \max_{X_t} \sum_{i=1}^T X_t, , \qquad \text{s.t.}\quad FDR^{(t)}\leq \alpha. \]
This is exactly the formulation of the average budget problem. Rephrase it into our setting, the reward and weight for each arrival in this case is one, and the cost for each arrival is its posterior probability of being normal. 

\subsubsection{Data description and model estimation.} 
\begin{figure}[ht]
    \centering
    \includegraphics[width=0.9\linewidth]{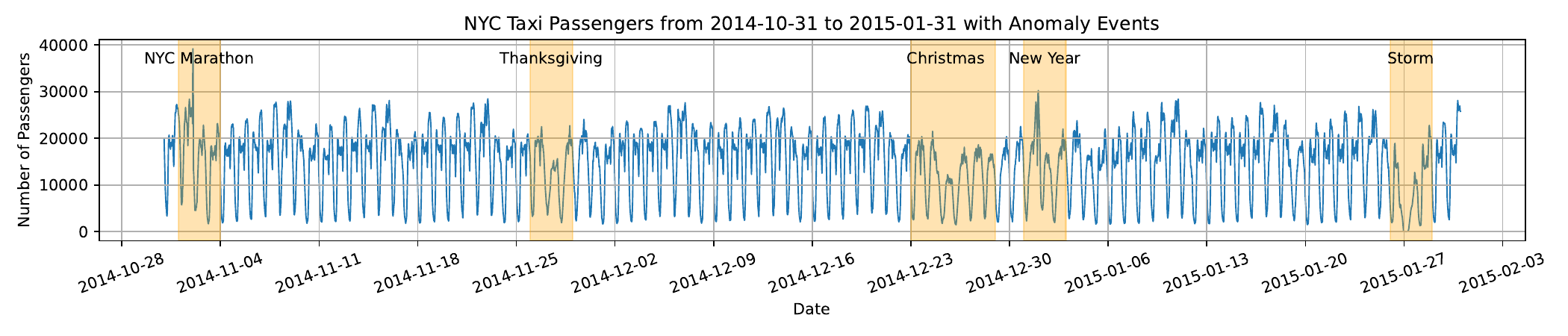}
    \caption{\centering Illustration of the NYC taxi passenger data with abnormal events highlighted in orange.}
    \label{fig:raw_data}
\end{figure}

We will use the New York City (NYC) taxi dataset downloaded from the Numenta Anomaly Benchmark (NAB) repository (\citealt{ahmad2017unsupervised}) to illustrate this application. This dataset, also employed in other studies (\citealt{gang2023structure}), records the number of taxi passengers in NYC every 30 minutes from July 1, 2014, to January 31, 2015. It captures fluctuations caused by five major events: the NYC marathon, Thanksgiving, Christmas, New Year's Eve, and a significant snowstorm. We plot the data from October 31, 2014 to January 31, 2015, a time window that encompasses all the specified abnormal occurrences, along with the highlighted events in Figure \ref{fig:raw_data}.

\begin{figure}[ht]
    \centering
    \includegraphics[width=0.9\linewidth]{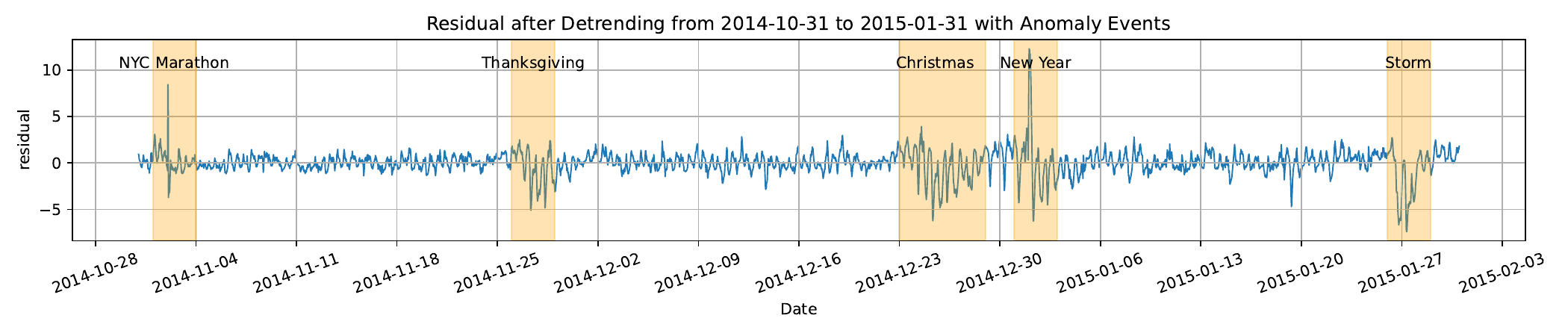}
    \caption{\centering Residual term after the STL decomposition of the NYC taxi passenger data.}
    \label{fig:residual}
\end{figure}

As one can see from Figure \ref{fig:raw_data}, the data has strong seasonality in a single day or week. To obtain a stationary input to the model, we first use the STL decomposition method (\citealt{cleveland1990stl}) to decompose the data into trend, seasonality, and residuals. We plot the residual term in Figure \ref{fig:residual}, from which we observe that the residual term is relatively stable except for the highlighted events. In the following, we will use the residual as the testing data instead of the raw count of the passengers.

\begin{figure}[ht]
    \centering
    \includegraphics[width=0.9\linewidth]{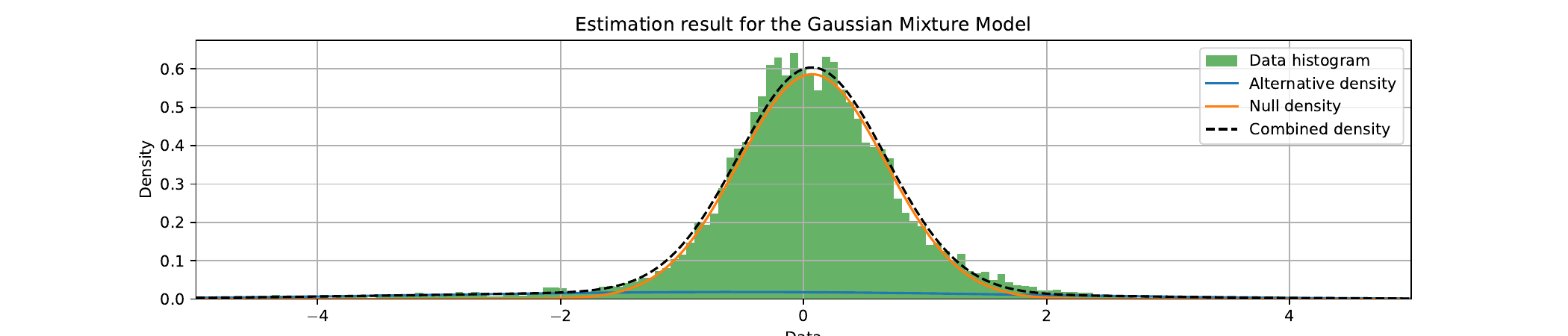}
    \caption{\centering Density of the residual and the estimation result for the Gaussian Mixture model.}
    \label{fig:density}
\end{figure}

To obtain the posterior for each data point being normal, we fit a two-group Bayesian model by estimating the distributions for the normal case (we call it the null density) and the abnormal case (the alternative density). To obtain a valid estimation of the two distributions, we fit a Gaussian Mixture model with two components to the residual data. The result is shown in Figure \ref{fig:density}. The orange line represents the fitted null distribution, which is $\mathcal{N}(0.07, 0.36)$, and the blue line stands for the alternative distribution, for which the result is $\mathcal{N}(-0.56, 5.74)$. The prior probabilities for the two distributions are $0.89$ and $0.11$ respectively. As one can see from Figure \ref{fig:density}, the fitted density aligns well with the data's true histogram marked in green.
After obtaining the estimated null and alternative distribution, we use it to obtain the posterior null probability $c^{(t)}$ for each time period $t$. Specifically, suppose the CDF for the null and alternative is $\hat{F}_0$ and $\hat{F}_1$ with prior probabilities $\pi_0$ and $\pi_1$. Then the $p$-value of the residual $z^{(t)}$ at time $t$ relative to the null and alternative are $p_0=2\hat{F}_0(-|z^{(t)}|)$ and $p_1=2(1-\hat{F}_1(|z^{(t)}|))$ respectively. Then, the posterior is calculated as $c^{(t)}=\P(\gamma^{(t)}=0\mid Z^{(t)}\; \mbox{being at least as extreme as}\; z^{(t)})=\alpha_0p_0/(\alpha_0p_0+\alpha_1p_1)$. Here, the posterior probability is calculated as conditioning on the event that the data $Z^{(t)}$ is at least as extreme as the observed data $z^{(t)}$, which is one of the approaches mentioned in \cite{efron2001empirical}.

\subsubsection{Testing policies and results.} 
We examine and compare two streams of FDR control policies. \textbf{(1) Bayesian approach.} The first stream employs the Bayesian principle and makes the decision based on the posterior null probability $c^{(t)}$, which is the methodology discussed in this paper. For this line of work, we test the SAST policy \citep{gang2023structure}, the proposed $\MLB$-AC policy (Algorithm \ref{alg:MLB-AC}), and the $\MLB$-AC-A policy, the alternative of the $\MLB$-AC policy without the knowledge of $T$. \textbf{(2) Frequentist approach.} The other group of strategies adopts a frequentist viewpoint and makes decisions based on $p$-values. For this line of work, we will evaluate the LOND \citep{javanmard2015online}, LORD++ \citep{ramdas2017online}, ADDIS \citep{tian2019addis}, and the offline BH \citep{benjamini1995controlling} policies. Our focus is to benchmark our policy against existing policies in the Bayesian domain, where policies take $w^{(t)}$ as an input. For comparisons with the frequentist approach where $p$-values are taken as input, the results should only be considered for illustration purposes. 

\begin{figure}[ht]
    \centering
    \includegraphics[width=0.9\linewidth]{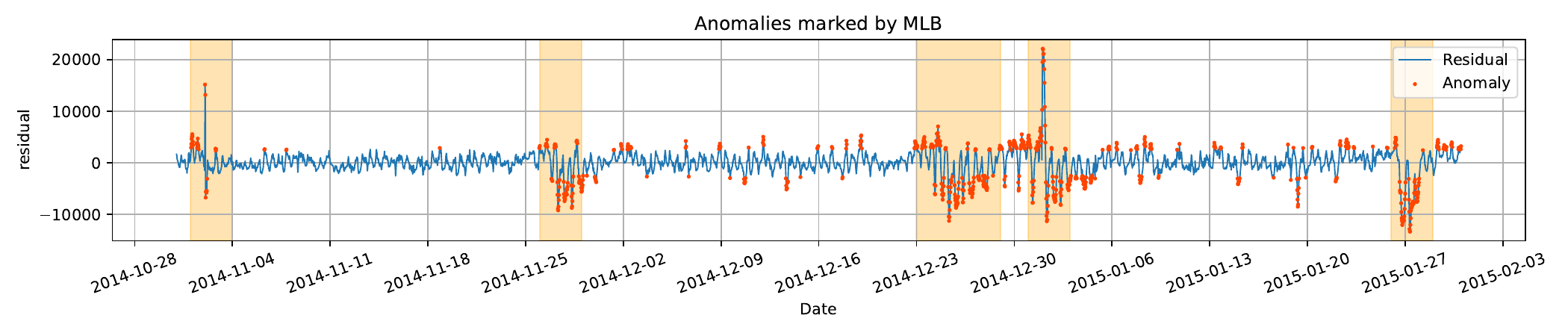}
    \caption{\centering Anomalies detected by $\MLB$-AC.}
    \label{fig:main_result}
\end{figure}

We run the policies with FDR level of 5\% and report the total number of discoveries for these policies. The result is presented in Table \ref{tab:result}. Here, besides the policies mentioned above, we also report the upper bound obtained by solving the corresponding linear programming. As one can see from the table, the proposed policy $\MLB$-AC attains the highest number of discoveries, demonstrating its superior performance. Note that our $\MLB$-AC policy only misses 20 discoveries compared to the linear programming upper bound, which only consists of 2\% of the total anomalies detected by the LP. Interestingly, the policy $\MLB$-AC-A, which does not require the information of $T$, also has a nice performance of detecting 858 anomalies, only missing 4 anomalies compared to the $\MLB$-AC policy. We also plot the anomalies detected by the $\MLB$-AC policy in Figure \ref{fig:main_result}. From the figure, we can see that the $\MLB$-AC policy detects most of the anomalies around the highlighted events.
\begin{table}[ht]
    \centering
    \begin{tabular}{c|c|c|c|c|c|c|c|c}
        Approach & \multicolumn{4}{c|}{Bayesian} & \multicolumn{4}{c}{Frequentist} \\\hline
        Policies & SAST & MLB-AC & MLB-AC-A & LP & LOND & LORD++ & ADDIS & BH \\ \hline
        \# Discoveries & 834 & 862 & 858 & 882 & 315 & 562 & 792 & 659
    \end{tabular}
    \caption{Number of discoveries for different policies.}
    \label{tab:result}
\end{table}

\section{Conclusion} \label{sec:conclusion}
In this paper, we study the problem of optimal policy design for average budget constraint problems. The goal for the decision maker is to maximize the cumulative reward while ensuring the average budget for the accepted task is below a pre-specified threshold. Such formulations has wide applications for problem requiring long-term stable control, including online FDR control, online advertising, and risk control in investment. To solve the problem, we investigate two main scenarios where the distribution of the incoming cost-reward pairs is either general or discrete.

For the general distribution, we propose the Static Greedy ($\SG$) policy, showing it achieves a $O(\sqrt{T})$ regret with respect to the offline upper bound. We complement the result by showing that in general any online policy will incur a $\Omega(\sqrt{T})$ regret. For the discrete distribution, we propose the Multilevel Logarithm Buffer ($\MLB$) policy that achieves a near-optimal regret of order $O(\ln^2T)$. Interestingly, existing state-of-the-art policies that achieve constant regrets in the network revenue management problem fail in our setting as they are shown to achieve $\Omega(\sqrt{T})$ or even $\Omega(T)$ regret. The cause is that these policies are too optimistic about future replenishment and can over claim discoveries. Built upon the insights from our design and analysis of $\MLB$, we propose $\MLB$-AC and $\MLB$-AC-A, amended versions of $\MLB$ that can handle practical problems with continuous cost-reward pair distributions as well as time-dependent information structures. In particular, $\MLB$-AC-A can handle the situation when $T$ is not known a priori. Numerical experiments on both synthetic and real data validate the superior performance of $\MLB$ (in the discrete setting) and $\MLB$-AC as well as $\MLB$-AC-A (in the continuous setting).

 There are also some interesting future work. It is worth investigating whether we can obtain better instance-dependent regret bounds in the discrete case. Also, it is intriguing to see if we can obtain either a $O(\text{polylog}(T))$ upper bound or a $\Omega(\sqrt{T})$ lower bound for a \textit{fixed} continuous cost distribution under some mild assumptions (e.g., the uniform distribution).

{
% \SingleSpacedXI
\bibliographystyle{informs2014}
\bibliography{main}
}

\newpage 
\setcounter{page}{1}
\begin{APPENDICES}
\small
% \fontsize{9}{10}

\section{More Applications of the Knapsack Formulation}\label{apx:examples}
Below we state several operations examples where our knapsack formulation may also accommodate.

\textbf{Inventory control with exogenous replenishment.} For retailers or manufacturers, product demand from customers represents different amount of consumption (representing positive weights), while shipments from suppliers or production batches can possibly be exogenous replenishment (representing negative weights). An online policy can help in adjusting inventory levels dynamically and decide how to satisfy different types of requests, with the goal of maximizing the total number of accepted requests.

\textbf{Online multiple experiments with budget replenishment.} Consider a group of people managing a series of sequential experiments on a digital platform. Different experiments can belong to different categories. The platform starts with an initial budget $B^{(1)}$, which is used to fund experiments. In each time period $t$, the group has a budget $B^{(t)}$ at hand, and faces either a replenishment ($a^{(t)} < 0$) that replenishes the budget or an experiment that, if conducted, costs $a^{(t)}>0$ from the budget. The group needs to irrevocably decide whether or not the experiment should be carried out. The objective is to maximize the total number of conducted experiments, denoted as the sum of $\mathds 1\{a^{(t)} > 0\}$, while maintaining the budget never drops below $0$. This involves making informed decisions on which experiments to conduct, taking into account the potential costs, as well as the fluctuating budget due to exogenous replenishment and endogenous expenditures.

\textbf{Task scheduling with sustainable energy consumption.} In a computing environment featuring a single computational resource powered by electricity, we encounter a dynamic scenario where, at each time period, the system may receive either an energy replenishment (representing negative weights) or a computation request that consumes a certain amount of energy (representing positive weights). The replenishment could come from various sources, including grid electricity, backup generators, or even renewable energy sources integrated into the system's supply chain. The requests, on the other hand, could range from data processing tasks, cloud-based services, to complex computational operations required by end-users or automated systems. The decision maker needs to effectively manage the balance between maximizing the total number of accepted computation requests and ensuring that the resource never runs out of energy.

\section{Proof of Results in Section \ref{sec:model}}
\subsection{Proof of Theorem \ref{thm:continuous-upper}}
For simplicity of notation, we write $x^*=x^*(\overline{\rho})$ and $p^*=\P(a^{(t)}=\overline{\rho})$. By the design of $\SG$, each time when $a^{(t)}/r^{(t)}=\overline{\rho}$, we sample an independent uniform random variable $u^{(t)}$ and accept $a^{(t)}$ if and only if $u^{(t)} \leq x^*$. Thus, 
\begin{equation} \label{eq:SGP-value}
\begin{aligned}
    f_\cD^T(\SG) & = \ex{}{\sum_{t=1}^T\mathds 1\{B^{(t)} \geq a^{(t)}\}\left(\mathds 1\{a^{(t)}/r^{(t)} < \overline{\rho}\} + \mathds 1\{a^{(t)}/r^{(t)} =\overline{\rho}, u^{(t)}\leq x^*\}\right)}
\end{aligned}
\end{equation}
Meanwhile, by the nature of $\DLP$, we know that
\begin{equation} \label{eq:DLP-value}
\begin{aligned}
    f_\cD^T(\DLP) = T\cdot\left(\P(a^{(t)}/r^{(t)} < \overline{\rho}) + p^*x^*\right) = \ex{}{\sum_{t=1}^T\left(\mathds 1\{a^{(t)}/r^{(t)} < \overline{\rho}\} + \mathds 1\{a^{(t)}/r^{(t)} =\overline{\rho}, u^{(t)}\leq x^*\}\right)}.
\end{aligned}
\end{equation}
Combining \eqref{eq:SGP-value} and \eqref{eq:DLP-value} we have
\begin{equation} \label{eq:continuous-gap}
\begin{aligned}
    f_\cD^T(\DLP) - f_\cD^T(\SG) & = \ex{}{\sum_{t=1}^T\mathds 1\{B^{(t)} < a^{(t)}\}\left(\mathds 1\{a^{(t)}/r^{(t)} < \overline{\rho}\} + \mathds 1\{a^{(t)}/r^{(t)} =\overline{\rho}, u^{(t)}\leq x^*\}\right)}.
\end{aligned}
\end{equation}

Define a new process $\{\tilde B^{(t)}\}_{t}$ as a ``coupled'' version of the process $\{B^{(t)}\}_{t}$ as follows. $\tilde B^{(1)} = B^{(1)} = 0$. For general $t\geq 1$, we define
\begin{equation} \label{eq:continuous-coupled}
\begin{aligned}
    \tilde B^{(t+1)} = \max\left\{\tilde B^{(t)} - a^{(t)}\left(\mathds 1\{a^{(t)}/r^{(t)} < \overline{\rho}\} + \mathds 1\{a^{(t)}/r^{(t)} =\overline{\rho}, u^{(t)}\leq x^*\}\right), 0\right\}.
\end{aligned}
\end{equation}
That is, for each sample path $\{a^{(t)}\}_t$ with $\{u^{(t)}\}_t$, in the ``coupled'' version regardless of the budget, in each time $t$ we always accept the arrival only it satisfies Line 5-8 in Algorithm \ref{alg:static} (here we also couple the random seed when we face $\overline{\rho}$). If the budget drops below $0$, we restart the budget level as $0$. We now show that for each sample path $\tilde B^{(t)} \leq B^{(t)}$ for any $t$ via induction. Apparently $\tilde B^{(1)} \leq B^{(1)}$. Suppose we have $\tilde B^{(t)} \leq B^{(t)}$. If at time $t$ the arrival $a^{(t)}$ is rejected by $\SG$, then it implies $a^{(t)} > 0$. We have $\tilde B^{(t+1)} \leq \tilde B^{(t)}\leq B^{(t)} = B^{(t+1)}$. If the arrival is accepted by $\SG$, then we also have
\begin{equation*}
\begin{aligned}
    \tilde B^{(t+1)} & = \max\left\{\tilde B^{(t)} - a^{(t)}\left(\mathds 1\{a^{(t)}/r^{(t)} < \overline{\rho}\} + \mathds 1\{a^{(t)}/r^{(t)} =\overline{\rho}, u^{(t)}\leq x^*\}\right), 0\right\} \\
    & \leq \max\left\{B^{(t)} - a^{(t)}\left(\mathds 1\{a^{(t)}/r^{(t)} < \overline{\rho}\} + \mathds 1\{a^{(t)}/r^{(t)} =\overline{\rho}, u^{(t)}\leq x^*\}\right), 0\right\} \\
    & = \max\left\{B^{(t+1)}, 0\right\} \\
    & = B^{(t+1)}.
\end{aligned}
\end{equation*}
This leads to
\begin{equation} \label{eq:continuous-upper-couple}
    \mathds 1\left\{B^{(t)} < a^{(t)}\right\} \leq \mathds 1\left\{\tilde B^{(t)} < a^{(t)}\right\}.
\end{equation}

Define $Z^{(t)} = - a^{(t)}\left(\mathds 1\{a^{(t)}/r^{(t)} < \overline{\rho}\} + \mathds 1\{a^{(t)}/r^{(t)} =\overline{\rho}, u^{(t)}\leq x^*\}\right)$, then we know that $\{Z^{(t)} - \ex{}{Z^{(t)}}\}_{t=1, \cdots}$ are i.i.d. random variables bounded within $[-1, 1]$. By Proposition 6.2 in \cite{asmussen2003applied}, we have that
\begin{equation*}
\begin{aligned}
    \tilde B^{(t)} = \max\{0, X_{t-1}, X_{t-1}+X_{t-2}, \cdots, X_{t-1}+\cdots+X_1\}
\end{aligned}
\end{equation*}
is the maximum of the first $t$ positions of a random walk with a non-negative trend. By Lemma \ref{lem:magic}, we know that for any $a > 0$, 
\begin{equation*}
    \P(\tilde B^{(t)} < a) \leq O\left(\frac{a}{\sigma\sqrt{t}} + \frac{\kappa}{\sqrt{t}}\right).
\end{equation*}
Therefore, 
\begin{equation} \label{eq:continuous-single-gap}
\begin{aligned}
    & \indent \ex{}{\mathds 1\{B^{(t)} < a^{(t)}\}\left(\mathds 1\{a^{(t)}/r^{(t)} < \overline{\rho}\} + \mathds 1\{a^{(t)}/r^{(t)} =\overline{\rho}, u^{(t)}\leq x^*\}\right)} \\
    & \leq \ex{}{\mathds 1\{\tilde B^{(t)} < a^{(t)}\}\left(\mathds 1\{a^{(t)}/r^{(t)} < \overline{\rho}\} + \mathds 1\{a^{(t)}/r^{(t)} =\overline{\rho}, u^{(t)}\leq x^*\}\right)} \\
    & = \ex{}{\ex{}{\mathds 1\{\tilde B^{(t)} < a^{(t)}\}\big|a^{(t)}}\left(\mathds 1\{a^{(t)}/r^{(t)} < \overline{\rho}\} + \mathds 1\{a^{(t)}/r^{(t)} =\overline{\rho}, u^{(t)}\leq x^*\}\right)} \\
    & = \ex{}{O\left(\frac{a^{(t)}}{\sigma\sqrt{t}} + \frac{\kappa}{\sqrt{t}}\right)\left(\mathds 1\{0 < a^{(t)}/r^{(t)} < \overline{\rho}\} + \mathds 1\{a^{(t)}/r^{(t)} =\overline{\rho}, u^{(t)}\leq x^*\}\right)} \\
    & = O\left(\frac{\ex{}{a^{(t)}\left(\mathds 1\{0 < a^{(t)}/r^{(t)} < \overline{\rho}\} + \mathds 1\{a^{(t)}/r^{(t)} =\overline{\rho}, u^{(t)}\leq x^*\}\right)}}{\sigma\sqrt{t}}\right) + O\left(\frac{\kappa}{\sqrt{t}}\right) \\
    & = O\left(\frac{\kappa}{\sqrt{t}}\right).
\end{aligned}
\end{equation}
Note that $a^{(t)}$ is independent with $\tilde B^{(t)}$. In the last inequality, we use the following inequality:
\begin{equation*}
\begin{aligned}
    \sigma & \geq \sqrt{\ex{}{|Z^{(t)} - \ex{}{Z^{(t)}}|^2\left(\mathds 1\{0 < a^{(t)}/r^{(t)} < \overline{\rho}\} + \mathds 1\{a^{(t)}/r^{(t)} =\overline{\rho}, u^{(t)}\leq x^*\}\right)}} \\
    & \geq \sqrt{\ex{}{|-a^{(t)}|^2\left(\mathds 1\{0 < a^{(t)}/r^{(t)} < \overline{\rho}\} + \mathds 1\{a^{(t)}/r^{(t)} =\overline{\rho}, u^{(t)}\leq x^*\}\right)}} \\
    & \geq \ex{}{a^{(t)}\left(\mathds 1\{0 < a^{(t)}/r^{(t)} < \overline{\rho}\} + \mathds 1\{a^{(t)}/r^{(t)} =\overline{\rho}, u^{(t)}\leq x^*\}\right)}
\end{aligned}
\end{equation*}
Combining \eqref{eq:continuous-gap} and \eqref{eq:continuous-single-gap} yields
\begin{equation*}
\begin{aligned}
    f_\cD^T(\DLP) - f_\cD^T(\SG) = \sum_{t=1}^T\ex{}{O\left(\frac{\kappa}{\sqrt{t}}\right)} = O(\kappa\sqrt{T}).
\end{aligned}
\end{equation*}

\subsection{Proof of Theorem \ref{thm:continuous-lower}}

Consider $\cD_T$ as the following discrete distribution:
\begin{equation}
\begin{aligned}
    a^{(t)} = \left\{
    \begin{matrix}
        & -1/5\triangleq a_0 & \text{w.p. } 1/2 + 1/\sqrt{T}, \\
        & 1/5\triangleq a_1 & \text{w.p. } 1/2 - 2/\sqrt{T}, \\
        & 3/5\triangleq a_2 & \text{w.p. } 1/\sqrt{T}.
    \end{matrix}
    \right.
\end{aligned}
\end{equation}
Solving $\DLP$ yields $x^*(-1/5) = x^*(1/5) = x^*(3/5) = 1$.

First, we point out that in the optimal online policy, every arrival of type $a_1$ will be accepted as long as the budget is positive. To prove this, it suffices to show that the probability of wrong accept is $0$. In fact, by Lemma \ref{lem:there_is_gap2_general}, the probability of wrongly accept $a^{(t)}$ when $a^{(t)} = 1$ can be written as
\begin{equation*}
\begin{aligned}
    \pwa_1^{(t)} & \leq \P\prn{B^{(t)} - 1 < \max_{s\in[t+1, T]}\sum_{r=t+1}^sa^{(r)}\mathds 1\{a^{(r)} < 1\}} \\
    & \leq \P\prn{0 < \max_{s\in[t+1, T]}\sum_{r=t+1}^sa^{(r)}\mathds 1\{a^{(r)} < 1\}} \\
    & = 0.
\end{aligned}
\end{equation*}
From now on we only consider policies that always accept $a^{(t)}$ if $a^{(t)}=a_1$. We now provide a property of $B^{(t)}$.
\paragraph{Claim 0.} There exists absolute constants $c_1$, $c_2$, $c_3$, $c_4$ such that for any fixed $t\in[T/4, T/2)$:
\begin{equation*}
\begin{aligned}
    \P\prn{a_2+c_1\sqrt{t} \leq B^{(t)} \leq c_2\sqrt{t}} \geq c_3 -c_4/\sqrt{t}.
\end{aligned}
\end{equation*}

Construct $\underline{B}^{(t)}$ as a ``coupled" version of $B^{(t)}$: $\underline{B}^{(1)} = B^{(1)} = 0, \underline{B}^{(t+1)} = \max\{\underline B^{(t)} - a^{(t)}, 0\}$. Following the similar argument in the proof of Theorem \ref{thm:continuous-upper}, we know that $\underline B^{(t)}$ is a sample path-wise lower bound of $B^{(t)}$, and that $$\underline B^{(t)} = \max\{0, -a^{(t-1)}, -a^{(t-1)}- a^{(t-2)}, \cdots, -a^{(t-1)} - \cdots - a^{(1)}\}.$$ Let $c_1>0$ be some small positive constant to be determined, From Lemma \ref{lem:magic}, we know that
\begin{equation} \label{eq:lower-bound-couple-lower}
    \P\prn{B^{(t)} \leq a_2 + c_1\sqrt{t}} \leq \P\prn{\underline B^{(t)} \leq a_2+c_1\sqrt{t}} = O\prn{\frac{a_2+c_1\sqrt{t}}{\sqrt{t-1}} + \frac{1}{\sqrt{t}}} = O(c_1) + O(1/\sqrt{t}),
\end{equation}
where in $O(\cdot)$ we hiding absolute constants. 

Construct $\overline B^{(t)}$ as a ``coupled" version of $B^{(t)}$: $\overline{B}^{(1)} = B^{(1)} = 0, \overline{B}^{(t+1)} = \max\{\overline B^{(t)} -a^{(t)}\mathds 1\{a^{(t)}\leq a_1\}, 3\}$. Following the similar argument in the proof of Theorem \ref{thm:continuous-upper}, we know $\overline B^{(t)}$ is a sample path-wise upper bound of $B^{(t)}$, and
\begin{equation*}
\begin{aligned}
    \overline B^{(t)} & = \max\left\{3, 3-a^{(t-1)}\mathds 1\{a^{(t-1)}\leq a_1\}, \cdots, 3-a^{(t-1)}\mathds 1\{a^{(t-1)}\leq a_1\} - \cdots - a^{(1)}\mathds 1\{a^{(1)}\leq a_1\}\right\} \\
    & \leq \max\left\{3, 3+\frac{a_2}{\sqrt{T}}-\prn{a^{(t-1)}\mathds 1\{a^{(t-1)}\leq a_1\}+\frac{a_2}{\sqrt{T}}}, \cdots, 3+\frac{(t-1)a_2}{\sqrt{T}}-\sum_{i=1}^{t-1}\prn{a^{(i)}\mathds 1\{a^{(i)}\leq a_1\}+\frac{a_2}{\sqrt{T}}}\right\} \\
    & \leq 3 + \frac{t}{\sqrt{T}} + \max\left\{0, -\prn{a^{(t-1)}\mathds 1\{a^{(t-1)}\leq a_1\}+\frac{a_2}{\sqrt{T}}}, \cdots, -\sum_{i=1}^{t-1}\prn{a^{(i)}\mathds 1\{a^{(i)}\leq a_1\}+\frac{a_2}{\sqrt{T}}}\right\} \\
    & \leq 4\sqrt{t} + \max\left\{0, -\prn{a^{(t-1)}\mathds 1\{a^{(t-1)}\leq a_1\}+\frac{a_2}{\sqrt{T}}}, \cdots, -\sum_{i=1}^{t-1}\prn{a^{(i)}\mathds 1\{a^{(i)}\leq a_1\}+\frac{a_2}{\sqrt{T}}}\right\}.
\end{aligned}
\end{equation*}
Let $c_2>4$ be some large positive constant to be determined. From Lemma \ref{lem:magic}, we know that
\begin{equation} \label{eq:lower-bound-couple-upper}
\begin{aligned}
    & \indent \P\prn{B^{(t)} \geq c_2\sqrt{t}} \\
    & \leq \P\prn{\overline B^{(t)} \geq c_2\sqrt{t}} \\
    & \leq \P\prn{\overline B^{(t)} - 4\sqrt{t}\geq (c_2-4)\sqrt{t}} \\
    & = O\prn{\bar{\Phi}^c\prn{\frac{(c_2-4)\sqrt{t}}{\sqrt{t-1}}} + \frac{1}{\sqrt{t}}} \\
    & = O(\bar{\Phi}^c(c_2-4)) + O(1/\sqrt{t}),
\end{aligned}
\end{equation}
where in $O(\cdot)$ we hiding absolute constants.

Therefore, combining \eqref{eq:lower-bound-couple-lower} and \eqref{eq:lower-bound-couple-upper} yields
\begin{equation*}
\begin{aligned}
    \P\prn{a_2+c_1\sqrt{t}\leq B^{(t)}\leq c_2\sqrt{t}} = 1 - O\prn{c_1+\bar{\Phi}^c(c_2-4)} - O\prn{1/\sqrt{t}},
\end{aligned}
\end{equation*}
where in $O(\cdot)$ we are hiding absolute constants. It suffices to take $c_1$ to be small enough and $c_2$ to be large enough.

We then consider the loss incurred by wrongly accepting or rejecting arrivals of type $a_2$. 

{\small\it\indent Claim 1.} For $1\leq t < T/2$, we have
\begin{equation*}
\begin{aligned}
    \pwa_2^{(t)} & \geq \P\prn{\left.2B^{(t)} < \sum_{r=t+1}^{t+\lfloor T/2\rfloor}a^{(r)}\mathds 1\{a^{(r)} < a_2\}, \pi\text{ accepts  }a^{(t)}\right| a^{(t)} = a_2}
\end{aligned}
\end{equation*}
In fact, let's assume the event in Claim 1 happens. This means that standing at time $t$ with budget $B^{(t)}$, from time $t+1$ to $t+\lfloor T/2\rfloor$, only accepting $-1/5$ and $1/5$ will reduce the budget to $-B^{(t)}$ (which is not feasible for an online policy). Now consider $\mix^{(t)}$ (remember in $\mix^{(t)}$ we follow $\pi$ until time $t$, and so in $\mix^{(t)}$ $a^{(t)}$ is accepted). The statement above indicates that in $\mix^{(t)}$ we accept a small number of arrivals of type $a_1$ during time $[t+1, t+\lfloor T/2\rfloor]$. In fact, at least $2$ of the arrivals of type $a_1$ during time $[t+1, t+\lfloor T/2\rfloor]$ must be rejected. Otherwise, the remaining budget at time $t+\lfloor T/2\rfloor + 1$ is at most
\begin{equation*}
\begin{aligned}
    B^{(t)} - \prn{\sum_{r=t+1}^{t+\lfloor T/2\rfloor}a^{(r)}\mathds 1\{a^{(r)} < a_2\} - 1/5} < 1/5 - B^{(t)} < 0,
\end{aligned}
\end{equation*}
where in the last inequality we have used the fact that $B^{(t)} > 1/5$ because $a^{(t)}$ is accepted. Now consider the following ``modification'' of $\mix^{(t)}$: instead of accepting $a^{(t)}=a_2$ at time $t$, we accept two more arrivals of type $a_1$ during time $[t+1, t+\lfloor T/2\rfloor]$. This will not violate the any-time constraint, since $a_2 > 2a_1$ and we postpone depleting the budget to later time periods. Apparently, this indicates that the total number of accepted requests induced by $\mix^{(t-1)}$ must be strictly larger than that of $\mix^{(t)}$ --- accepting $a^{(t)}$ is a wrong decision.

{\small\it\indent Claim 2.} For $1\leq t< T/2$, we have
\begin{equation*}
\begin{aligned}
    \pwr_2^{(t)} \geq \P\prn{\left.B^{(t)} - a_2 \leq \max_{s\in[t+1, T]}\sum_{r=t+1}^{s}a^{(r)}, \pi\text{ rejects  }a^{(t)}\right| a^{(t)} = a_2}
\end{aligned}
\end{equation*}
In fact, let's assume the event in Claim 2 happens. It indicates that always accepting the arrivals can never violate the any-time constraint. Therefore, rejecting $a^{(t)}$ is a wrong decision.

Now let's bound the terms in Claim 1 and 2 separately. Fix $t\in[T/4, T/2)$.
\begin{equation} \label{eq:lower-bound-ac}
\begin{aligned}
    \pwa_2^{(t)} & \geq \P\prn{\left.2B^{(t)} < \sum_{r=t+1}^{t+\lfloor T/2\rfloor}a^{(r)}\mathds 1\{a^{(r)} < a_2\}, \pi\text{ accepts  }a^{(t)}\right| a^{(t)} = a_2} \\
    & \geq \P\prn{\left.2B^{(t)} < \sum_{r=t+1}^{t+\lfloor T/2\rfloor}a^{(r)}\mathds 1\{a^{(r)} < a_2\}, B^{(t)}\leq c_2\sqrt{t}, \pi\text{ accepts  }a^{(t)}\right| a^{(t)} = a_2} \\
    & \geq \P\prn{\left.2c_2\sqrt{t} < \sum_{r=t+1}^{t+\lfloor T/2\rfloor}a^{(r)}\mathds 1\{a^{(r)} < a_2\}, B^{(t)}\leq c_2\sqrt{t}, \pi\text{ accepts  }a^{(t)}\right| a^{(t)} = a_2} \\
    & = \P\prn{2c_2\sqrt{t} < \sum_{r=t+1}^{t+\lfloor T/2\rfloor}a^{(r)}\mathds 1\{a^{(r)} < a_2\}}\P\prn{\left.B^{(t)}\leq c_2\sqrt{t}, \pi\text{ accepts  }a^{(t)}\right| a^{(t)} = a_2} \\
    & \geq \P\prn{\sum_{r=t+1}^{t+\lfloor T/2\rfloor}\prn{a^{(r)}\mathds 1\{a^{(r)} < a_2\} + a_2/\sqrt{T}} > 3c_2\sqrt{T}}\P\prn{\left.B^{(t)}\leq c_2\sqrt{t}, \pi\text{ accepts  }a^{(t)}\right| a^{(t)} = a_2} \\
    & = \prn{\Omega(1) - O(1/\sqrt{T})}\P\prn{\left.B^{(t)}\leq c_2\sqrt{t}, \pi\text{ accepts  }a^{(t)}\right| a^{(t)} = a_2}.
\end{aligned}
\end{equation}
Note that in the last equality we have applied the Berry-Esseen theorem to give a lower bound for deviation of sum of i.i.d. random variables. In $\Omega(\cdot)$ and $O(\cdot)$ we are hiding absolute constants.
\begin{equation} \label{eq:lower-bound-rj}
\begin{aligned}
    \pwr_2^{(t)} & \geq \P\prn{\left.B^{(t)} - a_2 \geq \max_{s\in[t+1, T]}\sum_{r=t+1}^{s}a^{(r)}, \pi\text{ rejects  }a^{(t)}\right| a^{(t)} = a_2} \\
    & \geq \P\prn{\left.B^{(t)} - a_2 \geq \max_{s\in[t+1, T]}\sum_{r=t+1}^{s}a^{(r)}, B^{(t)}\geq a_2+c_1\sqrt{t}, \pi\text{ rejects  }a^{(t)}\right| a^{(t)} = a_2} \\
    & \geq \P\prn{\left.c_1\sqrt{t} \geq \max_{s\in[t+1, T]}\sum_{r=t+1}^{s}a^{(r)}, B^{(t)}\geq a_2+c_1\sqrt{t}, \pi\text{ rejects  }a^{(t)}\right| a^{(t)} = a_2} \\
    & = \P\prn{c_1\sqrt{t} \geq \max_{s\in[t+1, T]}\sum_{r=t+1}^{s}a^{(r)}} \P\prn{\left.B^{(t)}\geq c_1\sqrt{t}, \pi\text{ rejects  }a^{(t)}\right| a^{(t)} = a_2} \\
    & \geq \P\prn{c_1\sqrt{T-t}/2 \geq \max_{s\in[t+1, T]}\sum_{r=t+1}^{s}a^{(r)}} \P\prn{\left.B^{(t)}\geq c_1\sqrt{t}, \pi\text{ rejects  }a^{(t)}\right| a^{(t)} = a_2} \\
    & = \prn{\Omega(1) - O(1/\sqrt{T-t})}\P\prn{\left.B^{(t)}\geq c_1\sqrt{t}, \pi\text{ rejects  }a^{(t)}\right| a^{(t)} = a_2}.
\end{aligned}
\end{equation}
Note that in the last equality we have applied Lemma \ref{lem:magic}. In $\Omega(\cdot)$ and $O(\cdot)$ we are hiding absolute constants.

Combining \eqref{eq:lower-bound-ac} and \eqref{eq:lower-bound-rj} we have for $t\in[T/4, T/2)$:
\begin{align*}
    & \indent \pwa_2^{(t)} + \pwr_2^{(t)} \\
    & \geq \prn{\Omega(1) - O(1/\sqrt{T})}\P\prn{\left.B^{(t)}\leq c_2\sqrt{t}, \pi\text{ accepts  }a^{(t)}\right| a^{(t)} = a_2} \\
    & \quad\quad\quad + \prn{\Omega(1) - O(1/\sqrt{T-t})}\P\prn{\left.B^{(t)}\geq a_2+c_1\sqrt{t}, \pi\text{ rejects  }a^{(t)}\right| a^{(t)} = a_2} \\
    & = \prn{\Omega(1) - O(1/\sqrt{T})}\left(\P\prn{\left.B^{(t)}\leq c_2\sqrt{t}, \pi\text{ accepts  }a^{(t)}\right| a^{(t)} = a_2} + \right.\\
    & \quad\quad\quad + \left.\P\prn{\left.B^{(t)}\geq a_2+c_1\sqrt{t}, \pi\text{ rejects  }a^{(t)}\right| a^{(t)} = a_2}\right) \\
    & = \prn{\Omega(1) - O(1/\sqrt{T})}\left(\P\prn{\left.\pi\text{ accepts  }a^{(t)}\right| a^{(t)} = a_2} - \P\prn{\left.B^{(t)} > c_2\sqrt{t}, \pi\text{ accepts  }a^{(t)}\right| a^{(t)} = a_2} \right. \\
    & \quad\quad\quad + \left.\P\prn{\left.\pi\text{ rejects  }a^{(t)}\right| a^{(t)} = a_2} - \P\prn{\left.B^{(t)}<a_2+c_1\sqrt{t}, \pi\text{ rejects  }a^{(t)}\right| a^{(t)} = a_2}\right) \\
    & \geq \prn{\Omega(1) - O(1/\sqrt{T})}\left(1 - \P\prn{\left.B^{(t)} > c_2\sqrt{t}\right| a^{(t)} = a_2} - \P\prn{\left.B^{(t)} < a_2+c_1\sqrt{t}\right| a^{(t)} = a_2}\right) \\
    & = \prn{\Omega(1) - O(1/\sqrt{T})}\P\prn{\left.a_2+c_1\sqrt{t} \leq B^{(t)} \leq c_2\sqrt{t}\right| a^{(t)} = a_2} \\
    & = \prn{\Omega(1) - O(1/\sqrt{T})}\P\prn{a_2+c_1\sqrt{t} \leq B^{(t)} \leq c_2\sqrt{t}} \\
    & = \prn{\Omega(1) - O(1/\sqrt{T})}\prn{\Omega(1) - O(1/\sqrt{t})} \\
    & = \Omega(1) - O(1/\sqrt{T}).
\end{align*}
In $\Omega(\cdot)$ and $O(\cdot)$ we are hiding absolute constants.

Now it's time to wrap up. The total expected loss incurred by wrongly accepting or rejecting arrivals of type $a_2$ is at least
\begin{equation*}
\begin{aligned}
    & \indent \sum_{\frac{T}{4}\leq t < T/2}p_2\prn{\pwa_2^{(t)} + \pwr_2^{(t)}} \\
    & = \Omega(T)\cdot\frac{1}{\sqrt{T}}\cdot\prn{\Omega(1) - O(1/\sqrt{T})} \\
    & = \Omega(\sqrt{T}).
\end{aligned}
\end{equation*}
In $\Omega(\cdot)$ and $O(\cdot)$ we are hiding absolute constants.

\section{Proof of Results in Section \ref{sec:discrete:resolve}}
\subsection{Proof of Proposition \ref{prop:sqrt}}
Consider the example $(a_{-1},a_1) = (-1,1)$ and $(p_{-1},p_1)=(-1,1)$, which is a simple random walk. In this case, solving \eqref{eq:DLP} directly leads to $x_\DLP^{(T)}=(T/2,T/2)$ and $\pow(\DLP) = T$. Solving \eqref{eq:HOany_discrete} leads to:
\begin{equation*}
\begin{aligned}
    x_\HOfix^{(T)}(-1|\W^{(T)}) & = \Lambda^{[1,T]}(-1|\W^{(T)}), \\
    x_\HOfix^{(T)}(1|\W^{(T)}) & = \min\{\Lambda^{[1,T]}(-1|\W^{(T)}), \\
    \Lambda^{[1,T]}(1|\W^{(T)})\} & = T - \max\{\Lambda^{[1,T]}(1|\W^{(T)}) - \Lambda^{[1,T]}(-1|\W^{(T)}),0\}.
\end{aligned}
\end{equation*}
Note that the quantity \(\left|\Lambda^{[1,T]}(-1|\W^{(T)}) - \Lambda^{[1,T]}(1|\W^{(T)})\right|  \) is the distance from zero at time $T$ of a simple random walk, which is well-known as $\Omega(\sqrt{T})$ \citep{durrett2019probability}.

Furthermore, in the context of a simple random walk, the first constraint in \eqref{eq:HOany_discrete} can be interpreted as follows: a walker starts at point 0 and, from time 1 to \( T \), receives steps from the set \(\{-1, +1\}\), deciding whether to accept each step. The walker cannot move right of 0, with the objective being to maximize the number of accepted steps. A greedy policy — where the walker rejects a step if and only if it is currently at zero and the step is \( +1 \) — simplifies the process to a simple random walk with a wall at zero. This is sometimes called The next lemma simplifies the difference \( \pow(\HOfix|\W^{(T)}) - \pow(\HOany|\W^{(T)}) \) to a property of the reflected simple random walks. We then show that \( \pow(\HOfix|\W^{(T)}) - \pow(\HOany|\W^{(T)}) \) has the same distribution with the distance of the walker from zero at time \( T \). 
\begin{lemma} \small
    \label{lem:distance1}
    Denote $ D_r(T)$ as the distance of the walker from zero at time $T$ and $S_r(T)$ as the total length of time of the walker stopping at zero by time $T$, respectively in the random walk described above. Then we have 
    \begin{itemize}
        \item [(a)]$ S_r(T)-D_r(T) = \Lambda^{[1,T]}(1|\W^{(T)}) - \Lambda^{[1,T]}(-1|\W^{(T)})$;
        \item [(b)]\( \pow(\HOfix|\W^{(T)}) - \pow(\HOany|\W^{(T)}) =\min\{D_r(T),S_r(T)\}. \)
    \end{itemize}
\end{lemma}
By Lemma \ref{lem:distance1}, it follows that 
\begin{equation*}
    \pow(\HOfix|\W^{(T)}) - \pow(\HOany|\W^{(T)})  = D_r(T) - \max\{\Lambda^{[1,T]}(-1|\W^{(T)}) - \Lambda^{[1,T]}(1|\W^{(T)}),0\}.
\end{equation*}
Therefore, it is enough to show that $\ex{}{D_r(T) - \max\{\Lambda^{[1,T]}(-1|\W^{(T)}) - \Lambda^{[1,T]}(1|\W^{(T)}),0\}}=\Omega(\sqrt{T})$.

Note that $D_r(t+1) = \max\{D_r(t)-a^{(t+1)},0\}=\max\{D_r(t)+\Lambda^{[t+1,t+1]}(-1|\WT)-\Lambda^{[t+1,t+1]}(1|\WT),0\}$ by definition. If follows that $\{D_r(t)\}_{t=0}^\infty$ is a Lindley process. By Proposition 6.3 in \cite{asmussen2003applied}, we have 
\begin{equation*}
    \begin{aligned}
        D_r(t+1) = \max\{&\Lambda^{[t+1,t+1]}(-1|\WT)-\Lambda^{[t+1,t+1]}(1|\WT),\Lambda^{[t,t+1]}(-1|\WT)-\Lambda^{[t,t+1]}(1|\WT),\dots,\\&\Lambda^{[1,t+1]}(-1|\WT)-\Lambda^{[1,t+1]}(1|\WT),0\}.
    \end{aligned}
\end{equation*}
Therefore, we only need to prove that 
\begin{equation*}
       \ex{}{ \max_{t\in[T]}\{\Lambda^{[t,T]}(-1|\WT)-\Lambda^{[t,T]}(1|\WT),0\}} - \ex{}{\max\{\Lambda^{[1,T]}(-1|\WT)-\Lambda^{[1,T]}(1|\WT),0\}} = \Omega(\sqrt{T}).
\end{equation*}

By the reflection principle of simple random walk \citep{durrett2019probability}, we have 
\begin{equation*}
\begin{aligned}
    & \indent \P\prn{ \max_{t\in[T]}\{\Lambda^{[t,T]}(-1|\WT)-\Lambda^{[t,T]}(1|\WT),0\} \ge a} \\
    & = 2\P\prn{\max\{\Lambda^{[1,T]}(-1|\WT)-\Lambda^{[1,T]}(1|\WT),0\}\ge a},\quad \forall a> 0.
\end{aligned}
\end{equation*}
Hence, it follows that 
\begin{equation*}
    \begin{aligned}
        \ex{}{ \max_{t\in[T]}\{\Lambda^{[t,T]}(-1|\WT)-\Lambda^{[t,T]}(1|\WT),0\}} &\overset{(*)}{=} \sum_{a=1}^\infty \P\prn{ \max_{t\in[T]}\{\Lambda^{[t,T]}(-1|\WT)-\Lambda^{[t,T]}(1|\WT),0\} \ge a a}\\
        &=\sum_{a=1}^\infty 2\P\prn{\max\{\Lambda^{[1,T]}(-1|\WT)-\Lambda^{[1,T]}(1|\WT),0\}\ge a}\\ 
        &=2\ex{}{ \max_{t\in[T]}\{\Lambda^{[t,T]}(-1|\WT)-\Lambda^{[t,T]}(1|\WT),0\}},
    \end{aligned}
\end{equation*}
where $(*)$ uses the equation 
\begin{equation*}
    \begin{aligned}
        \ex{}{A} = \sum_{a=0}^\infty a\P(A=a) = \sum_{a'=1}^\infty \P(A\ge a').
    \end{aligned}
\end{equation*}
As a result, we get 
\begin{equation*}
    \begin{aligned}
        &\ex{}{ \max_{t\in[T]}\{\Lambda^{[t,T]}(-1|\WT)-\Lambda^{[t,T]}(1|\WT),0\}} - \ex{}{\max\{\Lambda^{[1,T]}(-1|\WT)-\Lambda^{[1,T]}(1|\WT),0\}} \\ 
        &= \ex{}{\max\{\Lambda^{[1,T]}(-1|\WT)-\Lambda^{[1,T]}(1|\WT),0\}} \\ 
        &= \Omega(\sqrt{T}),
    \end{aligned}
\end{equation*}
where the last equation is derived by the Central Limit Theorem \citep{durrett2019probability}.

\subsection{Proof of Proposition \ref{prop:canonical-bad}}

We begin by giving a lemma that characterize the large deviation of i.i.d. random variables (i.e. probability of deviation of order $t$ from partial sum of $t$ i.i.d. random variables).

\begin{lemma}[Large deviation I] \small
\label{lem:large_deviation}
    Assume $\xi_1, \dots, \xi_t, \cdots$ are i.i.d. random variables on $[-1,1]$ with zero mean (i.e. $\ex{}{\xi_1}=0$). Then for any $\epsilon>0,B\geq 0$, we have 
    \begin{equation}
        \begin{aligned}
            \P\prn{\exists t\geq 1\text{ s.t. }\sum_{s=1}^t\xi_s \ge B+\epsilon t} \le \exp\prn{-2B\epsilon},\quad\P\prn{\exists t\geq 1\text{ s.t. }\sum_{s=1}^t\xi_s \le -B-\epsilon t} \le \exp\prn{-2B\epsilon}.
        \end{aligned}
    \end{equation}
\end{lemma}

{\small\it\indent Instance (i): $a=[-1/2,1/2,1],p=[0.6,0.2,0.2]$.} Solving $\DLP$ yields $x^*=[1,1,1]$. Then the four policies all degenerate to the greedy policy, i.e. accept all arrivals whenever the budget is available. Following the proof of Theorem \ref{thm:regret_discrete_general}, we define $\mix^{(t)}$ as the policy that applying greedy policy in time $[1,t]$ and applying offline optimal policy $\HOany^{([t+1,T])}$ to the remaining time periods. Specially, define $\mix^{(0)}=\HOany^{([1,T])}$ as the policy that applying hindsight optimal throughout the process and $\mix^{(T)} =\text{Greedy}^{[1,T]}$ as the policy that applying the greedy policy throughout the process. We restate \eqref{eq:regret_decomp}:
\begin{equation}
    \label{eq:regret_decomp1}
    f(\HOany) - f(\FR) = \ex{}{\sum_{t=1}^T\brk{f(\mix^{(t-1)}\big\vert \W^{(T)})-f(\mix^{(t)}\big\vert \W^{(T)})}}.
\end{equation}
It directly follows from the definition of $f$ that 
\begin{equation*}
    \begin{aligned}
        & \indent f(\mix^{(t-1)}\big\vert \W^{(T)})-f(\mix^{(t)}\big\vert \W^{(T)}) \\
        & = \mathds 1\{\mix^{(t)}\text{ wrongly accepts }a^{(t)}\} + \mathds 1\{\mix^{(t)}\text{ wrongly rejects }a^{(t)}\}.
    \end{aligned}
\end{equation*}
Therefore, we have 
\begin{equation*}
    \begin{aligned}
        &\quad\ex{}{f(\mix^{(t-1)}\big\vert \W^{(T)})-f(\mix^{(t)}\big\vert \W^{(T)})}\\ 
        &\ge p_2\cdot\P(\mix^{(t)}\text{ wrongly accepts }a^{(t)}|a^{(t)}=a_2)\\ 
        &= 0.2\cdot \underset{:=\pwr^{(t)}}{\underbrace{\P(\mix^{(t)}\text{ wrongly accepts }a^{(t)}|a^{(t)}=1)}}.
    \end{aligned}
\end{equation*}
In order to give lower bound to the Now we construct a coupling random process $\{\tilde B^{(t)}\}_{t=0}^T$ such that $\tilde B^{(0)}=1/2$ and $\tilde B^{(t+1)} = \max\{\tilde B^{(t)}-a^{(t)},1/2\}$. By induction it is easy to verify that $\tilde B^{(t)}\ge B^{(t)},\forall t\ge 0$. By induction, we know
\begin{align*}
    \tilde B^{(t)} = \max\left\{1/2-a^{(t-1)},1/2-\sum_{s=t-2}^{t-1}a^{(s)},\dots,1/2-\sum_{s=1}^{t-1}a^{(s)},1/2\right\}.
\end{align*}
For $t\ge 2$, applying Lemma \ref{lem:magic} yields a probability of $\Omega(1/\sqrt{t})$ such that $\tilde B^{(t)}\le 1$. Thus means with probability $\Omega(1/\sqrt{t})$, we have $B^{(t)}\le 1$. Note that
\begin{align*}
    & \indent \mbP(B^{(t)}=1) + \mbP(B^{(t-1)}=1) \\
    & \geq \mbP(B^{(t-2)}=0, a^{(t-2)}=-1/2, a^{(t-1)}=-1/2) + \mbP(B^{(t-2)}=1, a^{(t-2)}=-1/2, a^{(t-1)}=1/2) \\
    & \quad\quad\quad + \mbP(B^{(t-2)}=1/2, a^{(t-2)}=-1/2) \\
    & \geq 0.2^2\prn{\mbP(B^{(t-2)}=0) + \mbP(B^{(t-2)}=1/2) + \mbP(B^{(t-2)}=1)} \\
    & = \Omega(1/\sqrt{t}).
\end{align*}
Conditioned on $B^{(t)}=1$, $[a^{(t)},a^{(t+1)},a^{(t+2)}]=[1,1/2,1/2]$ happens with probability $0.2^3$. When this happens, accepting $a^{(t)}$ generates a wrong acceptance of $a^{(t)}=1$ because it follows a rejection of two arrivals of cost $1/2$. We then have $\pwr^{(t)} = \Omega(\mbP(B^{(t)}=1))$. Combining this with \eqref{eq:regret_decomp1} yields
\begin{equation*}
    f_\cD^T(\HOany) - f_\cD^T(\FR) = \sum_{t=1}^T\Omega(1/\sqrt{t})=\Omega(\sqrt{T}).
\end{equation*}

{\small\it\indent Instance (ii): $a=[-1,1,2],p=[0.5,0.4,0.1]$.} Solving $\DLP$ yields $x^*=[1,1,1/2]$. Then $\Bayes$ degenerates to the greedy policy, i.e. accept all arrivals when the budget is available. We construct the same sequence $\{\tilde B^{(t)}\}_{t=0}^T$ as in instance (i). In this case, since $a\cdot p'>0$, applying Lemma \ref{lem:large_deviation} with $B=1$ and $\epsilon=\mbE[a^{(t)}]$ leads to $\P(\tilde B^{(t-2)}\le 1)\geq c$ for some constant $c$ independent of $t$. Following a similar argument in instance (i), we have $\mbP(B^{(t)}=1) + \mbP(B^{(t-1)}=1) = \Omega(1)$ and $\pwr^{(t)} = \Omega(\mbP(B^{(t)}=1))$. Combining this result with \eqref{eq:regret_decomp1} yields
\begin{equation*}
    f_\cD^T(\HOany) - f_\cD^T(\Bayes) = \Omega(T).
\end{equation*}

% \input{pf-main_results}

% \subsection{Proof of Lemma \ref{lem:HOanyL}}
\section{Proof of Results in Section \ref{sec:discrete:MLB}}
\label{appendix:proof_general}

We begin by giving two additional lemmas that help us to build up the proof. The first is about the large deviation of i.i.d. random variables (i.e. probability of deviation of order $t$ from partial sum of $t$ i.i.d. random variables).

\begin{lemma}[Large deviation II] \small
\label{lem:large_deviation_2}
    Assume $\xi_1, \dots, \xi_t, \cdots$ are i.i.d. random variables on $[-1, 1]$ with zero mean (i.e. $\ex{}{\xi_1}=0$). Then for any $t\geq 1$ and $\epsilon>0$, we have 
    \begin{equation}
        \begin{aligned}
            \P\prn{\exists 1\leq s\leq t\text{ s.t. }\sum_{r=1}^s\xi_s \geq \epsilon t} \leq \exp\prn{-\frac{\epsilon^2t}{2}},\quad\P\prn{\exists 1\leq s\leq t\text{ s.t. }\sum_{r=1}^s\xi_s \leq -\epsilon t} \le \exp\prn{-\frac{\epsilon^2t}{2}}.
        \end{aligned}
    \end{equation}
\end{lemma}

We then introduce the following theorem proved in \cite{nagaev1970speed} concerning moderate deviation of i.i.d. random variables, which gives an efficient bound to the maximum of partial sum of zero-mean i.i.d. random variables.
\begin{theorem} \small
    \label{thm:magic}
    Assume $\xi_1,\dots,\xi_t$ are i.i.d. mean zero random variables with $\rho^3=\ex{}{|\xi_s|^3}<\infty$, $\sigma^2=\ex{}{\xi_s^2}<\infty$. Then there exists an absolute constant $K$ such that 
    \begin{equation*}
        \sup_{0\le x<\infty}\left|\P\prn{\max_{1\le s\le t}\sum_{r=1}^s\xi_r\le \sigma x\sqrt{t}} -\prn{\frac{2}{\pi}}^{1/2}\int_0^xe^{-u^2/2}du\right| < \frac{K\rho^6}{\sigma^6\sqrt{t}}.
    \end{equation*}
\end{theorem}

For $x \geq 0$, define
\begin{equation*}
\begin{aligned}
    \bar{\Phi}(x) = \prn{\frac{2}{\pi}}^{1/2}\int_0^xe^{-u^2/2}du, \quad\bar{\Phi}^c(x) = 1 - \bar{\Phi}(x) = \prn{\frac{2}{\pi}}^{1/2}\int_x^{+\infty}e^{-u^2/2}du.
\end{aligned}
\end{equation*}
By setting $x = B/(\sigma\sqrt{t})$ in Theorem \ref{thm:magic}, we get 
\begin{equation*}
\begin{aligned}
    \frac{K\rho^6}{\sigma^6\sqrt{t}}&>\left|\P\prn{\max_{1\le s\le t}\sum_{r=1}^s\xi_r\le B} -\prn{\frac{2}{\pi}}^{1/2}\int_0^{B/(\sigma\sqrt{t})}e^{-u^2/2}du\right| \\ 
    &\ge \P\prn{\max_{1\le s\le t}\sum_{r=1}^s\xi_r\le B}-\bar{\Phi}\prn{\frac{B}{\sigma\sqrt{t}}} \\
    &\ge \P\prn{\max_{1\le s\le t}\sum_{r=1}^s\xi_r\le B}-\prn{\frac{2}{\pi}}^{1/2}B/(\sigma\sqrt{t}).
\end{aligned}
\end{equation*}
Meanwhile, 
\begin{equation*}
\begin{aligned}
    \frac{K\rho^6}{\sigma^6\sqrt{t}}
    & > \left|\P\prn{\max_{1\le s\le t}\sum_{r=1}^s\xi_r\le B} -\prn{\frac{2}{\pi}}^{1/2}\int_0^{B/(\sigma\sqrt{t})}e^{-u^2/2}du\right| \\
    & = \left|\prn{\frac{2}{\pi}}^{1/2}\int_{B/(\sigma\sqrt{t})}^{+\infty}e^{-u^2/2}du - \P\prn{\max_{1\le s\le t}\sum_{r=1}^s\xi_r > B}\right| \\
    & \geq \bar{\Phi}^c\prn{\frac{B}{\sigma\sqrt{t}}} - \P\prn{\max_{1\le s\le t}\sum_{r=1}^s\xi_r\le B}.
\end{aligned}
\end{equation*}
It immediately leads to the following lemma.
\begin{lemma} \small
\label{lem:magic}
Assume $\xi_1,\dots,\xi_t$ are i.i.d. random variables on $[-1,1]$ with zero mean (i.e. $\ex{}{\xi_1}=0$). Let $\rho^3=\ex{}{|\xi_s|^3}<\infty$, $\sigma^2=\ex{}{\xi_s^2}<\infty$, $\kappa:=\rho^6/\sigma^6$. Then for any $B>0$, it holds that
\begin{equation*}
\begin{aligned}
    \P\prn{\max_{1\le s\le t}\sum_{r=1}^{s}\xi_r\le B} & = O\prn{\frac{B}{\sigma\sqrt{t}} + \frac{\kappa}{\sqrt{t}}}, \\
    \P\prn{\max_{1\le s\le t}\sum_{r=1}^{s}\xi_r\le B} & = \Omega\prn{\bar{\Phi}\prn{\frac{B}{\sigma\sqrt{t}}} - \frac{\kappa}{\sqrt{t}}}, \quad\P\prn{\max_{1\le s\le t}\sum_{r=1}^{s}\xi_r > B} & = \Omega\prn{\bar{\Phi}^c\prn{\frac{B}{\sigma\sqrt{t}}} - \frac{\kappa}{\sqrt{t}}}, 
\end{aligned}
\end{equation*}
where in $O(\cdot)$ and $\Omega(\cdot)$ we are hiding absolute constants.
\end{lemma}

Let's additionally define
\begin{equation*}
\begin{aligned}
    T_0 = \left\lfloor T - \frac{1 + 16\ln T}{a_i^2p_i^2}\right\rfloor \vee \left\lfloor T - 4\frac{1+C_{\midd}\ln T}{a_ip_i}\right\rfloor.
\end{aligned}
\end{equation*}

\subsection*{Proof of Theorem \ref{thm:regret_discrete_general}}

{\small\it\indent Case I: $i\leq i_0$.} In this case, the arrival is of ``low cost'' type ($\sum_{j=-m}^ia_jp_j<0$). 

\underline{(i). Let's first bound $\pwa_i^{(t)}$.} We note that when $i=1$, following the proof of Theorem \ref{thm:continuous-lower}, we know that always accepting the lowest cost does no harm. We only consider the case when $i > 1$. Recall the definition $a_{-1}/r_{-1}\le\dots\le a_{-m}/r_{-m}<0<a_n/r_n\le\dots\le a_1/r_1.$. Note that $a^{(l)}\mathds 1\{a^{(l)}/r^{(l)}<a_i/r_i\}$, $l=t+1,\cdots,T$ are i.i.d. random variables with expectation
\begin{align*}
    \sum_{j=-m}^{i-1}a_jp_j\triangleq \Delta_{i-1} < 0.
\end{align*}
Then $a^{(l)}\mathds 1\{a^{(l)}/r^{(l)}<a_i/r_i\}-\Delta_{i-1}$ are zero-mean i.i.d. random variables in $[-1,1]$. We can bound $\pwa_i^{(t)}$ by
\begin{equation} \label{eq:ac-discrete-low_general}
    \begin{aligned}
        & \quad\ \P\prn{B^{(t)} - a_i < \max_{s\in[t+1, T]}\sum_{l=t+1}^sa^{(l)}\mathds 1\{a^{(l)}/r^{(l)}<a_i/r_i\}, B^{(t)}\geq C_{\low}\ln (T-t+1)} \\
        & \leq \P\prn{C_{\low}\ln (T-t+1) - a_i < \max_{s\in[t+1, T]}\sum_{l=t+1}^sa^{(l)}\mathds 1\{a^{(l)}/r^{(l)}<a_i/r_i\}} \\
        & = \P\prn{\exists s\in[t+1, T]: \sum_{l=t+1}^s\prn{a^{(l)}\mathds 1\{a^{(l)}/r^{(l)}<a_i/r_i\}-\Delta_{i-1}} > C_{\low}\ln (T-t+1) - a_i - \Delta_{i-1}\cdot(s-t)} \\
        & \leq \exp\prn{2(C_{\low}\ln (T-t+1) - a_i)\Delta_{i-1}} \\
        & \leq e^2 (T-t+1)^{2C\Delta_{i-1}}
    \end{aligned}
\end{equation}
where in the last inequality we use Lemma \ref{lem:large_deviation}. When $C_{\low} \geq -\frac{1}{\Delta_{i_0-1}}$, we know that
\begin{align*}
    e^2 (T-t+1)^{2C\Delta_{i-1}} \leq e^2 (T-t+1)^{-2\frac{\Delta_{i-1}}{\Delta_{i_0-1}}} = O\prn{(T-t+1)^{-2}},
\end{align*}
where in $O(\cdot)$ we are hiding an absolute constant.

\underline{(ii). Let's first bound $\pwr_i^{(t)}$.} It suffices to bound $\P\prn{B^{(t)} < C_{\low}\ln(T-t+1)}$. We cover the event by two parts: (a) $B^{(s)} < C_{\midd}\ln (T-s+1)$ for $s=1,\dots, t$; (b) there exists $1\le s\le t$ such that $B^{(s)}\geq C_{\midd}\ln (T-s+1)$, and at any time $s'\in(s, t]$, the budget is always below $C_{\midd}\ln(T-s')$. Then
\begin{equation*}
    \begin{aligned}
        & \quadd \P(B^{(t)} < C_{\low}\ln (T-t+1)) \\
        & = \P(B^{(t)} < C_{\low}\ln (T-t+1), \text{ (a) happens})+\P(B^{(t)} < C_{\low}\ln (T-t+1), \text{ (b) happens}).
    \end{aligned}
\end{equation*}

Consider the situation when event (a) holds. It follows that only type $j\leq i_0$ will be accepted throughout time $1$ to $t$. Therefore, we have 
\begin{equation*}
    \begin{aligned}
       & C_{\low}\ln(T-t+1) > B^{(t)} = B^{(1)} -\sum_{s=1}^ta^{(s)}\mathds 1\{\text{accept $a^{(s)}$}\}  
         \geq -\sum_{s=1}^ta^{(s)}\mathds 1\{a^{(s)}\leq a_{i_0}\}.
    \end{aligned}
\end{equation*}
Using the fact that $a^{(s)}\mathds 1\{a^{(s)}\leq a_{i_0}\}\ (s=1, \cdots, t)$ are i.i.d. random variables in $[-1,1]$ with expectation $\Delta_{i_0}$, we get  
\begin{equation}
    \label{eq:rj-discrete-low-1}
    \begin{aligned}
        & \quadd \P(B^{(t)} < C_{\low}\ln (T-t+1), \text{ (a) happens}) \\
        & \leq \P\prn{C_{\low}\ln(T-t+1)\geq - \sum_{s=1}^ta^{(s)}\mathds 1\{a^{(s)}\leq a_{i_0}\}} \\
        & \leq \P\prn{\sum_{s=1}^t\prn{a^{(s)}\mathds 1\{a^{(s)}/r^{(s)}\leq a_{i_0}/r_{i_0}\} - \Delta_{i_0}}\geq - \Delta_{i_0}t - C_{\low}\ln(T-t+1)} \\
        &\leq \exp\prn{-\frac{\prn{-\Delta_{i_0}t - C_{\low}\ln T}_+^2}{2t}}.
    \end{aligned}
\end{equation}
The last inequality holds by the Hoeffding's inequality.

Consider the situation when event (b) holds, without loss of generality, let $s<t$ be the largest time such that $B^{(s)}\geq C_{\midd}\ln(T-t+1)$. Then $B^{(r)}<C_{\midd}\ln(T-r+1)$ and for $s < r \leq t$. It is not difficult to observe that $B^{(s+1)}\geq C_{\midd}\ln(T-s) - a_{i_0+1} \geq C_{\midd}\ln(T-t+1) - a_{i_0+1}$. Thus, 
\begin{equation*}
    \begin{aligned}
        (C_{\midd}-C_{\low})\ln(T-t+1) - a_{i_0+1}\leq B^{(s+1)} - B^{(t)} \leq \sum_{r=s+1}^ta^{(l)}\mathds 1\{\text{accept $a^{(l)}$}\} \le \sum_{r=s+1}^ta^{(l)}\mathds 1\{a^{(l)}\leq a_{i_0}\}.
    \end{aligned}
\end{equation*}
Therefore, 
\begin{equation}
    \label{eq:rj-discrete-low-2}
    \begin{aligned}
        & \quadd \P(B^{(t)} < C_{\low}\ln(T-t+1), \text{ (b) happens}) \\
        & \le \P\prn{\exists s\in[1, t): \sum_{r=s+1}^ta^{(l)}\mathds 1\{a^{(l)}/r^{(l)}\leq a_{i_0}/r_{i_0}\} \geq (C_{\midd}-C)\ln(T-t+1) - a_{i_0+1}} \\ 
        & \leq \P\prn{\exists s: \sum_{r=1}^sa^{(t+1-r)}\mathds 1\{a^{(t+1-r)}/r^{(t+1-r)}\leq a_{i_0}/r_{i_0}\} \geq (C_{\midd}-C)\ln(T-t+1) - a_{i_0+1}} \\
        & \leq \P\prn{\exists s: \sum_{r=1}^s(a^{(t+1-r)}\mathds 1\{a^{(t+1-r)}/r^{(t+1-r)}\leq a_{i_0}/r_{i_0}\} - \Delta_{i_0}) \geq (C_{\midd}-C)\ln(T-t+1) - a_{i_0+1} - \Delta_{i_0}s} \\
        & \leq \exp\prn{2\prn{(C_{\midd}-C)\ln(T-t+1) - 1}\Delta_{i_0}} \\
        & \leq e^2(T-t+1)^{2(C_{\midd} - C)\Delta_{i_0}}
    \end{aligned}
\end{equation}
where in the last inequality we use Lemma \ref{lem:large_deviation}. When $C_{\midd}\geq C_{\low} - \frac{1}{\Delta_{i_0}}$, we know that
\begin{align*}
    e^2 (T-t+1)^{2(C_{\midd}-C)\Delta_{i_0}} \leq e^2 (T-t+1)^{-2\frac{\Delta_{i_0}}{\Delta_{i_0}}} = O\prn{(T-t+1)^{-2}},
\end{align*}
where in $O(\cdot)$ we are hiding an absolute constant.

Combining Lemma \ref{lem:there_is_gap2_general}, \eqref{eq:ac-discrete-low_general}, \eqref{eq:rj-discrete-low-1}, \eqref{eq:rj-discrete-low-2} and letting
\begin{equation*}
\begin{aligned}
    t_0 = \left\lfloor\frac{4C_{\low}\ln T}{-\Delta_{i_0}}\right\rfloor,
\end{aligned}
\end{equation*}
the total loss incurred by wrongly accepting or rejecting the low-cost type of arrivals can be bounded by
\begin{equation} \label{eq:discrete-low}
\begin{aligned}
    & \quadd \sum_{t=1}^T\sum_{i=1}^{i_0-1}p_i\rmax(T-t)\cdot\pwa_i^{(t)} + \sum_{t=1}^T\sum_{i=1}^{i_0-1}p_i\rmax\cdot\pwr_i^{(t)} \\
    & \leq \sum_{t=1}^T \rmax(T-t)\cdot O\prn{(T-t+1)^{-2}} + t_0\rmax + r_i\sum_{t=t_0+1}^T \left(\exp\prn{-\frac{\prn{-\Delta_{i_0}t - C_{\low}\ln T}_+^2}{2t}} + O\prn{(T-t+1)^{-2}}\right) \\
    & \leq O(\rmax\ln T) + O\prn{\frac{\rmax\ln T}{\Delta_{i_0}\Delta_{i_0-1}}} + \rmax\sum_{t=t_0+1}^{+\infty}\exp\prn{-\frac{\Delta_{i_0}^2t}{8}} + O(1) \\
    & \leq O\prn{\frac{\rmax\ln T}{\Delta_{i_0}\Delta_{i_0-1}} + \frac{\rmax}{\Delta_{i_0}^2}}.
    \end{aligned}
\end{equation}
Note that in $O(\cdot)$ we are hiding absolute constant terms.

{\small\it\indent Case II: $i=i_0+1$.} In this case, the arrival is of ``boundary'' type ($\Delta_{i_0}<0$ while $\Delta_{i_0+1}\geq 0$). 

\underline{(i). Let's first bound $\pwa_i^{(t)}$.}
\begin{equation} \label{eq:ac-discrete-mid}
    \begin{aligned}
        & \quadd \P\prn{B^{(t)} - a_{i_0+1} < \max_{s\in[t+1, T]}\sum_{l=t+1}^sa^{(l)}\mathds 1\{a^{(l)}/r^{(l)}<a_{i_0+1}/r_{i_0+1}\}, B^{(t)} \geq C_{\midd}\ln(T-t+1)} \\
        & \leq \P\prn{\max_{s\in[t+1, T]}\sum_{l=t+1}^sa^{(l)}\mathds 1\{a^{(l)}/r^{(l)}<a_{i_0+1}/r_{i_0+1}\} > C_{\midd}\ln(T-t+1) - a_{i_0+1}} \\
        & = \P\prn{\exists s\in(t, T]: \sum_{l=t+1}^s\prn{a^{(l)}\mathds 1\{a^{(l)}/r^{(l)}<a_{i_0+1}/r_{i_0+1}\}-\Delta_{i_0}} > C_{\low}\ln(T-t+1) - a_{i_0+1} - \Delta_{i_0}\cdot(s-t)} \\
        & \leq \exp\prn{2\prn{C_{\midd}\ln(T-t+1) - 1)\Delta_{i_0}}} \\
        & \leq e^2(T-t+1)^{2C_{\midd}\Delta_{i_0}},
    \end{aligned}
\end{equation}
where in the last inequality we use Lemma \ref{lem:large_deviation}. When $C_{\midd} \geq C_{\low} -\frac{1}{\Delta_{i_0}} \geq -\frac{1}{\Delta_{i_0}}$, we know that
\begin{align*}
    e^2 (T-t+1)^{2C_{\midd}\Delta_{i_0}} \leq e^2 (T-t+1)^{-2\frac{\Delta_{i_0}}{\Delta_{i_0}}} = O\prn{(T-t+1)^{-2}},
\end{align*}
where in $O(\cdot)$ we are hiding an absolute constant.

\underline{(ii). Let's then bound $\pwr_i^{(t)}$, which is a more complicated case.} By Lemma \ref{lem:there_is_gap2_general} we know that $\pwr_i^{(t)}$ can be bounded as follows:
\begin{equation} \label{eq:rj-discrete-mid}
\begin{aligned}
    & \quadd\P\prn{B^{(t)}-a_i\geq\max_{s\in[t+1,T]}\sum_{l=t+1}^sa^{(l)}\mathds 1\{a^{(l)}/r^{(l)}\leq a_i/r_i\}, B^{(t)} < C_{\midd}\ln (T-t+1)} \\
    & \leq \P\prn{C_{\midd}\ln (T-t+1) -a_i\geq\max_{s\in[t+1,T]}\sum_{l=t+1}^sa^{(l)}\mathds 1\{a^{(l)}/r^{(l)}\leq a_i/r_i\}, B^{(t)} < C_{\midd}\ln (T-t+1)} \\
    & = \P\prn{C_{\midd}\ln (T-t+1) -a_i\geq\max_{s\in[t+1,T]}\sum_{l=t+1}^sa^{(l)}\mathds 1\{a^{(l)}/r^{(l)}\leq a_i/r_i\}}\cdot\P\prn{B^{(t)} < C_{\midd}\ln (T-t+1)}.
\end{aligned}
\end{equation}
We need to differentiate between two different scenarios.

\underline{(a). $\Delta_{i_0+1}>0$.}

\begin{equation} \label{eq:rj-discrete-mid-nondegenerate}
\begin{aligned}
    & \quadd \P\prn{C_{\midd}\ln (T-t+1) -a_i\geq\max_{s\in[t+1,T]}\sum_{l=t+1}^sa^{(l)}\mathds 1\{a^{(l)}/r^{(l)}\leq a_i/r_i\}} \\
    & \leq \P\prn{C_{\midd}\ln (T-t+1) -a_i\geq\sum_{l=t+1}^Ta^{(l)}\mathds 1\{a^{(l)}/r^{(l)}\leq a_i/r_i\}} \\
    & \leq \P\prn{C_{\midd}\ln (T-t+1) -\Delta_{i_0+1}(T-t)\geq\sum_{l=t+1}^T\left(a^{(l)}\mathds 1\{a^{(l)}/r^{(l)}\leq a_i/r_i\} - \Delta_{i_0+1}\right)} \\
    & \leq \exp\prn{-\frac{2(\Delta_{i_0+1}(T-t)-C_{\midd}\ln T)_+^2}{T-t}}
\end{aligned}
\end{equation}
Combining Lemma \ref{lem:there_is_gap2_general}, \eqref{eq:ac-discrete-mid}, \eqref{eq:rj-discrete-mid}, \eqref{eq:rj-discrete-mid-nondegenerate} and letting
\begin{equation*}
    t_{\midd} = \left\lfloor T - \frac{2C_{\midd}\ln T}{\Delta_{i_0+1}}\right\rfloor
\end{equation*}
yields
\begin{equation} \label{eq:discrete-mid-nondegenrate}
\begin{aligned}
    & \quadd p_{i_0}\rmax\sum_{t=1}^T(T-t)\cdot\pwa_i^{(t)} + p_{i_0}\rmax\sum_{t=t_{\midd+1}}^{T}\pwr_i^{(t)} + p_{i_0}\rmax\sum_{t=1}^{t_{\midd}}\pwr_i^{(t)} \\
    & \leq p_{i_0}\rmax\sum_{t=1}^T (T-t)\cdot O\prn{(T-t+1)^{-2}} + p_{i_0}\rmax O\prn{\frac{C_{\midd}}{\Delta_{i_0+1}}\ln T} + \\
    & \quad\quad p_{i_0}\rmax\sum_{t=1}^{t_\midd}O\prn{\exp\prn{-\frac{2(\Delta_{i_0+1}(T-t)-C_{\midd}\ln T)_+^2}{T-t}}} \\
    & = p_{i_0}\rmax\prn{O(\ln T) + O\prn{\frac{ C_{\midd}\ln T}{\Delta_{i_0+1}}} + O\prn{\frac{1}{\Delta_{i_0+1}^2}}} \\
    & = O\prn{\frac{\rmax\ln T}{|\Delta_{i_0}\Delta_{i_0+1}|} + \frac{\rmax}{\Delta_{i_0+1}^2}}.
    \end{aligned}
\end{equation}
Note that in $O(\cdot)$ we are hiding absolute constant terms.

\underline{(b). $\Delta_{i_0+1}=0$.}
By Lemma \ref{lem:magic}, we know that
\begin{equation} \label{eq:rj-discrete-mid-1}
\begin{aligned}
    & \quadd \P\prn{C_{\midd}\ln (T-t+1) -a_i\geq\max_{s\in[t+1,T]}\sum_{l=t+1}^sa^{(l)}\mathds 1\{a^{(l)}/r^{(l)}\leq a_i/r_i\}} \\
    & \leq \P\prn{C_{\midd}\ln (T-t+1) -a_i\geq\max_{s\in[t+1,T]}\sum_{l=t+1}^sa^{(l)}\mathds 1\{a^{(l)}/r^{(l)}\leq a_i/r_i, u^{(r)}\leq x^*\}} \\
    & \leq O\prn{\frac{1}{\sqrt{T-t}}\left(\frac{C_{\midd}\ln(T-t+1)}{\sigma} + \kappa\right)}
\end{aligned}
\end{equation}
It suffices to bound $\P\prn{B^{(t)} < C_{\midd}\ln (T-t+1)}$. Note that for any high cost type $a_j\ (j > i_0+1)$, its buffer is lower bounded by
\begin{equation*}
\begin{aligned}
    \frac{p_ja_j}{2}(T-t+1) + C_{\high}\ln (T-t+1)\geq \frac{p_ja_j}{2}(T-t+1) + \frac{1}{p_ja_j}\ln (T-t+1) \geq \sqrt{2(T-t+1)\ln(T-t+1)}
\end{aligned}
\end{equation*}
We cover the event by two parts: (a) $B^{(s)} < \sqrt{2(T-s+1)\ln(T-s+1)}$ for $s=1,\dots, t$; (b) there exists $1\le s\le t$ such that $B^{(s)}\geq \sqrt{2(T-s+1)\ln(T-s+1)}$, and at any time $s'\in(s, t]$, the budget is always below $\sqrt{2(T-s'+1)\ln(T-s'+1)}$. Then
\begin{equation*}
    \begin{aligned}
        & \quadd \P(B^{(t)} < C_{\midd}\ln (T-t+1)) \\
        & = \P(B^{(t)} < C_{\midd}\ln (T-t+1), \text{ (a) happens})+\P(B^{(t)} < C_{\midd}\ln (T-t+1), \text{ (b) happens}).
    \end{aligned}
\end{equation*}

Consider the situation when event (a) holds. It follows that only type $j\leq i_0+1$ will be accepted throughout time $1$ to $t$. Similar to the proof in Theorem \ref{thm:continuous-upper}, construct a new process $\{\tilde B^{(t)}\}_{t}$ as a ``coupled'' version of the process $\{B^{(t)}\}_{t}$ as follows. $\tilde B^{(1)} = B^{(1)} = 0$. For general $t\geq 1$, we define
\begin{equation*} 
\label{eq:discrete-coupled_general}
\begin{aligned}
    \tilde B^{(t+1)} = \max\left\{\tilde B^{(t)} - a^{(t)}\left(\mathds 1\{a^{(t)}/r^{(t)} < a_i/r_i\} + \mathds 1\{(a^{(t)},r^{(t)}) = (a_i,r_i), u^{(t)}\leq x^*\}\right), 0\right\}.
\end{aligned}
\end{equation*}
That is, for each sample path $\{a^{(t)}\}_t$ with $\{u^{(t)}\}_t$, in the ``coupled'' version regardless of the budget and the buffer, in each time $t$ we always accept the arrival as long as it is of low or middle type (here we also couple the random seed when we face $a_{i}$). If the budget drops below $0$, we restart the budget level as $0$. We can show that for each sample path $\tilde B^{(t)} \leq B^{(t)}$ for any $t$ via induction. Apparently $\tilde B^{(1)} \leq B^{(1)}$. Suppose we have $\tilde B^{(t)} \leq B^{(t)}$. If at time $t$ the arrival $a^{(t)}$ is rejected by $\MLB$, then it implies $a^{(t)} > 0$. We have $\tilde B^{(t+1)} \leq \tilde B^{(t)}\leq B^{(t)} = B^{(t+1)}$. If the arrival is accepted by $\MLB$, then we also have
\begin{equation*}
\begin{aligned}
    \tilde B^{(t+1)} & = \max\left\{\tilde B^{(t)} - a^{(t)}\left(\mathds 1\{a^{(t)} < a_i\} + \mathds 1\{(a^{(t)},r^{(t)}) = (a_i,r_i), u^{(t)}\leq x^*\}\right), 0\right\} \\
    & \leq \max\left\{B^{(t)} - a^{(t)}\left(\mathds 1\{a^{(t)} < a_i\} + \mathds 1\{(a^{(t)},r^{(t)}) = (a_i,r_i), u^{(t)}\leq x^*\}\right), 0\right\} \\
    & = B^{(t+1)}.
\end{aligned}
\end{equation*}
This leads to
\begin{equation} \label{eq:rj-discrete-mid-2-1}
\begin{aligned}
    & \quadd \P(B^{(t)} < C_{\midd}\ln (T-t+1), \text{ (a) happens}) \\
    & \leq \P(\tilde B^{(t)} < C_{\midd}\ln (T-t+1)) \\
    & \leq O\prn{\frac{C_{\midd}\ln(T-t+1)}{\sigma\sqrt{t}} + \frac{\kappa}{\sqrt{t}}}.
\end{aligned}
\end{equation}
The last inequality holds by Lemma \ref{lem:magic}. Note that here we have utilized the fact that $\tilde B^{(t)}$ is the maximum of the first $t$ positions of a random walk $Z^{(t-1)}, Z^{(t-2)}, \cdots$, where $$Z^{(t)} = - a^{(t)}\left(\mathds 1\{a^{(t)} < a_i\} + \mathds 1\{(a^{(t)},r^{(t)}) = (a_i,r_i), u^{(t)}\leq x^*\}\right)$$ is zero-mean, independent, and bounded within $[-1, 1]$.

Consider the situation when event (b) holds, without loss of generality, let $s<t$ be the largest time such that $B^{(s)}\geq \sqrt{2(T-s+1)\ln(T-s+1)}$. Then $B^{(r)}<\sqrt{2(T-r+1)\ln(T-r+1)}$ for $s < r \leq t$. It is not difficult to observe that $B^{(s+1)}\geq \sqrt{2(T-s+1)\ln(T-s+1)} - a_{i_0+1} \geq \sqrt{2(T-s)\ln(T-s)} - 1$. Thus, 
\begin{equation*}
    \begin{aligned}
        & \quadd \sqrt{2(T-s)\ln(T-s)} - 1 - C_{\midd}\ln(T-t+1) \\
        & \leq B^{(s+1)} - B^{(t)} \leq \sum_{r=s+1}^ta^{(l)}\mathds 1\{\text{accept $a^{(l)}$}\} \leq \sum_{r=s+1}^ta^{(l)}\mathds 1\{a^{(l)}\leq a_{i_0}\}.
    \end{aligned}
\end{equation*}
Meanwhile, when $t \leq T - (1+\sqrt{2})^2C_{\midd}^2\ln T$, we can observe that for any $s<t$:
\begin{equation*}
\begin{aligned}
    & \quadd \sqrt{2(T-s)\ln(T-s)} - 1 - C_{\midd}\ln(T-t+1) \\
    & \geq \sqrt{2(T-s)\ln(T-s)} - 1 - (\sqrt{2}-1)\sqrt{\frac{T-t}{\ln T}}\ln(T-t+1) \\
    & \geq \sqrt{(t-s)\ln(T-s)} - 1
\end{aligned}
\end{equation*}
Therefore, 
\begin{equation}
    \label{eq:rj-discrete-mid-2-2}
    \begin{aligned}
        & \quadd \P(B^{(t)} < C_{\midd}\ln(T-t+1), \text{ (b) happens}) \\
        & \le \P\prn{\exists s\in[1, t): \sum_{r=s+1}^ta^{(l)}\mathds 1\{a^{(l)}\leq a_{i_0}, u^{(r)}\leq x^*\} \geq \sqrt{(t-s)\ln(T-s)} - 1} \\ 
        & \leq \sum_{s=1}^{+\infty}\P\prn{\sum_{r=1}^s(a^{(t+1-r)}\mathds 1\{a^{(t+1-r)}\leq a_{i_0}, u^{(t+1-r)}\leq x^*\}) \geq \sqrt{s\ln(s+T-t)} - 1} \\
        & \leq \sum_{s=1}^t\exp\prn{-\frac{2(\sqrt{s\ln(s+T-t)}-1)_+^2}{s}} \\
        & = \sum_{s=1}^t O\left((s+T-t)^{-2}\right) \\
        & = O((T-t)^{-1})
    \end{aligned}
\end{equation}
where in $O(\cdot)$ we are hiding absolute constants. In the last inequality we use Hoeffding's inequality by noticing that $a^{(t+1-r)}\mathds 1\{a^{(t+1-r)}\leq a_{i_0}, u^{(t+1-r)}\leq x^*\}$ is bounded within $[-\alpha, 1-\alpha]$.

Combining Lemma \ref{lem:there_is_gap2_general}, \eqref{eq:ac-discrete-mid}, \eqref{eq:rj-discrete-mid}, \eqref{eq:rj-discrete-mid-1}, \eqref{eq:rj-discrete-mid-2-1}, \eqref{eq:rj-discrete-mid-2-2} and letting
\begin{equation*}
\begin{aligned}
    t_{\midd} = \left\lfloor T - (1+\sqrt{2})^2C_{\midd}^2\ln T\right\rfloor,
\end{aligned}
\end{equation*}
the total loss incurred by wrongly accepting or rejecting the middle-cost type of arrivals can be bounded by
\begin{equation} \label{eq:discrete-mid}
\begin{aligned}
    & \quadd p_{i_0}\rmax\sum_{t=1}^T(T-t)\cdot\pwa_i^{(t)} + p_{i_0}\rmax\sum_{t=t_{\midd+1}}^{T}\pwr_i^{(t)} + p_{i_0}\rmax\sum_{t=1}^{t_{\midd}}\pwr_i^{(t)} \\
    & \leq p_{i_0}\rmax\sum_{t=1}^T (T-t)\cdot O\prn{(T-t+1)^{-2}} + p_{i_0}\rmax O\prn{C_{\midd}^2\ln T} + \\
    & \quad\quad p_{i_0}\rmax\sum_{t=1}^{t_\midd}O\prn{\frac{1}{\sqrt{T-t}}\frac{C_{\midd}\ln(T-t+1)}{\sigma}\cdot\prn{ \frac{C_{\midd}\ln(T-t+1)}{\sigma\sqrt{t}} + \frac{\kappa}{\sqrt{t}} + \frac{1}{T-t}}} \\
    & = p_{i_0}\rmax\prn{O(\ln T) + O(\rmax C_{\midd}^2\ln T) + O\prn{\frac{ C_{\midd}^2\ln^2T}{\sigma^2}} + O\prn{\frac{\kappa  C_{\midd}\ln T}{\sigma}}} \\
    & = O\prn{\frac{p_{i_0}\ln^2T}{\sigma^2\Delta_{i_0}^2} + \frac{p_{i_0}\rmax\kappa\ln T}{\sigma\Delta_{i_0}}} \\
    & = O\prn{\frac{p_{i_0}\rmax\ln^2T}{\sigma^2\Delta_{i_0}^2}}.
    \end{aligned}
\end{equation}
Note that in $O(\cdot)$ we are hiding absolute constant terms. The last equality holds because
\begin{align*}
    \kappa = \frac{\ex{}{|Z^{(t)} - \E[Z^{(t)}]|^3}^2}{\ex{}{|Z^{(t)} - \E[Z^{(t)}]|^2}^3} = \frac{\ex{}{|Z^{(t)}|^3}^2}{\ex{}{|Z^{(t)}|^2}^3} \leq \frac{1}{\ex{}{|Z^{(t)}|^2}} \leq \frac{1}{\sigma\ex{}{|Z^{(t)}|}} \leq \frac{1}{\sigma\Delta}.
\end{align*}

{\small\it\indent Case III: $i>i_0+1$.} In this case, the arrival is of high-cost type. Define
\begin{equation*}
\begin{aligned}
    t_i = \left\lfloor T - \frac{16\ln T}{a_i^2p_i^2}\right\rfloor
\end{aligned}
\end{equation*}

\underline{(i). Let's first bound $\pwa_i^{(t)}$.} We have when $t \leq t_i$:
\begin{equation} \label{eq:ac-discrete-high}
    \begin{aligned}
        & \quadd \P\prn{B^{(t)} - a_i<\max_{s\in[t+1,T]}\sum_{l=t+1}^sa^{(l)}\mathds 1\{a^{(l)}/r^{(l)}<a_i/r_i\}, B^{(t)} > \prn{\Delta_{i-1} + \frac{a_ip_i}{2}}(T-t+1) + C_i\ln (T-t+1)} \\
        & \leq \P\prn{\max_{s\in[t+1, T]}\sum_{l=t+1}^sa^{(l)}\mathds 1\{a^{(l)}/r^{(l)}<a_i/r_i\} > \prn{\Delta_{i-1} + \frac{a_ip_i}{2}}(T-t+1) + C_i\ln (T-t+1) - a_i} \\
        & \leq \P\prn{\max_{s\in[t+1, T]}\sum_{l=t+1}^s\prn{a^{(l)}\mathds 1\{a^{(l)}/r^{(l)}<a_i/r_i\} - \Delta_{i-1}} > \frac{a_ip_i}{2}(T-t+1)} \\
        & \leq \exp\prn{-\frac{a_i^2p_i^2(T-t)}{8}}, 
    \end{aligned}
\end{equation}
where in the last inequality we use Lemma \ref{lem:large_deviation_2}.

\underline{(ii). Let's then bound $\pwr_i^{(t)}$.} We have when $t \leq t_i$:
\begin{equation} \label{eq:rj-discrete-high}
\begin{aligned}
    & \quadd \P\prn{B^{(t)}-a_i>\max_{s\in[t+1,T]}\sum_{l=t+1}^sa^{(l)}\mathds 1\{a^{(l)}/r^{(l)}\leq a_i/r_i\}, B^{(t)} \leq \prn{\Delta_i - \frac{a_ip_i}{2}}(T-t+1) + \frac{\ln(T-t+1)}{a_ip_i}} \\
    & \leq \P\prn{\sum_{l=t+1}^Ta^{(l)}\mathds 1\{a^{(l)}/r^{(l)}\leq a_i/r_i\} < \prn{\Delta_i - \frac{a_ip_i}{2}}(T-t+1) + \frac{\ln(T-t+1)}{a_ip_i}} \\
    & \leq \P\prn{\sum_{l=t+1}^T\prn{a^{(l)}\mathds 1\{a^{(l)}/r^{(l)}\leq a_i/r_i\} - \Delta_i} < \prn{- \frac{a_ip_i}{4}}(T-t)} \\
    & \leq \exp\prn{-\frac{a_i^2p_i^2(T-t)}{8}},
\end{aligned}
\end{equation}
where in the last inequality we use Hoeffding's inequality.

Therefore, combining \eqref{eq:ac-discrete-high} and \eqref{eq:rj-discrete-high}, the total expected loss incurred by wrongly accepting or rejecting a high-cost type $a_i$ can be bounded by
\begin{equation} \label{eq:discrete-high}
\begin{aligned}
    & \quadd p_{i}\rmax\sum_{t=1}^{t_i}(T-t)\cdot\pwa_i^{(t)} + p_{i}\sum_{t=t_i+1}^{T}\pwa_i^{(t)} + p_{i}\rmax\sum_{t=1}^{t_i}\pwr_i^{(t)} + p_{i}\rmax\sum_{t=t_i+1}^{T}\pwr_i^{(t)} \\
    & \leq p_{i}\rmax\sum_{t=1}^T (T-t)\cdot O\prn{T^{-2}} + p_{i}\cdot O\prn{\frac{\ln T}{p_i^2a_i^2}} + p_{i}\rmax\sum_{t=1}^{t_i}\exp\prn{-\frac{a_i^2p_i^2(T-t)}{8}} + p_i\rmax\cdot O\prn{\frac{\ln T}{p_i^2a_i^2}} \\
    & = O\prn{\frac{\rmax\ln T}{p_ia_i^2}}.
    \end{aligned},
\end{equation}
where in $O(\cdot)$ we are hiding absolute constant terms.

{\small\it\indent Wrap-up.} To summarize:

\noindent 1. If $\Delta_{i_0+1}>0$, combining \eqref{eq:discrete-low}, \eqref{eq:discrete-mid-nondegenrate}, \eqref{eq:discrete-high} yields
\begin{equation*}
\begin{aligned}
    f_\cD^T(\HOany) - f_\cD^T(\MLB) = \rmax O\prn{\frac{\ln T}{|\Delta_{i_0-1}\Delta_{i_0}|} + \frac{1}{\Delta_{i_0}^2} + \frac{\ln T}{|\Delta_{i_0}\Delta_{i_0+1}|} + \frac{1}{\Delta_{i_0+1}^2}+\sum_{i>i_0+1}\frac{\ln T}{p_ia_i^2}},
\end{aligned}
\end{equation*}

\noindent 2. If $\Delta_{i_0+1}=0$, combining \eqref{eq:discrete-low}, \eqref{eq:discrete-mid}, \eqref{eq:discrete-high} yields
\begin{equation*}
\begin{aligned}
    f_\cD^T(\HOany) - f_\cD^T(\MLB) = \rmax O\prn{\frac{\ln^2 T}{\sigma^2\Delta_{i_0}^2}+\sum_{i>i_0+1}\frac{\ln T}{p_ia_i^2}},
\end{aligned}
\end{equation*}
In $O(\cdot)$ we are always hiding absolute constants.

\section{Proof of Lemmas}

\subsection{Proof of Lemma \ref{lem:HOanyL}}
For $\{X^{(t)}\}_{t=1}^T\subset [0,1]^T$ in $\HOanyL,$ we construct a coupling solutions in $\{0,1\}^T$ with constant loss of total rewards. Let
\begin{align*}
    Y^{(t)} = \min\left\{1,\left\lfloor\sum_{l=1}^tX^{(l)}\mathds 1\{(a^{(l)},r^{(l)})=(a^{(t)},r^{(t)})\}-\sum_{l=1}^{t-1}Y^{(l)}\mathds 1\{(a^{(l)},r^{(l)})=(a^{(t)},r^{(t)})\}\right\rfloor\right\}.
\end{align*}
Then we have $\sum_{l=1}^tY^{l}\mathds 1\{(a^{(l)},r^{(l)})=(a_i,r_i)\}\le \sum_{l=1}^tX^{l}\mathds 1\{(a^{(l)},r^{(l)})=(a_i,r_i)\}$ for all $t\in[1,T]$ and all $i$. As a result, we have $\sum_{l=1}^tY^{(l)}a^{(l)}\le \sum_{l=1}^tX^{(l)}a^{(l)}\le 0$ for all $t\in[1,T].$ Since $Y^{(t)}\in\{0,1\},$ we get a feasible solution to integer programming offline problem \eqref{eq:HOany_discrete}. The remain is to examine that: 
\begin{equation*}
    \sum_{t=1}^Tr^{(t)}X^{(t)}-\sum_{t=1}^Tr^{(t)}Y^{(t)}\le \sum_{i=1}^nr_i.
\end{equation*}
We first note that $X^{(t)}=Y^{(t)}=1$ for $a^{(t)}\le 0$. We the only need to verify that
\begin{align*}
    \sum_{l=1}^tY^{l}\mathds 1\{(a^{(l)},r^{(l)})= (a_i,r_i)\}\ge \sum_{l=1}^tX^{l}\mathds 1\{(a^{(l)},r^{(l)})=(a_i,r_i)\}-1
\end{align*}
holds for all $1\le i\le n$. We prove this by induction. For $t=1$, this holds trivially. Assume it holds for $t-1$, then if it does not hold for $t$, we must have $\sum_{l=1}^tX^{l}\mathds 1\{(a^{(l)},r^{(l)})= (a_i,r_i)\}- \sum_{l=1}^{t-1}Y^{l}\mathds 1\{(a^{(l)},r^{(l)})=(a_i,r_i)\}>1.$ Then by definition, we have $Y^{(t)} = 1\ge X^{(t)}$, which leads to $\sum_{l=1}^{t-1}X^{l}\mathds 1\{(a^{(l)},r^{(l)})= (a_i,r_i)\}- \sum_{l=1}^{t-1}Y^{l}\mathds 1\{(a^{(l)},r^{(l)})=(a_i,r_i)\}\ge \sum_{l=1}^{t}X^{l}\mathds 1\{(a^{(l)},r^{(l)})= (a_i,r_i)\}- \sum_{l=1}^{t}Y^{l}\mathds 1\{(a^{(l)},r^{(l)})=(a_i,r_i)\} > 1,$ a contradiction! Hence we have completed the induction process and get $\sum_{l=1}^tY^{l}\mathds 1\{(a^{(l)},r^{(l)})= (a_i,r_i)\}\ge \sum_{l=1}^tX^{l}\mathds 1\{(a^{(l)},r^{(l)})=(a_i,r_i)\}-1$ for all $1\le i\le n.$ Then the result follows since $f_\cD^T(\HOany)\ge \ex{}{\sum_{t=1}^Tr^{(t)}Y^{(t)}}$.

\subsection{Proof of Lemma \ref{lem:there_is_gap1_general}}

When $\mix^{(t-1)}$ and $\mix^{(t)}$ do the same action at time $t$, there is no gap between them since they follow the same policy after time $t$. Therefore, we only need to consider two cases at time $t$: (I) $\mix^{(t)}$ accepts, $\mix^{(t-1)}$ has $0<X^{(t)}<1$; (II) $\mix^{(t)}$ rejects, $\mix^{(t-1)}$ has $0<X^{(t)}<1$. In this case, note that either our Threshold policy or the hindsight optimal policy will always fully accept arrivals with non-positive weights, WLOG we assume $a^{(t)} > 0$. 

{\small\it\indent Case I: $\mix^{(t)}$ rejects, $\mix^{(t-1)}$ has $X^{(t)}>0$.} In this case, since the budget $B^{(t+1)}= B^{(t)}-X^{(t)}a^{(t)}<B^{(t)}$ for $\mix^{(t-1)}$ and $B^{(t+1)}=B^{(t)}$ for $\mix^{(t)}$, in the remaining time $t+1,\dots,T$, the one starting from $B^{(t+1)}$in $\mix^{(t)}$ can always choose the same $X^{(l)},l\ge t+1$ as that in $\mix^{(t-1)}$. Hence the gap in $\pow(\mix^{(t-1)}|\W^T)-\pow(\mix^{(t)}|\W^T)$ can only generate by the wrongly rejection of $a^{(t)}$. Hence
\begin{equation*}
\begin{aligned}
    \pow(\mix^{(t-1)}|\W^T)-\pow(\mix^{(t)}|\W^T)\le r^{(t)}\le\rmax.
\end{aligned}
\end{equation*}

{\small\it\indent Case II: $\mix^{(t)}$ accepts, $\mix^{(t-1)}$ has $X^{(t)}<1$.} In this case, following the similar construction strategy above, starting from $B^{(t+1)}$ generated by $\mix^{(t)},$ one can slightly modify $\mix^{(t-1)}$ as follows: suppose $X^{(l)}>0$ in $\mix^{(t-1)}$ for the coming arrivals with $a^{(l)}/r^{(l)}\ge a^{(t)}/r^{(t)}$. One can set the decision maker's action at time $l$ as $Y^{(l)} = \max\{0,\sum_{s\le l,a^{(s)}/r^{(s)}\ge a^{(t)}/r^{(t)}}^{l-1}Y^{(l)}a^{(l)}-a^{(t)}\}/a^{(l)} $ until the first time $l:\sum_{s\le l,a^{(s)}/r^{(s)}\ge a^{(t)}/r^{(t)}}^l Y^{(l)}a^{(l)}\ge a^{(t)}$ (If no such $l$ exists the proof is done). Then by definition the first one now has budget now less than the hindsight optimal one in $\mix^{(t-1)}$. Then it can follow the same action as the later and the gap
\begin{equation*}
\begin{aligned}
    \pow(\mix^{(t-1)}|\W^T)-\pow(\mix^{(t)}|\W^T)\le L,
\end{aligned}
\end{equation*}
generated by the adjustment until time $l$ (which costs buffer of size $a^{(t)}$ and will generate loss of size at most $\max_{i}(r_i/a_i)a^{(t)}$) minus the one extra acception of $a^{(t)}$.

\subsection{Proof of Lemma \ref{lem:there_is_gap2_general}}
We follow the similar streamline in the proof of Lemma \ref{lem:there_is_gap1_general}. We only need to consider two cases at time $t$: (I) $\mix^{(t-1)}$ has $X^{(t)}>0$, $\mix^{(t)}$ rejects; (II) $\mix^{(t-1)}$ has $X^{(t)}<1$, $\mix^{(t)}$ accepts. In this case, note that either our policy or the hindsight optimal policy will never reject arrivals with non-positive weights, WLOG we assume $a^{(t)} > 0$. For notation brevity, we will hide $\W^T$ in $f(\cdot)$, but keep in mind that $f(\cdot)$ is dependent on the sample path $\W^T$.

{\small\it\indent Case I: $\mix^{(t-1)}$ has $X^{(t)}<1$, $\mix^{(t)}$ accepts.} In this case, we have to prove that 
\begin{equation}
    \label{eq:prob1_general}
    \begin{aligned}
        &\quadd\P\prn{\pow(\mix^{(t-1)})> \pow(\mix^{(t)})\big\vert a^{(t)}=a_i}\\ 
        &\le \P\prn{B^{(t)} - a_i<\max_{s\in[t+1,T]}\sum_{r=t+1}^sa^{(l)}\mathds 1\{a^{(l)}/r^{(l)}<a_i/r_i\}, B^{(t)}\geq \Buffer_i}.
    \end{aligned}
\end{equation}
It is enough to show that, when the event $\{\pow(\mix^{(t-1)})> \pow(\mix^{(t)}), \mix^{(t)}\text{ accepts }a^{(t)}\big\vert a^{(t)}=a_i\}$ happens, at least one of the equations
\begin{equation*}
    \begin{aligned}
        B^{(t)}-a_i<\sum_{r=t+1}^sa^{(l)}\mathds 1\{a^{(l)}/r^{(l)}<a_i/r_i\},s\in[t+1,T]
    \end{aligned}
\end{equation*}
holds. Otherwise, consider the strategy induced by $\mix^{(t-1)}$. Denote $t'+1\ge t+1$ as the first time that $\mix^{(t-1)}$ accumulates $\sum_{l=t,a^{(l)}/r^{(l)}\ge a_i/r_i}^{t'+1}X^{(l)}a^{(l)}\ge a_i$ (if $\mix^{(t-1)}$ never reaches such status, we take $t'=T$). Then we have
\begin{equation*}
    \max_{s\in[t+1,t']} \sum_{r=t+1}^sa^{(l)}\mathds 1\{\mix^{(t-1)} \text{ accepts }a^{(l)}\} \leq a_i +\max_{s\in[t+1,t']} \sum_{r=t+1}^sa^{(l)}\mathds 1\{a^{(l)}/r^{(l)}<a_i/r_i\} \leq B^{(t)}.
\end{equation*}
Now given $a^{(t)}=a_i$, the decision maker can make the same decision as $\mix^{(t-1)}$ from time $t+1$ to $t'-1$ and set action as $Y^{(s)}=\max\{0,\sum_{l=t}^{s-1}X^{(l)}a^{(l)}-a_i\}/a^{(s)}$ for all $s$ such that $a^{(s)}/r^{(s)}\ge a_i/r_i$ and $s\ge t'+1$. Such a policy is valid because the any-time constraints before and at time $t'$ are guaranteed by the equation above, and the any-time constraints after time $t'$ are guaranteed by the fact that $a^{(t'+1)}\geq a_i$. A contradiction. Therefore, we have 
\begin{equation*}
    \begin{aligned}
        &\quadd \P\prn{\pow(\mix^{(t-1)})> \pow(\mix^{(t)}), \mix^{(t)} \text{ accepts }a^{(t)}\big\vert a^{(t)}=a_i}\\ 
        &\le \P\prn{ B^{(t)} - a_i < \max_{s\in[t+1,t']} \sum_{r=t+1}^sa^{(l)}\mathds 1\{a^{(l)}/r^{(l)} < a_i/r_i\}, B^{(t)}\geq\Buffer_i \big\vert a^{(t)}=a_i} \\ 
        & = \P\prn{B^{(t)} - a_i < \max_{s\in[t+1,t']} \sum_{r=t+1}^sa^{(l)}\mathds 1\{a^{(l)}/r^{(l)} < a_i/r_i\}, B^{(t)}\geq\Buffer_i}
    \end{aligned}
\end{equation*}
In the last equality we use the fact that $a^{(t)}$ is independent of $B^{(t)}$ and $a^{(s)}\ (\forall s > t)$.

{\small\it\indent Case II: $\mix^{(t-1)}$ has $X^{(t)}>0$, $\mix^{(t)}$ rejects.} In this case, we have to prove that 
\begin{equation}
    \label{eq:prob2_general}
    \begin{aligned}
        &\quadd \P\prn{\pow(\mix^{(t-1)})> \pow(\mix^{(t)})\big\vert a^{(t)}=a_i}\\ 
        &\le \P\prn{B^{(t)} - a_i\ge\max_{s\in[t+1,T]}\sum_{r=t+1}^sa^{(l)}\mathds 1\{a^{(l)}/r^{(l)}\le a_i/r_i\}, B^{(t)}<\Buffer_i}.
    \end{aligned}
\end{equation}
We show that, when there is a gap generated by rejecting $a^{(t)}=a_i$, then the hindsight optimal policy from $t$ to $T$ in $\mix^{(t-1)}$ will set $X^{(s)}=1$ for $a^{(s)}/r^{(s)}\le a_i/r_i,\forall s\in[t,T]$. Otherwise, assume that $t'>t$ is the first time the hindsight optimal $\mix^{(t-1)}$ set $\sum_{l=t+1,a^{(l)}/r^{(l)}\le a_i/r_i}^{t'}(1-X^{(l)})a^{(l)}\ge a_i$. If no such $t'$ we set $t'=T$. Knowing this, we can construct a new offline strategy by following the same decisions with $\mix^{(t-1)}$ except setting actions $Y^{(s)} = 1-\max\{0,\sum_{l=t+1,a^{(l)}/r^{(l)}\le a_i/r_i}^{t'}(1-X^{(l)})a^{(l)}-a_i\}/a^{(s)}$ for $s\in [t+1,t'] $ such that $a^{(s)}/r^{(s)}\ge a_i/r_i$. By definition of $Y^{(s)}$ the actions are always valid. After time $t'$ the new offline strategy can follow the same policy as $\mix^{(t-1)}$. There will be no gap, a contradiction. Therefore, $\mix^{(t-1)}$ cannot reject arrivals $a^{(s)}/r^{(s)}\le a_i/r_i$ of size more than buffer $a_i$ and it holds that 
\begin{equation*}
    \begin{aligned}
        B^{(t)} - a_i\ge\max_{s\in[t+1,T]}\sum_{r=t+1}^sa^{(l)}\mathds 1\{a^{(l)}/r^{(l)}\le a_i/r_i\}
    \end{aligned}.
\end{equation*}

Therefore, we have
\begin{equation*}
    \begin{aligned}
        &\quadd \P\prn{\pow(\mix^{(t-1)})> \pow(\mix^{(t)}), \mix^{(t)} \text{ rejects }a^{(t)}\big\vert a^{(t)}=a_i}\\ 
        &\le \P\prn{ B^{(t)} - a_i \geq \max_{s\in[t+1,t']} \sum_{r=t+1}^sa^{(l)}\mathds 1\{a^{(l)}/r^{(l)} \leq a_i/r_i\}, B^{(t)}<\Buffer_i \big\vert a^{(t)}=a_i} \\ 
        & = \P\prn{B^{(t)} - a_i \ge \max_{s\in[t+1,t']} \sum_{r=t+1}^sa^{(l)}\mathds 1\{a^{(l)}/r^{(l)} \le a_i/r_i\}, B^{(t)}<\Buffer_i}
    \end{aligned}
\end{equation*}
In the last equality we again use the fact that $a^{(t)}$ is independent of $B^{(t)}$ and $a^{(s)}\ (\forall s > t)$.

\subsection{Proof of Lemma \ref{lem:distance1}}
We consider two cases: $\Lambda^{[1,T]}(-1|\W^T) \ge \Lambda^{[1,T]}(1|\W^T)$ and $\Lambda^{[1,T]}(-1|\W^T) < \Lambda^{[1,T]}(1|\W^T).$ To begin with, we point out the basic fact that $\pow(\HOany|\W^T) = T - S_r(T)$, because under the simple random walk with ``wall'', the only rejection happens when the walker is stopped by the wall at zero.

\paragraph{Case I: $\Lambda^{[1,T]}(-1|\W^T) \ge \Lambda^{[1,T]}(1|\W^T)$.} 
\begin{figure}[ht]
\begin{tikzpicture} \small
    % Draw axes
    \draw[thin,->] (3,0) -- (-10,0) node[anchor=east] {Position};
    \node at (-2,0) (0) [anchor=north]{0};
    \draw[thick, blue] (3,0) -- (2,0) -- (1,0) -- (0,0) -- (-1,0) -- (-2,0) -- (-3,0) -- (-4,0) -- (-5,0)-- (-6,0)  -- (-7,0) -- (-8,0) -- (-9,0)  -- (-10,0);
    \foreach \Point in {(3,0), (2,0), (1,0), (0,0), (-1,0), (-2,0), (-3,0), (-4,0), (-5,0), (-6,0), (-7,0), (-8,0), (-9,0), (-10,0)}{
        \node at \Point {\textbullet};
    }

    \node at (1,0.5)[font=\large] (A) {$\overset{\HOfix}{\uparrow}$};

    \node at (-6,0.5)[font=\large] (B) {$\overset{\HOany}{\uparrow}$};

    \node at (-6,-0.5) (C) {};
    \node at (-2,-0.5) (D) {};
    \node at (1,-0.5) (E) {};

    \draw[-{Latex[length=3mm,width=3mm]}] (B) -- (A);

    \draw [decorate, decoration={brace, amplitude=10pt, mirror, raise=4pt}]
    (A.north) -- (B.north) node[anchor=south,midway, yshift=10pt] {$S_r(T)$};

    \draw [decorate, decoration={brace, amplitude=10pt, mirror, raise=4pt}]
    (C.south) -- (D.south) node[anchor=north,midway, yshift=-10pt] { $D_r(T)$};

    \draw [decorate, decoration={brace, amplitude=10pt, mirror, raise=4pt}]
    (D.south) -- (E.south) node[anchor=north,midway, yshift=-10pt] { $\Lambda^{[1,T]}(1|\W^T)-\Lambda^{[1,T]}(-1|\W^T)$};

\end{tikzpicture}
\caption{\centering Random walk when $\Lambda^{[1,T]}(1|\W^T)>\Lambda^{[1,T]}(-1|\W^T)$\label{fig:segment1}.}
\end{figure}

We use coupling to prove the result. Consider two walkers $A,F$ starting from zero at time $0$, representing the policy \HOany,\HOfix, respectively. We then generate sample path $W{(T)}=(w^{(1)},\dots,w^{(T)})\in\{-1,+1\}^{T}$. Both walkers try to go right for one step at time $t$ if $w^{(t)}=1$ and go left otherwise. However, there is a wall at zero for $A$ and it must stay at zero when it aims to go right at zero. For $t\in[T]$, denote $Y_\HOany^{(t)}, Y_\HOfix^{(t)}$ as their position at time $t$. It follows that $Y_\HOfix^{(T)}= \Lambda^{[1,T]}(1|\W^T)-\Lambda^{[1,T]}(-1|\W^T)$ and $Y_\HOany^{(T)}=-D_r(T)$. In this case, note that $Y_\HOfix^{(t)}-Y_\HOany^{(t)}$ is nondecreasing with $t$ by definition. The event $Y_\HOfix^{(t+1)}-Y_\HOany^{(t+1)} = Y_\HOfix^{(t)}-Y_\HOany^{(t)}+1$ happens if and only if $Y_\HOany^{(t)} = 0$ and $w^{(t+1)}=1$. Therefore, we have 
\begin{equation*}
    Y_\HOfix^{(T)}-Y_\HOany^{(T)} = \sum_{t=0}^{T-1}\mathds 1\{Y_\HOany^{(t)}=0,w^{(t+1)=1}\} =S_r(T)
\end{equation*}
by definition of $S_r(T)$. Hence, $S_r(T)-D_r(T)=\Lambda^{[1,T]}(1|\W^T)-\Lambda^{[1,T]}(-1|\W^T)\ge 0$. Furthermore, note that $\pow(\HOfix|\W^T)=T- (\Lambda^{[1,T]}(1|\W^T)-\Lambda^{[1,T]}(-1|\W^T)) $ when $\Lambda^{[1,T]}(1|\W^T)\ge\Lambda^{[1,T]}(-1|\W^T)$. Then it follows that 
\begin{equation*}
    \begin{aligned}
        &\quadd \pow(\HOfix|\W^T)-\pow(\HOany|\W^T) \\
        &= S_r(T)-(\Lambda^{[1,T]}(1|\W^T)-\Lambda^{[1,T]}(-1|\W^T)) \\
        &= D_r(T) = \min\{D_r(T),S_r(T)\}.
    \end{aligned}.
\end{equation*}

\paragraph{Case II: $\Lambda^{[1,T]}(1|\W^T)<\Lambda^{[1,T]}(-1|\W^T) $.}

\begin{figure}[ht]
\begin{tikzpicture} \small
    % Draw axes
    \draw[thin,->] (3,0) -- (-10,0) node[anchor=east] {Position};
    \node at (-2,0) (0) [anchor=north]{0};

    \draw[thick, blue] (3,0) -- (2,0) -- (1,0) -- (0,0) -- (-1,0) -- (-2,0) -- (-3,0) -- (-4,0) -- (-5,0)-- (-6,0)  -- (-7,0) -- (-8,0) -- (-9,0)  -- (-10,0);
    \foreach \Point in {(3,0), (2,0), (1,0), (0,0), (-1,0), (-2,0), (-3,0), (-4,0), (-5,0), (-6,0), (-7,0), (-8,0), (-9,0), (-10,0)}{
        \node at \Point {\textbullet};
    }

    \node at (-3,0.5)[font=\large] (A) {$\overset{\HOfix}{\uparrow}$};

    \node at (-8,0.5)[font=\large] (B) {$\overset{\HOany}{\uparrow}$};

    \node at (-8,-0.5) (C) {};
    \node at (-3,-0.5) (D) {};
    \node at (-2,-0.5) (E) {};
    \node at (-2,0.82) (F) {};

    % \draw[->,linewidth=1mm] (B) -- (A);?
    \draw[-{Latex[length=3mm,width=3mm]}] (B) -- (A);

    \draw [decorate, decoration={brace, amplitude=10pt, mirror, raise=4pt}]
    (A.north) -- (B.north) node[anchor=south,midway, yshift=10pt] {$S_r(T)$};

    \draw [decorate, decoration={brace, amplitude=10pt, mirror, raise=4pt}]
    (C.south) -- (E.south) node[anchor=north,midway, yshift=-10pt] { $D_r(T)$};

    \draw [decorate, decoration={brace, amplitude=10pt, mirror, raise=4pt}]
    (F.north) -- (A.north) node[anchor=north,midway, yshift=40pt] { $\Lambda^{[1,T]}(-1|\W^T)-\Lambda^{[1,T]}(1|\W^T)$};

\end{tikzpicture}
\caption{\centering Random walk when $\Lambda^{[1,T]}(1|\W^T)<\Lambda^{[1,T]}(-1|\W^T)$\label{fig:segment2}.}
\end{figure}

In this case, we do the same coupling and following the same deduction, we get
\begin{equation*}
    S_r(T)-D_r(T)=\Lambda^{[1,T]}(-|\W^T)-\Lambda^{[1,T]}(-1|\W^T) <0.
\end{equation*}
Note that $\pow(\HOfix|\W^T) = T $ in this case, we get 
\begin{equation*}
    \pow(\HOfix|\W^T) - \pow(\HOany|\W^T) = S_r(T) = \min\{S_r(T),D_r(T)\}.
\end{equation*}
Combining the results above completes the proof.

\subsection{Proof of Lemma \ref{lem:large_deviation}}
It suffices to prove the first inequality. Let $S_t = \sum_{s=1}^t(\xi_s - \epsilon)$. We first show that $M_t = \exp\prn{2\epsilon S_t}$ is a super-martingale with $\mathcal F_t = \sigma(\xi_1, \cdots, \xi_t)$. In fact, 
\begin{align*}
    \ex{}{M_t|\mathcal F_{t-1}} & = M_{t-1}\cdot\ex{}{\exp\prn{2\epsilon(\xi_t-\epsilon)}} \\
    & \leq M_{t-1}\exp\prn{-2\epsilon^2}\cdot\ex{}{\exp\prn{\frac{1}{2}(2\epsilon)^2\cdot\frac{(1-(-1))^2}{4}}} \\
    & = M_{t-1}.
\end{align*}
Here, in the inequality we use the fact that a random variable bounded by $[a, b]$ is $\frac{(b-a)^2}{4}$-subGaussian. Define $\tau$ as the stopping time that $S_t$ first arrives at or above $B$. It suffices to bound $\P(\tau<+\infty)$. By optional sampling theorem, for any $t\geq 1$, we have
\begin{align*}
    1 = \ex{}{M_{\tau\wedge 0}} \geq \ex{}{M_{\tau\wedge t}} \geq \P(\tau\leq t)\cdot\exp\prn{2\epsilon B}.
\end{align*}
Since $t$ can be arbitrary, we can get $\P(\tau<+\infty) \leq \exp\prn{-2B\epsilon}$.

\subsection{Proof of Lemma \ref{lem:large_deviation_2}}
It suffices to prove the first inequality. Let $S_t = \sum_{s=1}^t\xi_s$. We first show that $M_t = \exp\prn{\epsilon S_t - \epsilon^2t/2}$ is a super-martingale with $\mathcal F_t = \sigma(\xi_1, \cdots, \xi_t)$. In fact, 
\begin{align*}
    \ex{}{M_t|\mathcal F_{t-1}} & = M_{t-1}\cdot\ex{}{\exp\prn{\epsilon\xi_t-\frac{\epsilon^2}{2}}} \\
    & \leq M_{t-1}\exp\prn{-\frac{\epsilon^2}{2}}\cdot\ex{}{\exp\prn{\frac{1}{2}\epsilon^2\cdot\frac{(1-(-1))^2}{4}}} \\
    & = M_{t-1}.
\end{align*}
Here, in the inequality we use the fact that a random variable bounded by $[a, b]$ is $\frac{(b-a)^2}{4}$-subGaussian. Define $\tau$ as the stopping time that $S_s$ first arrives at or above $\epsilon t$. It suffices to bound $\P(\tau\leq t)$. By optional sampling theorem, for any $t\geq 1$, we have
\begin{align*}
    1 = \ex{}{M_{\tau\wedge 0}} \geq \ex{}{M_{\tau\wedge t}} \geq \P(\tau\leq t)\cdot\exp\prn{\epsilon^2t - \frac{\epsilon^2}{2}t} \geq \P(\tau\leq t)\cdot\exp\prn{\frac{\epsilon^2}{2}t}.
\end{align*}
Thus, we can get $\P(\tau\leq t) \leq \exp\prn{-\epsilon^2t/2}$.

\section{More Experiments}

\subsection{SG Policy in the Continuous Case}
We first examine the performance of the SG policy proposed in Algorithm \ref{alg:static}. We compare the SG policy with the greedy policy which always accepts an arrival whenever there is enough budget (i.e., whenever $B^{(t)}\geq a^{(t)}$). We take the distribution for the $\alpha$-cost $a^{(t)}$ to be a uniform distribution in $[-0.05, 0.95]$, which represents the case when we require the local FDR to be less than $\alpha=0.05$ and the posterior null probability is uniform in $[0,1]$. We plot the regret of the each policy averaged across 100 sample paths with respect to the $\DLP$ upper bound. The result is shown in Figure \ref{fig:cont_upper}.
\begin{figure}[ht]
    \centering
    \includegraphics[width=0.65\linewidth]{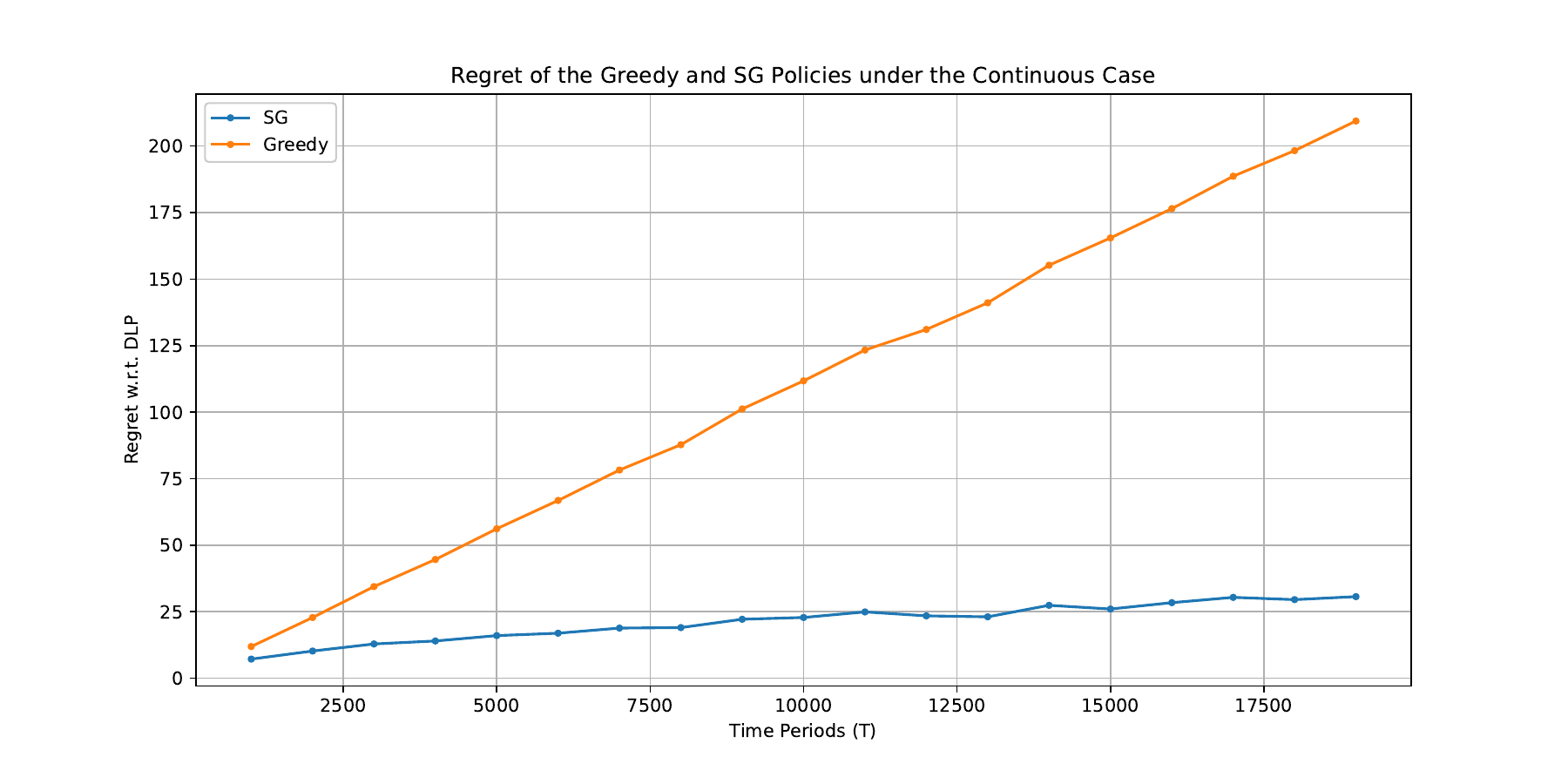}
    \caption{\centering Regret with respect to $\DLP$ for the Greedy and $\SG$ policies.}
    \label{fig:cont_upper}
\end{figure}
As one can see from Figure \ref{fig:cont_upper}, the $\SG$ policy achieves a significant lower regret compared to the greedy policy. Moreover, while the regret of the greedy policy grows linearly with time horizon $T$, the regret of the $\SG$ policy grows much slower, which echoes with Theorem \ref{thm:continuous-upper} that the regret upper bound should grow at a $O(\sqrt{T})$ rate.

\subsection{Lower Bound Validation}
Finally, we conduct a small experiment to validate the lower bound we provide in Theorem \ref{thm:continuous-lower}. In particular, for each distribution $\cD_T$ indexed by $T$, we let it take values in $[-1, 1, 3]$ with probability $[1/2+1/\sqrt{T}, 1/2-2/\sqrt{T}, 1/\sqrt{T}]$. Then we calculate the gap between the optimal dynamic programming solution \eqref{eq:DP} and the \eqref{eq:HOany_discrete} upper bound. Essentially, $\DP$ is the best online policy one can get without referring to the future information, and so the gap demonstrate the inherent difficulty of the online problem with respect to the offline counterpart. We take the average of 1000 sample paths and plot the log-log plot of the gap relative to the time horizon in Figure \ref{fig:lower_bound}. From Figure \ref{fig:lower_bound}, we can directly see that the $\log(\mbox{Regret})$ and the $\log t$ grows in a linear fashion with the slope being roughly $1/2$. This validates our proof in Theorem \ref{thm:continuous-lower} that there exists a sequence of distribution such that the gap between \eqref{eq:DP} and \eqref{eq:HOany_discrete} is $\Omega(\sqrt{T})$.

\begin{figure}[ht]
    \centering
    \includegraphics[width=0.65\linewidth]{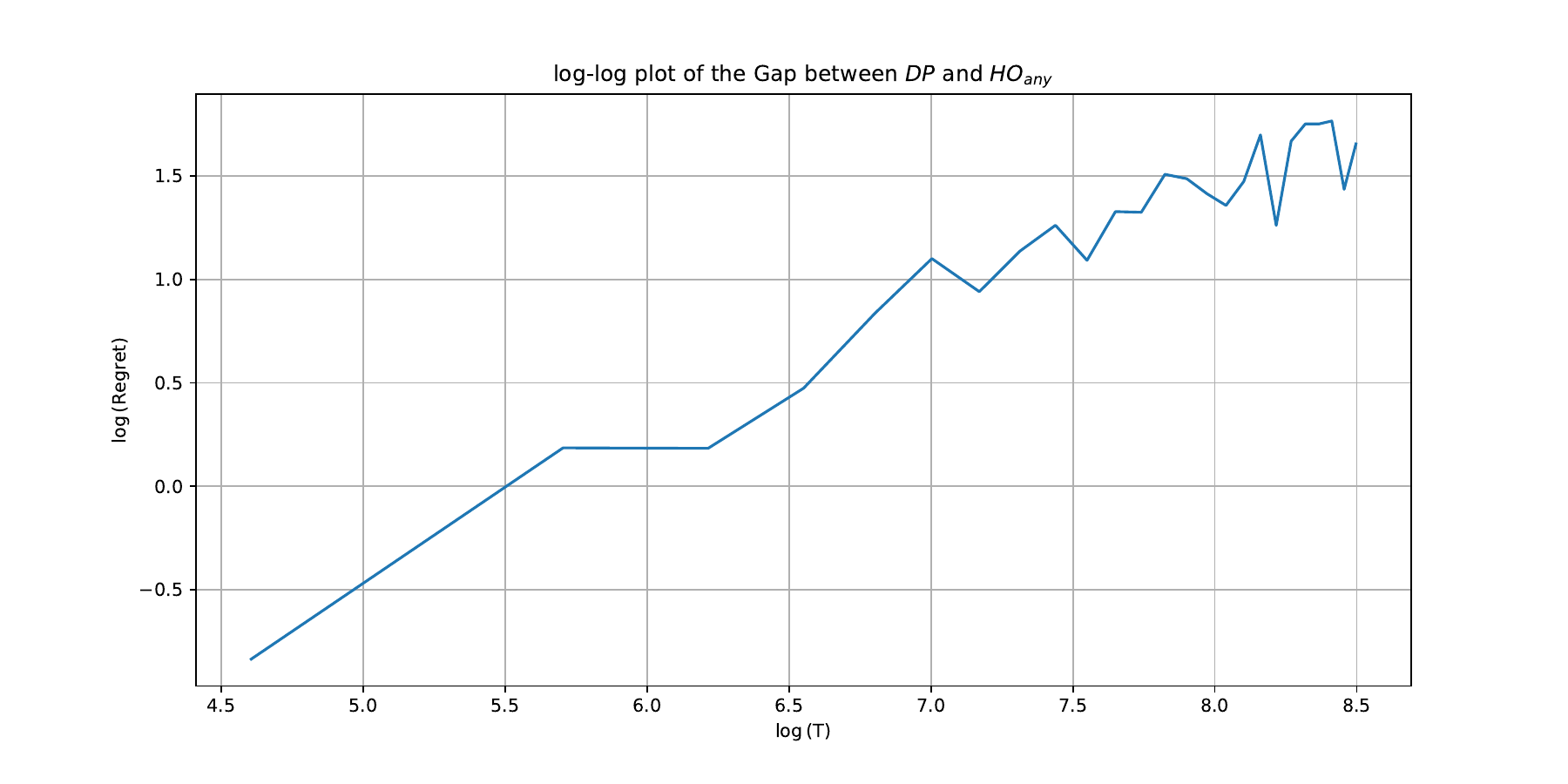}
    \caption{Gap between \eqref{eq:HOany_discrete} and \eqref{eq:DP}. Here, the $T$th distribution is $\cD_T$, which takes values in $[-1, 1, 3]$ with probability $[1/2+1/\sqrt{T}, 1/2-2/\sqrt{T}, 1/\sqrt{T}]$.}
    \label{fig:lower_bound}
\end{figure}

\end{APPENDICES}
%
%   or
%
% \begin{APPENDICES}
% \section{<Title of Section A>}
% \section{<Title of Section B>}
% etc
% \end{APPENDICES}

%%

%%%%%%%%%%%%%%%%%
\end{document}